\documentclass[aps,preprint,preprintnumbers,amsmath,amssymb,floatfix,noshowpacs]{revtex4}

\usepackage{graphicx}
\usepackage{epsfig}

\begin{document}
\title{
Analytical and numerical calculations of spectral and optical characteristics of spheroidal quantum dots
\footnote{Submitted to Physics of Atomic Nuclei}}

\author{\firstname{A.A.}~\surname{Gusev}}\email{gooseff@jinr.ru}
\affiliation{Joint Institute for Nuclear Research, Dubna, Russia}
\author{\firstname{L.L.}~\surname{Hai}}
\affiliation{Joint Institute for Nuclear Research, Dubna, Russia}
\author{\firstname{S.I.}~\surname{Vinitsky}}
\affiliation{Joint Institute for Nuclear Research, Dubna, Russia}
\author{\firstname{O.}~\surname{Chuluunbaatar}}
\affiliation{Joint Institute for Nuclear Research, Dubna, Russia}
\author{\firstname{V.L.}~\surname{Derbov}}
\affiliation{Saratov State University, Saratov, Russia}
\author{\firstname{A.S.}~\surname{Klombotskaya}}
\affiliation{Saratov State University, Saratov, Russia}
\author{\firstname{K.G.}~\surname{Dvoyan}}
\affiliation{Russian-Armenian (Slavonic) University, Yerevan, Armenia}
\author{\firstname{H.A.}~\surname{Sarkisyan}}
\affiliation{Russian-Armenian (Slavonic) University, Yerevan, Armenia}

\begin{abstract}
In the effective mass approximation for electronic (hole) states of
a spheroidal quantum dot with and without external fields the
perturbation theory schemes are constructed in the framework of the
Kantorovich and  adiabatic methods. The eigenvalues and eigenfunctions
of the problem, obtained in both analytical and numerical forms, were applied for
the analysis of spectral and optical characteristics of spheroidal quantum dots in
homogeneous electric fields.
\end{abstract}

\maketitle

\section{Introduction}
Quantum dots(QDs) are considered to be promising as the elementary basis for the new generation of semiconductor devices \cite{1A,Harrison}. The unique opportunity to perform the energy level control and flexible manipulation in QDs is due to the full quantization of charge carrier energy spectra in these systems.  This allows design and manufacturing of artificial structures with prescribed quantum physical characteristics  \cite{2A}. That is why the scope of QDs potential applications is very wide,  from heterostructure lasers to nanomedicine and nanobiology. An impressive example of such application is represented by
 QD lasers possessing low threshold
current and  high efficiency \cite{2A}.

The peculiarities of physical processes in QDs are caused by both their
composition and geometry.
Electronic, kinetic, optical and other properties of QDs have been
investigated experimentally and theoretically in many papers
 \cite{11A,LL94,Hayk02,79a,79,13A,14A,15,Suslov1,Suslov2}.
Particularly, the optical
absorption characteristics of QDs have been shown to be strongly correlated with their geometry, on one hand, and with their
physical--chemical
properties, on the other hand.  In one of the first publications on optical transitions in QD \cite{Efros1982} the
interband absorption of light was considered in the
ensemble of weakly interacting spherical QDs implanted in a
dielectric matrix. The dispersion of QD sizes was characterized
in the framework of Lifshitz--Slezov theory \cite{LS1958}. It was shown that in
the  absence of size dispersion, due to the full quantization
of charge carriers energy spectra in QD, the absorption coefficient behaves like a delta function,  and the absorption threshold frequencies
depend on the peculiarities of electron and hole
energy spectra. When the QD size dispersion is taken into account, the averaging procedure yields the absorption profile having
 finite width and height.

{Recently several reports concerning the experimental implementation of narrow-band InSb QDs have appeared \cite{17A,18A}, in which the
dispersion law for electrons and light holes is non-parabolic and
described according to the double-band mirror Kane model \cite{Kane,Askerov}.
 For non-interacting band of heavy holes the dispersion law is considered as quadratic.
The investigation of optical absorption
peculiarities  in InSb QDs with the transitions from light and heavy hole bands to the conduction band taken  into
account is an interesting problem.
Interband transitions in an ensemble of  cylindrical or  spherical InSb   QDs  were   considered theoretically  in the  dipole  approximation with and without magnetic field, including exciton effects, by means of the
perturbation theory and the adiabatic methods\cite{20A,21A,Hayk11}.}
 In our earlier work we elaborated the calculation schemes, symbolic-numerical algorithms (SNAs) and programs,
 based on the generalized Kantorovich method (KM)
 for numerical solving with required accuracy the boundary-value problems (BVPs)
 of discrete and continuous spectra describing the
 axial-symmetric models of quantum wells(QWs), quantum wires(QWrs) and quantum dots(QDs) in external fields  within the framework of the effective mass approximation
 ~\cite{Yu2,CASC10,JPCONF,Yaf10,Yaf12,kantbp,ODPEVP,Yu3,Yu4,Yu5,Yu8,Yu9,progr07}.
Meanwhile, for the analysis and estimations of the appropriate range of
material parameters, spectral and optical characteristic of quantum
dots  at the first stage of investigation, conventionally,  approximate
eigenvalues and eigenfunctions evaluated in the analytical form were
applied ~\cite{Efros1982,Hayk02,Hayk11,79,79a}.
However, it is a real challenge
to specify the range of applicability of such approximations in the
problems, depending on a few parameters \cite{Harrison}, e.g.,  for
impurity states of quantum wires in a homogeneous magnetic field
~\cite{JPCONF}.

With this aim in the present paper we report the formulation and MAPLE- environment implementation of
 algebraic schemes of the perturbation theory (PT)
of the Lennard-Jones (LJ) and Rayleigh-Schr\"odinger (RS) \cite{MottSneddon},
permissive in the nondiagonal and diagonal adiabatic approximations,
respectively, to evaluate  in numerical and in analytic forms the
eigenvalues and eigenfunctions of models of spheroidal QDs in
homogeneous magnetic and electric fields. To construct the required
perturbation schemes, we choose such models of spheroidal QDs, in which
the basis functions depending upon fast variables can be expressed in the analytic form.
The region of the model parameters, for which the PT asymptotic
series are applied, is estimated using the results of numerical calculations
carried out with required accuracy.
The efficiency of the schemes is
demonstrated by the analysis of spectral characteristics of oblate and
prolate spheroidal QDs and also spherical QDs with corresponding
shape of confinement well with walls of infinite height under the
influence of homogeneous electric fields (HEFs).
We apply the developed approach
to the analysis of spectral characteristics of oblate and prolate
spheroidal QDs with parabolic and non-parabolic dispersion laws
under the influence of
HEFs, i.e., the quantum-confined Stark effect.

The paper is organized as follows. In Section 2 the calculation
scheme for solving elliptic BVP describing  spheroidal QDs in
homogeneous electric fields using the Kantorovich method is
presented. Section 3 is devoted to the description of the PT schemes by
slow variables
in nondiagonal adiabatic approximation and the comparison of the results with those of numerical  calculation with given accuracy.
In section 4  the explicit PT scheme for evaluation of the basis
functions of the fast variable for oblate spheroidal QDs in a
homogeneous electric field is derived. Section 5 is devoted to
the description of PT schemes by slow variables in the diagonal
adiabatic approximation for spheroidal QDs in electric fields.
The results evaluated here in the analytic form are compared with
numerical ones to establish the range of their applicability.
In Section 6 the absorption coefficient for an ensemble of spheroidal QDs with random dimensions of minor semiaxis
and with parabolic and non-parabolic dispersion laws for holes and electrons
under the influence of
HEFs is found using the calculated eigenvalues and eigenfunctions.
In
conclusion we summarize the results and discuss further
applications.

\section{
Statement of the problem
}

Let us consider an impurity localized in the center of a quantum dot and take the electron-hole interaction into account. Then in the effective mass approximation of the
${\bf{k}} \cdot {\bf{p}}$-theory the Schr\"odinger equation for  the {slow}-varying envelope
wave function
$\tilde{\Psi}( \tilde {\mathbf{r}_e},{\tilde {\mathbf{r}_h}})$
of an electron (e) and a hole (h)
in a uniform magnetic field $\bf{H}$
with the vector-potential
${\mathbf{A}}=\frac{1}{2}{\mathbf{H}}\times\tilde{{\mathbf{r}}}$
and electric field
$\mathbf{F}$
in \textit{oblate} and \textit{prolate} QDs reads as \cite{79}:

\begin{eqnarray}&&
\label{sp01} \left\{\tilde H( \tilde {\mathbf{r}_e},{\tilde {\mathbf{r}_h}})-{\tilde E} \right\}
\tilde{\Psi}( \tilde {\mathbf{r}_e},{\tilde {\mathbf{r}_h}})=0,\\&&\nonumber
\tilde H( \tilde {\mathbf{r}_e},{\tilde {\mathbf{r}_h}})= \sum_{i=e,h}\left\{
\frac{1}{2\mu_i }
\left( \tilde {\hat {\mathbf{p}_i}} -\frac{q_i}{c}{\mathbf {A}}
\right)^2 -
{q_{i}} (\mathbf{F}\cdot \tilde{ \mathbf{r}_i})
+ \tilde {U}_{conf}(\tilde {\mathbf{r}_i})
- \frac{q_iq_c}{\kappa |{\mathbf{\tilde r}_i}|}\right\}
+ \frac{q_eq_h}{\kappa |{\mathbf{\tilde r}_e-\mathbf{\tilde r}_h}|}
   .
\end{eqnarray}
Here
$\tilde{{\mathbf{r}_i}}$
is the radius-vector,
$|\tilde{{\mathbf{r}_i}}|=\sqrt{\tilde{x_i}^2+\tilde{y_i}^2+\tilde{z_i}^2}$,
$ \tilde{\hat {{\bf p}_i}}=-\imath \hbar \nabla_{\tilde{{\mathbf{r}_i}}}$
is the momentum, $\tilde{E}$ is the energy of the particles,
  $q_e=-e$, $q_h=+e$, and $q_c$ are the Coulomb charges of the electron, the  hole,
and the impurity center,
 $\kappa $ is the dc permittivity,
$\mu_{i} = \beta_{e(h)} m_0$ is the effective mass of electron
or hole, $m_{0}$ is the mass of electron. For the model under consideration,
$\tilde U (\tilde {\bf{r}})$
is the potential of a spherical or axially-symmetric well
\begin{eqnarray}
\label{sp03}
\tilde U (\tilde {{\bf r}_i})=\{0,S(\tilde {{\bf r}_i}) < 0 ;
\tilde U_0, S(\tilde {{\bf r}_i}) \ge 0 \},
\end{eqnarray}
bounded by the surface
$S(\tilde {\bf{r}_i})=0$
with  walls of infinite height (infinite potential barrier model,
IPBM) or finite height $1\ll \tilde U_0 <\infty$ {(finite potential barrier model,
FPBM)}.
In Eq. (\ref{sp03})
$S(\tilde {{\bf r}_i})$
depends on the parameters
 $\tilde a$, $\tilde c$, which are semiaxes of a spheroidal QD,
\begin{eqnarray}   \label{eq99a}
S(\tilde {{\bf r}_i})
\equiv ({\tilde x_i^2+\tilde y_i^2})/{\tilde a^2}
+ {\tilde z_i^2}/{\tilde c^2}-1.
\end{eqnarray}

Below we restrict ourselves to IPBMs of  spheroidal quantum dots
with possible influence of the
uniform electric field ${\bf F}=(0,0,F)$, the magnetic field being switched off, ${\bf H}=0$,
and the Coulomb interaction of the electron and the hole with the impurity center being absent, $q_{c}=0$.
In this case the wave function $\tilde{\Psi}( \tilde {\mathbf{r}_e},{\tilde {\mathbf{r}_h}})=
\tilde{\Psi}^e( \tilde {\mathbf{r}_e})\tilde{\Psi}^h({\tilde {\mathbf{r}_h}})$ is factorized.
So, we arrive at the 3D BVPs for unknowns
$\tilde{\Psi}^e( \tilde {\mathbf{r}_e})$ and ${\tilde E}^e$ or  $\tilde{\Psi}^h({\tilde {\mathbf{r}_h}})$ and ${\tilde E}^h$.
The eigenvalues and eigenfunctions needed to evaluate the
absorption coefficients (ACs) were calculated with prescribed accuracy by means
of the program  packages ODPEVP  and KANTBP \cite{kantbp,ODPEVP,Yu3}. The models
with nonzero values of these parameters were announced in
\cite{JPCONF,79}.
Throughout the paper we make use of the reduced atomic
units~\cite{Harrison,LL94}: $a_B^*= {\kappa \hbar ^2}/{\mu _p e^2}$
is the reduced Bohr radius,
 $\tilde E_R \equiv Ry^*= {\hbar ^2}/({2\mu _p {a_B^*}^2 })$
is the reduced Rydberg unit of energy, and the following
dimensionless quantities are introduced: $\tilde \Psi(\tilde{\bf
r})= {a_B^*}^{-3/2}\Psi( {\bf r})$, $2 {\hat H}= \tilde {\hat
H}/{Ry^*}$,
${\cal E}\equiv 2 {E}={\tilde E}/{Ry^*}$, $2 {U({\bf r})}= {\tilde U (\tilde {\bf r})}/{Ry^*}$, ${\bf r}=\tilde {\bf r} /a_B^*$, $a=\tilde a/a_B^*$,
$c=\tilde c/a_B^*$, $2\gamma_{F} = F/F_{0}^*$, $F_{0}^{*}=Ry^*/(ea_B^{*})=e/(2\kappa (a_B^{*})^{2})$.

 \subsection{The BVP for SQDs in the effective mass approximation}
 In cylindrical coordinates $z,\rho,\varphi$ the solution of Eq.
(\ref{sp01}),  periodical with respect to the azimuthal angle
$\varphi$,  is sought in the form of a product
$\Psi(\rho,z,\varphi) = \Psi^{m} (\rho,z ){\exp{(im\varphi
)}}/{\sqrt {2\pi}}$, where $m = 0,\pm 1,\pm 2,...$
 is the magnetic quantum number.
 The 3D BVP for SQDs at fixed values of $m$
 is reduced to 2D BVP with respect to
\textit{fast} $x_f$ and \textit{slow} $x_s$ variables:
\textit{oblate}  $x_f=z$ (\textit{minor} axis), $x_s=\rho$ (major axis)
and  \textit{prolate} $x_f=\rho$ (\textit{minor} axis), $x_s=z$ (major axis) \cite{Yaf12}:
\begin{eqnarray} \label{2dbvp}
\left( \hat H_f (x_{f};x_s) + \hat H_s (x_{s})+ {\check
V_{fs}(x_{f},{x_s})}  - {{\cal E}_t^{m}}\right) \Psi_t^{m}
(x_{f},x_{s} ) = 0.
\end{eqnarray}
Here  $\hat H_s(x_s)$ is the operator of slow subsystem
\begin{eqnarray}
 \hat H_s(x_s)
 = - \frac{1}{g_{1s}(x_{s})}\frac{\partial }{\partial x_{s}}g_{2s}(x_{s})
 \frac{\partial }{\partial x_{s} }+ \check V_{s}(x_{s}),
\end{eqnarray}
and  $ \hat H_f(x_{f};{x_s})$ is the operator of fast subsystem
\begin{eqnarray}\label{sp09xf}
  \hat H_f(x_{f};{x_s})=
  - \frac{1}{g_{1f}(x_{f})}\frac{\partial }{\partial x_{f}}g_{2f}(x_{f})
 \frac{\partial }{\partial x_{f} }+\check V_{f}(x_{f};{x_s}).
\end{eqnarray}
For OSQD $g_{1s}(x_{s})=g_{2s}(x_{s})=1$,
$g_{1f}(x_{s})=g_{2f}(x_{s})=\rho$, $\check V_{f}(x_{f};{x_s})=0$,
$\check V_{s}(x_{s})=m^2/\rho^2$, $ {\check
V_{fs}(x_{f},{x_s})=2\gamma_Fz}$, while for PSQD
$g_{1s}(x_{s})=g_{2s}(x_{s})=\rho$, $g_{1f}(x_{s})=g_{2f}(x_{s})=1$,
$\check V_{f}(x_{f};{x_s})=m^2/\rho^2$,   $\check
V_{s}(x_{s})=2\gamma_Fz$, $ {\check V_{fs}(x_{f},{x_s})=0}$.
From (\ref{sp03}) the boundary conditions for the eigenfunctions $\Psi_{t}^{m} (x_{f},x_{s}
)$ of SQDs, corresponding  to a well with  walls of infinite height,
 have  the form
$$
\mathrel{\mathop{\lim}\limits_{\rho\rightarrow 0}}
\left(\rho\frac{\partial \Psi^{m}_t(\rho,z)}{\partial
\rho}\delta_{0m} +\Psi^{m}_t(\rho,z)(1-\delta_{0m})\right)=0,
\Psi^{m}_t(\rho,z)\biggl|_{\partial\Omega_2}=0,$$
$$\Omega_2=\left(\{\rho,z\}\biggl|\displaystyle\frac{\rho^2}{a^2}+
\displaystyle\frac{z^2}{c^2}< 1\right), \quad \partial\Omega_2
=\left(\{\rho,z\}\biggl|\displaystyle\frac{\rho^2}{a^2}+
\displaystyle\frac{z^2}{c^2}=1\right).$$ The  eigenfunctions
$\Psi_{t}^{m} (x_{f},x_{s} )$  corresponding to the eigenvalues $ {\cal
E}_t^{m}={\cal E}_1^{m}<{\cal E}_2^{m},...$ are subject to the normalization
and orthogonality conditions
$$\int_{\Omega_2}\rho d\rho dz \Psi_{t}^{m} (\rho,z)\Psi_{t'}^{m} (\rho,z)
=\delta_{tt'}.
$$
 Note, that at $\gamma_F=0$ the solutions are separated by the z-parity
$\sigma=\pm1$ into two invariant subspaces $\Psi_{t}^{m\sigma}$
corresponding to the eigenvalues ${\cal E}_t^{m\sigma}={\cal
E}_1^{m\sigma}<{\cal E}_2^{m\sigma},...$ , while at $\gamma_F\neq0$
the z-parity is broken.
\subsection{Kantorovich or adiabatic reduction of the BVP}
The solution  $\Psi^{}_t( x_{f},x_{s})\equiv\Psi^{m}_t(
x_{f},x_{s})$ of the above problem at fixed $m$  is sought in the
form of Kantorovich expansion
\begin{eqnarray}\label{KE}
  \Psi^{}_t( x_{f},x_{s})=\sum_{j=1}^{j_{\max}}
 B^{}_j(x_{f}; x_{s})\chi_{jt}^{}(x_{s}).
 \label{sp15}
\end{eqnarray}
The set of appropriate trial functions is chosen as the set of
eigenfunctions   $B^{}_j(x_{f}; x_{s}) $  corresponding to the
eigenvalues $\hat  E_j (x_s )$  of the Hamiltonian $\hat {H}_f (x_f
;x_s )$, Eq. (\ref{sp09xf}), depending parametrically on
$x_{s}\in\Omega(x_s)$:
$$ \hat H_f(x_{f};{x_s})B^{}_j(x_{f}; x_{s})=\hat  E_j (x_s )B^{}_j(x_{f}; x_{s}).$$
The eigenfunctions  $B^{}_j(x_{f}; x_{s})$ corresponding to the
eigenvalues $\hat E_j^{}(x_s)=\hat E_1^{}(x_s)<\hat E_2^{}(x_s),...$
are subject to the normalization and orthogonality conditions
with the weighting function $g_{1f}(x_f)$ in the same interval
$x_{f}\in\Omega_{x_f}(x_s)$:
\begin{equation}
\label{sp19}
\int\nolimits_{x_f^{\min}(x_{s})}^{x_f^{\max}(x_{s})}B _i^{} (x_f
;x_s ) B _j^{} (x_f ;x_s )
 g_{1f}(x_f)dx_f=\delta_{ij}.
\end{equation}
The BVP for a set of ODEs of the slow subsystem with respect to the
unknown vector functions ${\mbox{\boldmath $\chi$}}_t(x_s ) = (\chi
_{1;t} (x_s ),...,\chi _{j_{\max;t } } (x_s ))^T$ corresponding to the
unknown  eigenvalues $2E_t\equiv{\cal E}_t$,
\begin{eqnarray}&&
\nonumber
\biggl(
\mathbf{D}
+   {\bf E} ({ x_{s}})+{\rm {\bf W}} ({ x_{s}})
-  {\rm {\bf I}} {\cal E}_t
\biggr){\mbox{\boldmath $\chi$}}_t({ x_{s}})=0,
\\&&
 \mathbf{D}=-\frac{1}{g_{1s}(x_{s})}{\rm {\bf I}}
\frac{d}{d{ x_{s}}}g_{2s}(x_{s})\frac{d}{d{ x_{s}}}
+{\rm {\bf I}}\check{V}_{s}(x_s),\label{sp23}
\\&&
{\rm {\bf W}} ({ x_{s}})=
 {{\rm {\bf U}} ({ x_{s}})}+
\frac{g_{2s}(x_s)}{g_{1s}(x_s)} {\rm {\bf H}} ({ x_{s}})+
\frac{1}{g_{1s}(x_{s})}\frac{dg_{2s}(x_{s}){\rm {\bf Q}}({x_s})}{d{x_s}}
+\frac{g_{2s}(x_{s})}{g_{1s}(x_{s})}{\rm {\bf Q}}({ x_{s}})
\frac{d}{d x_{s}}\nonumber
\end{eqnarray}
satisfy the orthogonality and normalization conditions
\begin{equation}
\label{sp19a}
\int\nolimits_{x_s^{\min}}^{x_s^{\max}}
({\mbox{\boldmath $\chi$}}_t (x_s))^T{\mbox{\boldmath $\chi$}}_{t'}  (x_s)
 g_{1s}(x_s)dx_s=\delta_{tt'} .
\end{equation}
Here the effective potentials $H_{ij}(x_s)$ and $Q_{ij}(x_s)$ are defined by the formula
\begin{eqnarray}
U_{ij}(x_{s})=U_{ji}( x_{s})=
\int\nolimits_{x_f^{\min}(x_{s})}^{x_f^{\max}(x_{s})} B_{i}(x_{f};
x_{s}) {\check V_{fs}(x_{f},{x_s})} B_{j}(x_{f};
x_{s})g_{1f}(x_{f})dx_{f},
\nonumber  \\
H_{ij}(x_{s})=H_{ji}( x_{s})=
\int\nolimits_{x_f^{\min}(x_{s})}^{x_f^{\max}(x_{s})} \frac{\partial
B_{i}(x_{f}; x_{s})}{\partial x_{s}} \frac{\partial B_{j}(x_{f};
x_{s})}{\partial x_{s}}g_{1f}(x_{f})dx_{f},
\label{sp23a}  \\
Q_{ij}(x_{s})=-Q_{ji}( x_{s})=
-\int\nolimits_{x_f^{\min}(x_{s})}^{x_f^{\max}(x_{s})} B_{i}(x_{f};
x_{s}) \frac{\partial B_{j}(x_{f}; x_{s})}{\partial
x_{s}}g_{1f}(x_{f})dx_{f}.
 \nonumber
\end{eqnarray}
Here the basis functions
of the fast subsystem and the matrix elements are calculated  analytically.
For oblate spheroidal QDs ($x_f=z$, $x_s=\rho$)
\begin{eqnarray}&&\nonumber
B_{i}^{} \left( {x_f;x_s}\right)=B_{i}^{\sigma} \left(
{x_f;x_s}\right)\!=\!\sqrt{\frac{a}{c\sqrt {{a^2} - {x_s^2}}}}
\sin\left(\frac{\pi n_{o}}{2} \left(\frac{x_f}{ c\sqrt {1 - {x_s^2}/{a^2}}}-1\right)\right),\\
&&
  E_{i}(x_{s})= E_{i}^{\sigma}(x_{s})= E_{i;0}
\frac{ a^2}{(a^2-x_{s}^2)},\quad  E_{i;0}=\frac{ \pi^2i^2}{4c^2}
,\quad  {U_{ii}(x_{s})=0,}\label{eq10o}\\&&\nonumber
  {U_{ij}(x_{s})=U_{ij;0}(x_{s})\frac{\sqrt{a^2-x_{s}^2}}{ a},
\quad U_{ij;0}(x_{s})=
\frac{8\gamma_Fcij{(-1+(-1)^{i+j})}}{ (i^2-j^2)^2\pi^2}
,} \\&&
H_{ii}(x_{s})=H_{ii;0}(x_{s})\frac{ a^2 x_{s}^2}{(a^2-x_{s}^2)^2},\quad H_{ii;0}(x_{s})=\frac{3+\pi^2i^2}{12 a^2} ,\nonumber\\
&&H_{ij}(x_{s})=H_{ij;0}(x_{s})\frac{ a^2 x_{s}^2}{(a^2-x_{s}^2)^2},\quad H_{ij;0}(x_{s})=\frac{2ij(i^2+j^2){(1+(-1)^{i+j})}}{ a^2(i^2-j^2)^2} ,\nonumber\\
&&Q_{ij}(x_{s})=Q_{ij;0}(x_{s})\frac{ a  x_{s} }{ a^2-x_{s}^ 2},\quad Q_{ij;0}(x_{s})= \frac{ij{(1+(-1)^{i+j})}}{a(i^2-j^2)} \nonumber ,\quad j\neq i. \end{eqnarray}
For prolate spheroidal QDs ($x_f=\rho$, $x_s=z$) (at $m=0$ for nondiagonal potentials $i\neq j$)
\begin{eqnarray}&&
\nonumber B^{m}_{n_{\rho p}}(x_{s})=\frac{\sqrt{2}c}{  a\sqrt
{{c^2}- {x_s^2} }} \frac{ J_{|m|}(\sqrt{2  E _{n_{\rho p}+1,|m|}
\left( x_s \right)}x_{f}) }{|J_{|m|+1}(\alpha_{n_{\rho p}+1,|m|})|},
\\\label{eq10p}
&&   E_i \left( x_s \right) = E_{i;0}
\frac{ c^2}{(c^2-x_{s}^2)},\quad  E_{i;0}=\frac{(\bar J_{|m|}^i)^2 }{a^2 },
\\&&  {U_{ii}(x_{s})=0,} \quad  {U_{ij}(x_{s})=0,}\nonumber
\\&&
H_{ii}(x_{s})=H_{ii;0}(x_{s})\frac{ c^2 x_{s}^2}{(c^2-x_{s}^2)^2},\quad H_{ii;0}(x_{s})=\frac{ (1+(\bar J_{|m|}^i)^2)}{3 c^2 },
\nonumber\\
&&H_{ij}(x_{s})=H_{ij;0}(x_{s})\frac{ c^2 x_{s}^2}{(c^2-x_{s}^2)^2},\quad H_{ij;0}(x_{s})=
\frac{2 }{ c^2 }
\left(\bar J_0^i\bar J_0^j\int_0^1 \frac{J_1(\bar J_0^ix)}{J_1(\bar J_0^i)}
 \frac{J_1(\bar J_0^jx)}{J_1(\bar J_0^j)}x^3dx\right.\nonumber
 \\&&\left.\qquad\qquad
-
\bar J_0^i\int_0^1 \frac{J_1(\bar J_0^ix)}{J_1(\bar J_0^i)}
 \frac{J_0(\bar J_0^jx)}{J_1(\bar J_0^j)}x^2dx
-\bar J_0^j\int_0^1 \frac{J_0(\bar J_0^ix)}{J_1(\bar J_0^i)}
 \frac{J_1(\bar J_0^jx)}{J_1(\bar J_0^j)}x^2dx
\right) ,\nonumber\\
&&Q_{ij}(x_{s})=Q_{ij;0}(x_{s})\frac{ c  x_{s} }{c^2-x_{s}^2},\quad Q_{ij;0}(x_{s})=\frac{2 }{c }\bar J_0^j\int_0^1 \frac{J_0(\bar J_0^ix)}
{J_1(\bar J_0^i)}\frac{J_1(\bar J_0^jx)}{J_1(\bar J_0^j)}x^2dx,\quad j\neq i, \nonumber
\end{eqnarray}
where $\alpha _{n_{\rho p}+ 1,|m|}=\bar J^{n_{\rho p}+ 1}_{|m|}$
are positive zeros of the Bessel function of the first kind \cite{stigun}.

For the interesting lower part of the spectrum  ${\cal E}_t:~ {\cal
E}_1 < {\cal E}_2 < ... $, the number $j_{max}$ of the equations
solved should be at least not less than the number of the energy
levels of the problem (\ref{sp23}) at $a = c = r_0$. To ensure the
prescribed accuracy of calculation of the lower part of the spectrum
discussed below with eight significant digits we used $j_{max} = 16$
basis functions in the expansion (8) and the discrete approximation
of the desired solution by Lagrange finite elements of the fourth
order with respect to the  {grid pitch} $\Omega^p_{h_s}(x_s) =
[x_{s;\min}; x_{s;k} = x_{s;k-1} + h_s; x_{s;\max}]$. The details of
the corresponding computational scheme are given in \cite{CASC10}.

\begin{table}\caption{
The convergence of eigenenergy ${\cal E}_{t}$ vs number $j_{\max } $ of basis functions { at  $\gamma_F=0$  }.
Fast and slow variables $x_{f}=z$ and $x_{s}=\rho$ (\textit{oblate} SQD and spherical QD),
number of nodes $i=(n_{zo}=n_{o}-1,n_{\rho o})$,{ $^*$ notes diagonal approximation at $j=2$}}\label{fullo}
\begin{tabular}{|l||l|l|l||l|l|l|}\hline
 $j_{\max } $  & \multicolumn{3}{c||}{     $a=2.5$,           $c=0.5$    } &          \multicolumn{3}{c|}{     $a=2.5$,           $c=2.5$    }          \\ \hline
$(n_{zo},n_{\rho o})$ &(0,0)&(0,1)&(2,0)&(0,0)&(0,1)&(2,0) \\ \hline
  C & 12.737\;41 & 19.936\;21 & 96.696\;83$^*$& 1.468\;496 & 5.445\;665$^*$& 5.589\;461 \\ \hline
  1 & 12.765\;48 & 20.046\;02 & 96.753\;17$^*$& 1.590\;238 & 5.766\;612$^*$& 6.004\;794 \\ \hline
  2 & 12.764\;90 & 20.041\;33 & 96.754\;27    & 1.580\;243 & 5.340\;214    & 6.329\;334 \\ \hline
   4 &  12.764\;82 & 20.040\;74 & 96.752\;15    & 1.579\;273 & 5.316\;872    & 6.317\;204 \\ \hline
 16 & 12.764\;81 & 20.040\;65 & 96.752\;01    & 1.579\;140 & 5.314\;832    & 6.316\;562 \\ \hline
 Exact  &        \;    &       \;    &       \;    & 1.579\;136 & 5.314\;793    & 6.316\;546 \\ \hline
\end{tabular} 
\end{table}

\begin{table}\caption{
The convergence of eigenenergy ${\cal E}_{t}$ vs number $j_{\max } $ of basis functions { at  $\gamma_F=0$  }.
Fast and slow variables $x_{f}=\rho$ and $x_{s}=z$ (\textit{prolate} SQD and spherical QD),
number of nodes $i=(n_{\rho p},n_{zp})$,{ $^*$ notes diagonal approximation at $j=2$}}\label{fullp}
\begin{tabular}{|l||l|l|l||l|l|l|} \hline
 $j_{\max } $  & \multicolumn{3}{c||}{     $c=2.5$,           $a=0.5$    } &          \multicolumn{3}{c|}{     $c=2.5$,           $a=2.5$    }          \\  \hline
$(n_{\rho p},n_{zp})$ &(0,0)&(0,2)&(1,0)&(0,0)&(0,2)&(1,0)\\ \hline
  C & 25.184\;73 & 34.428\;85 & 126.424\;5$^*$& 1.493\;612 & 5.131\;784 & 5.898\;668$^*$ \\ \hline
  1 & 25.201\;74 & 34.530\;30 & 126.456\;5$^*$& 1.584\;433 & 5.680\;831 & 6.071\;435$^*$ \\ \hline
  2 & 25.201\;29 & 34.525\;78 & 126.457\;3    & 1.579\;860 & 5.331\;101 & 6.324\;717     \\ \hline
  4 & 25.201\;21 & 34.525\;12 & 126.456\;1    & 1.579\;239 & 5.316\;732 & 6.317\;058     \\ \hline
 16 & 25.201\;20 & 34.525\;02 & 126.456\;1    & 1.579\;138 & 5.314\;828 & 6.316\;554     \\ \hline
 Exact  &       \;    &       \;    &        \;   & 1.579\;136 & 5.314\;793 & 6.316\;546     \\ \hline
\end{tabular}  \end{table}

The convergence of eigenenergies ${\cal E}_{t}$ vs number $j_{\max }
$ of basis functions  for oblate and prolate spheroidal QDs, and for
spherical QD is shown on Tables \ref{fullo} and \ref{fullp} at
$\gamma_{F}=0$ and  $m=0$.
The considered QDs having the size comparable with De Broglie wavelength of composed particles with small effective masses are referred as
quantum-size systems. In the spheroidal QDs having different
length of minor and major axes the quantization procedure leads to different
transversal and longitudinal spectra.
Moreover, for PSQD ($c=2.5$, $a=0.5)$ the confinement in two variables
($xy$) with the minor semiaxis $a=0.5$ leads to greater eigenvalues, than the
confinement in one variable ($z$) with the size-for-size minor semiaxis $a=0.5$ for PSQD ($c=2.5$, $a=0.5)$.
Tables \ref{fullo} and \ref{fullp} show that the expansions in basis functions
(\ref{eq10o}) and (\ref{eq10p}) in cylindrical coordinates have
better rate of convergence in the adiabatic limit of strongly oblate and
prolate QDS, than for the benchmark  spherical QDs with the known
spectrum, which is not surprising.
For lower states the crude adiabatic approximation (without
$H_{jj}(x_s)$) (CAA) provides a lower estimate, while the adiabatic
approximation (AA) (with $H_{jj}(x_s)$) (1) gives an upper estimate,
such that at the ratio of minor to major semiaxis equal to  1/5
the bracket is approximated with the accuracy of $\sim 0.1$\%.

Below we present the analysis of the spectrum under the
variation of parameters, which opens the questions about the additional
symmetry of the problem, connected with the existence of exact and
approximate integrals of motion\cite{HELFRICH1972,Yaf12}.

\begin{figure}[h]
\noindent\includegraphics[width=0.31\textwidth]{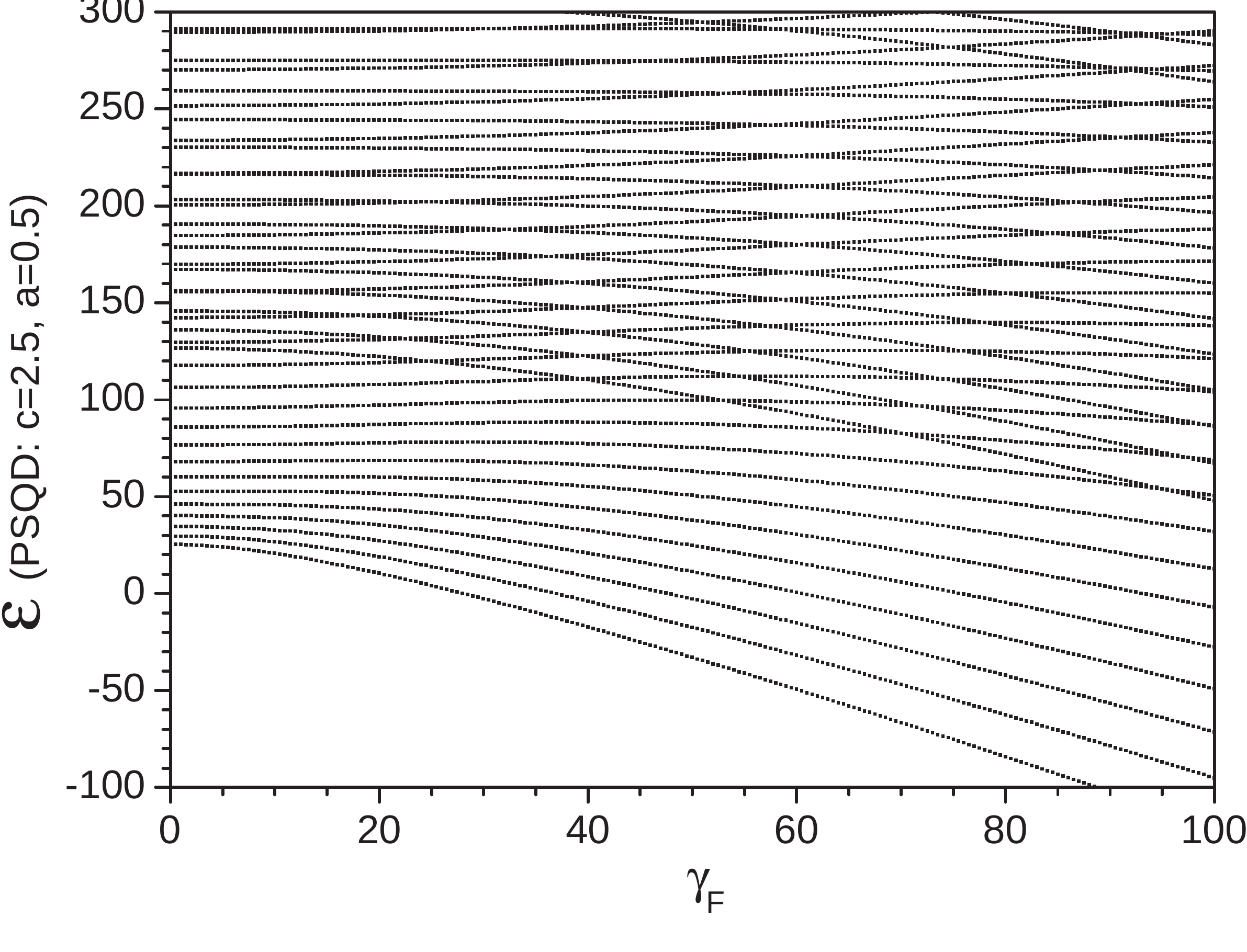}
\includegraphics[width=0.31\textwidth]{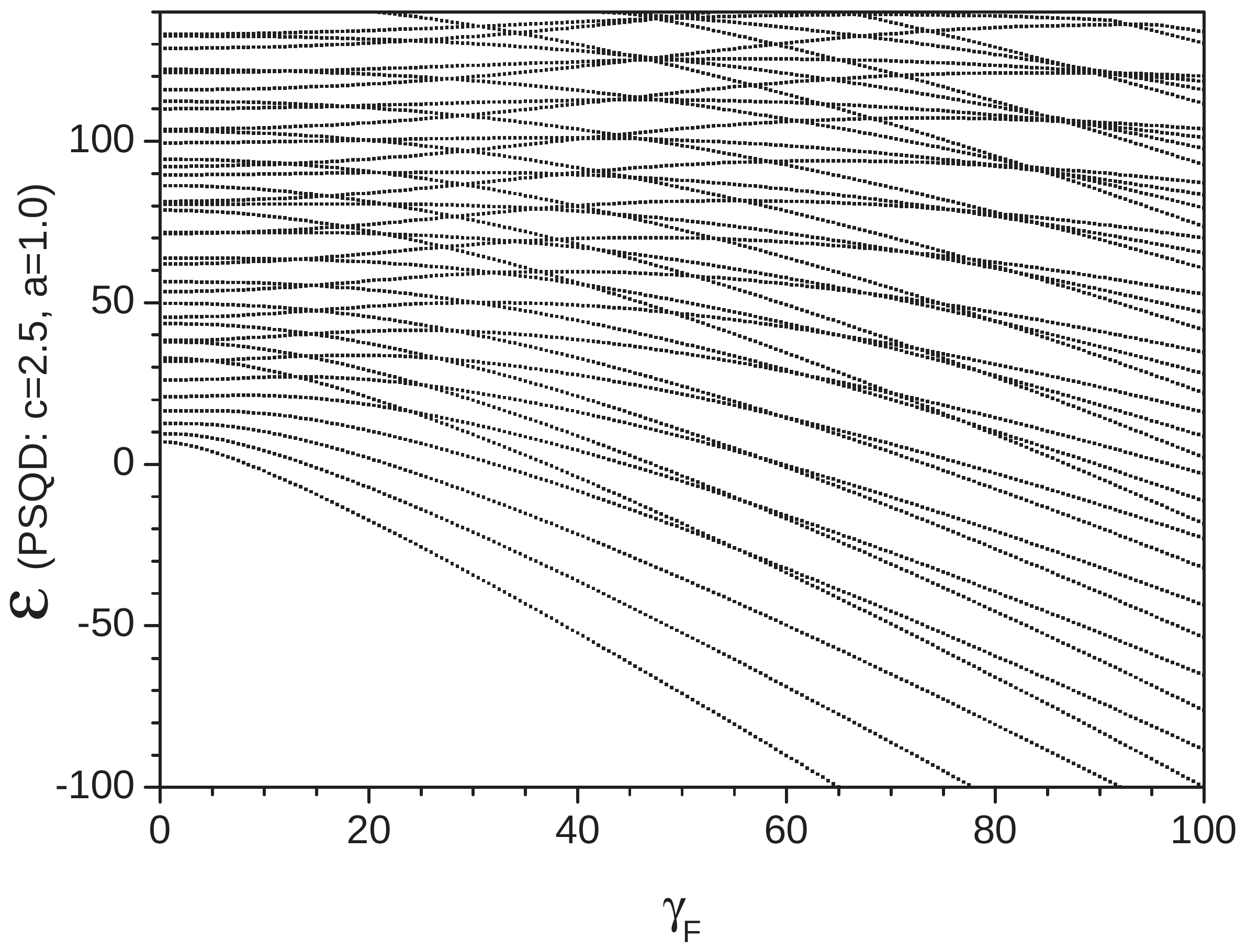}
\includegraphics[width=0.31\textwidth]{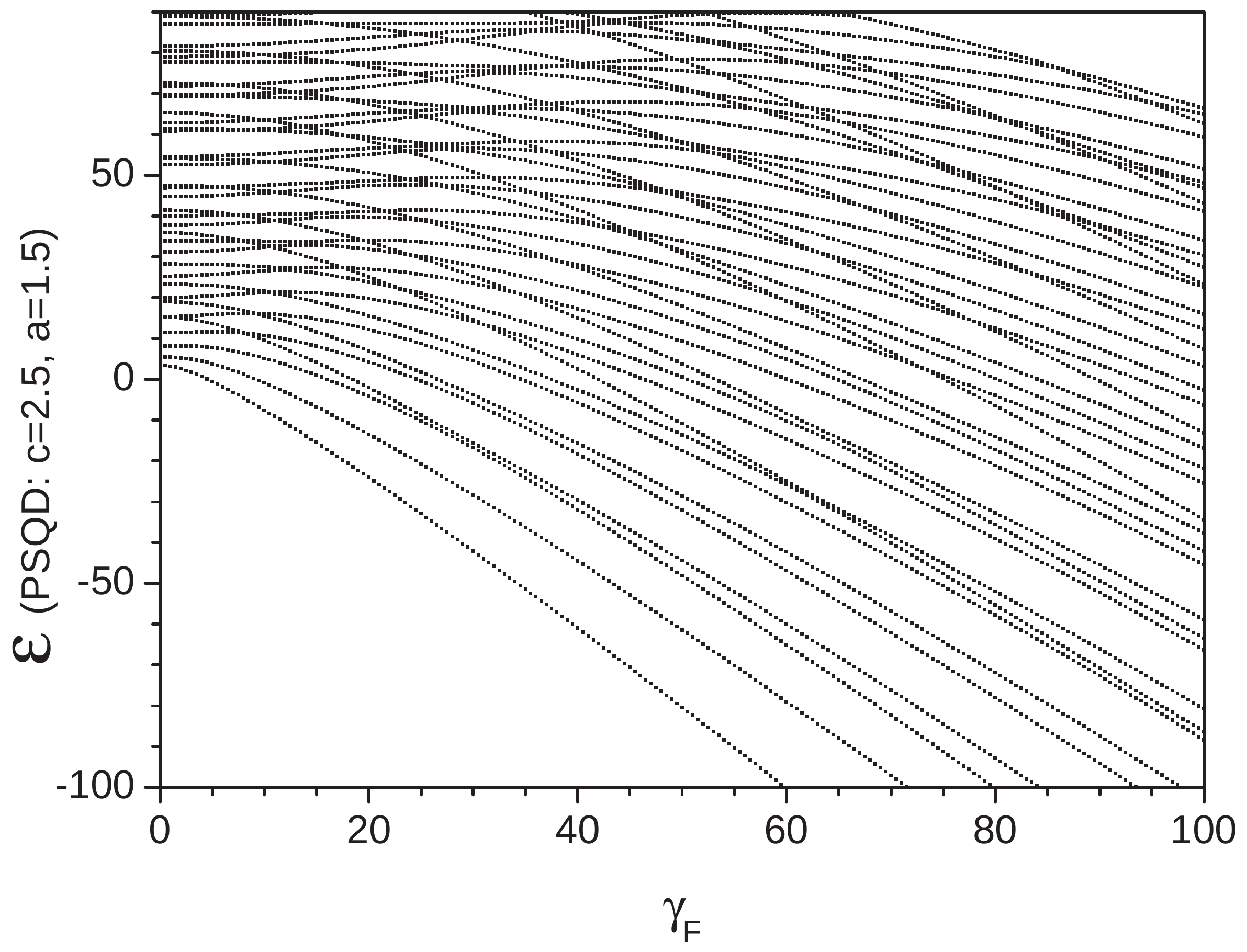}
\includegraphics[width=0.31\textwidth]{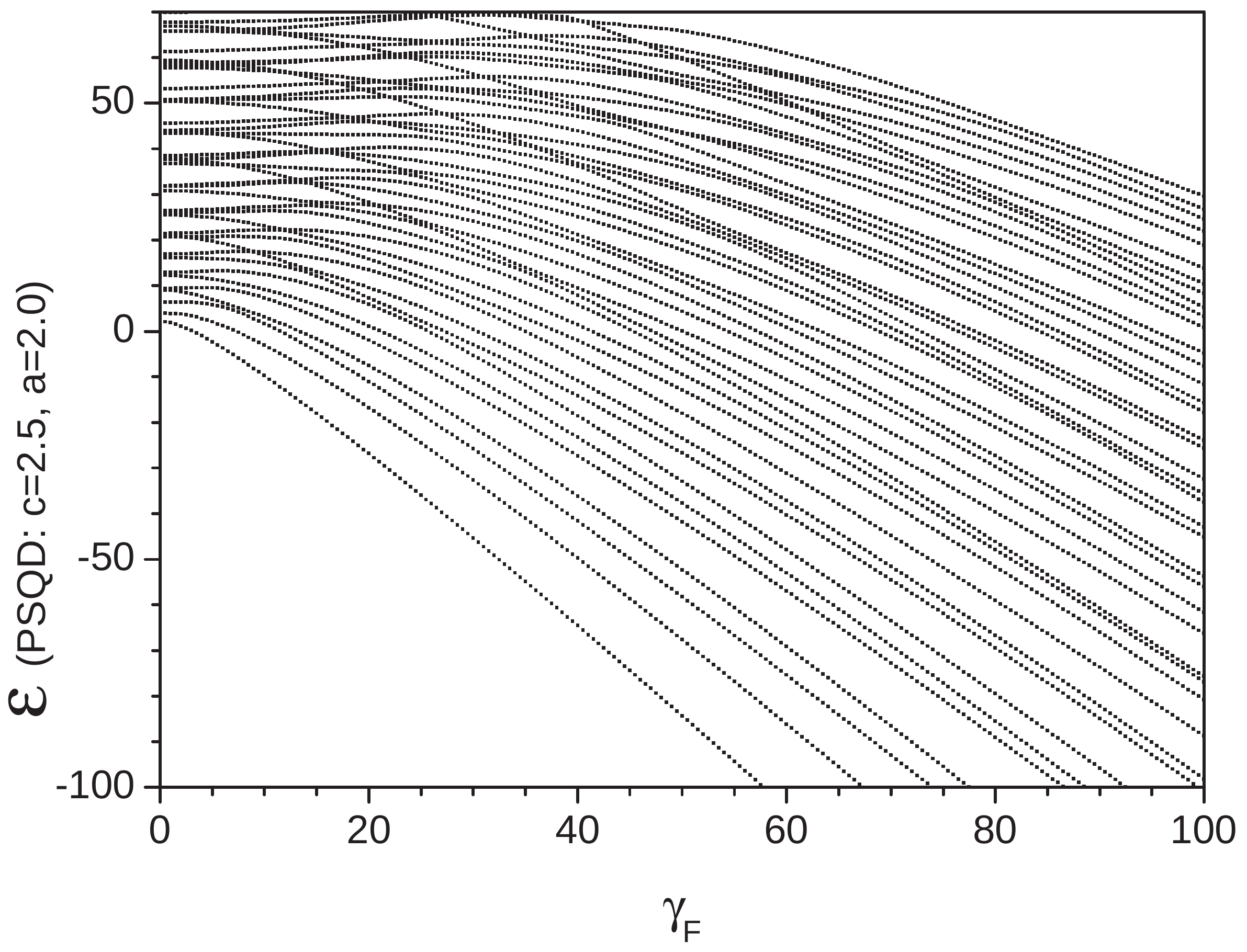}
\includegraphics[width=0.31\textwidth]{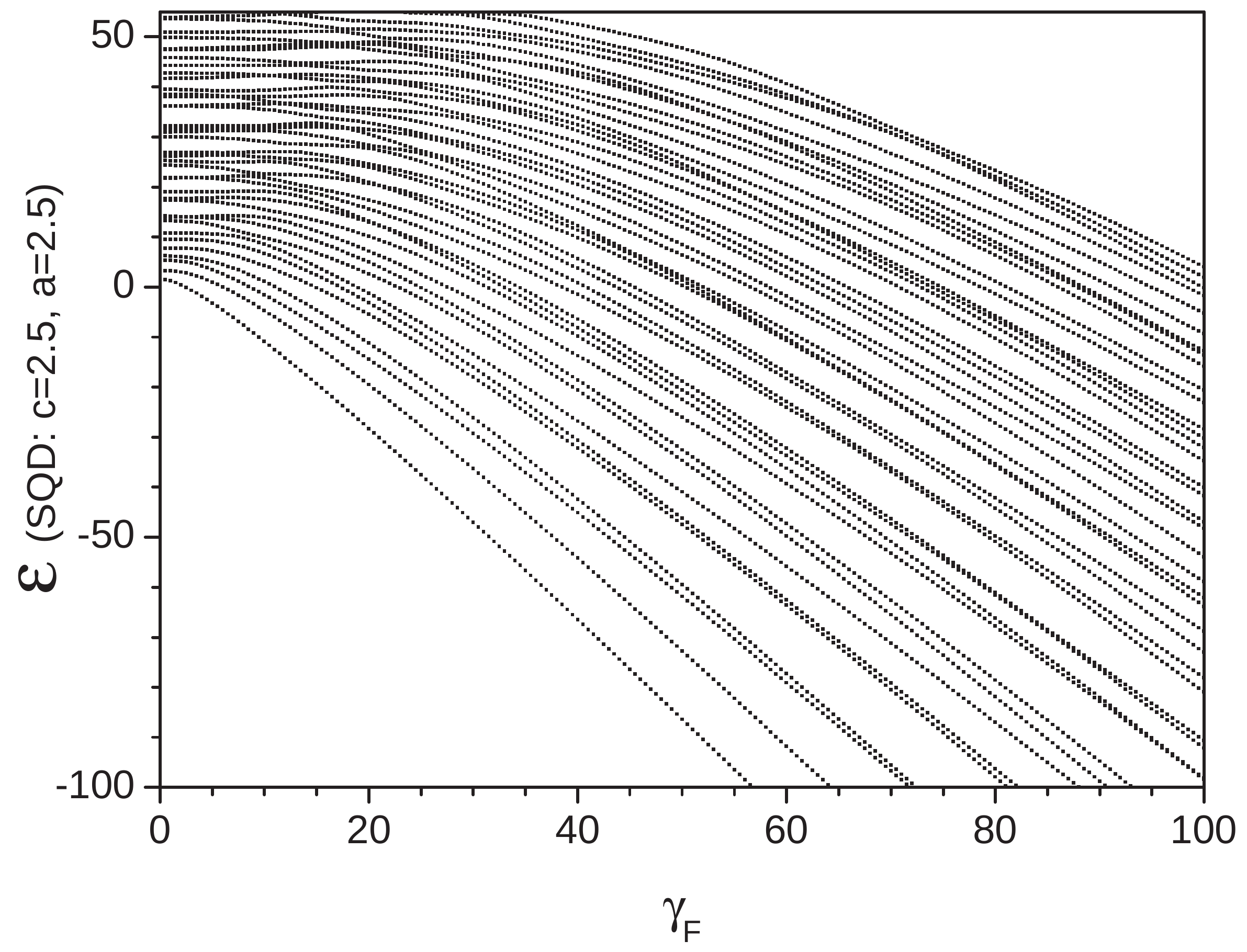}
\includegraphics[width=0.31\textwidth]{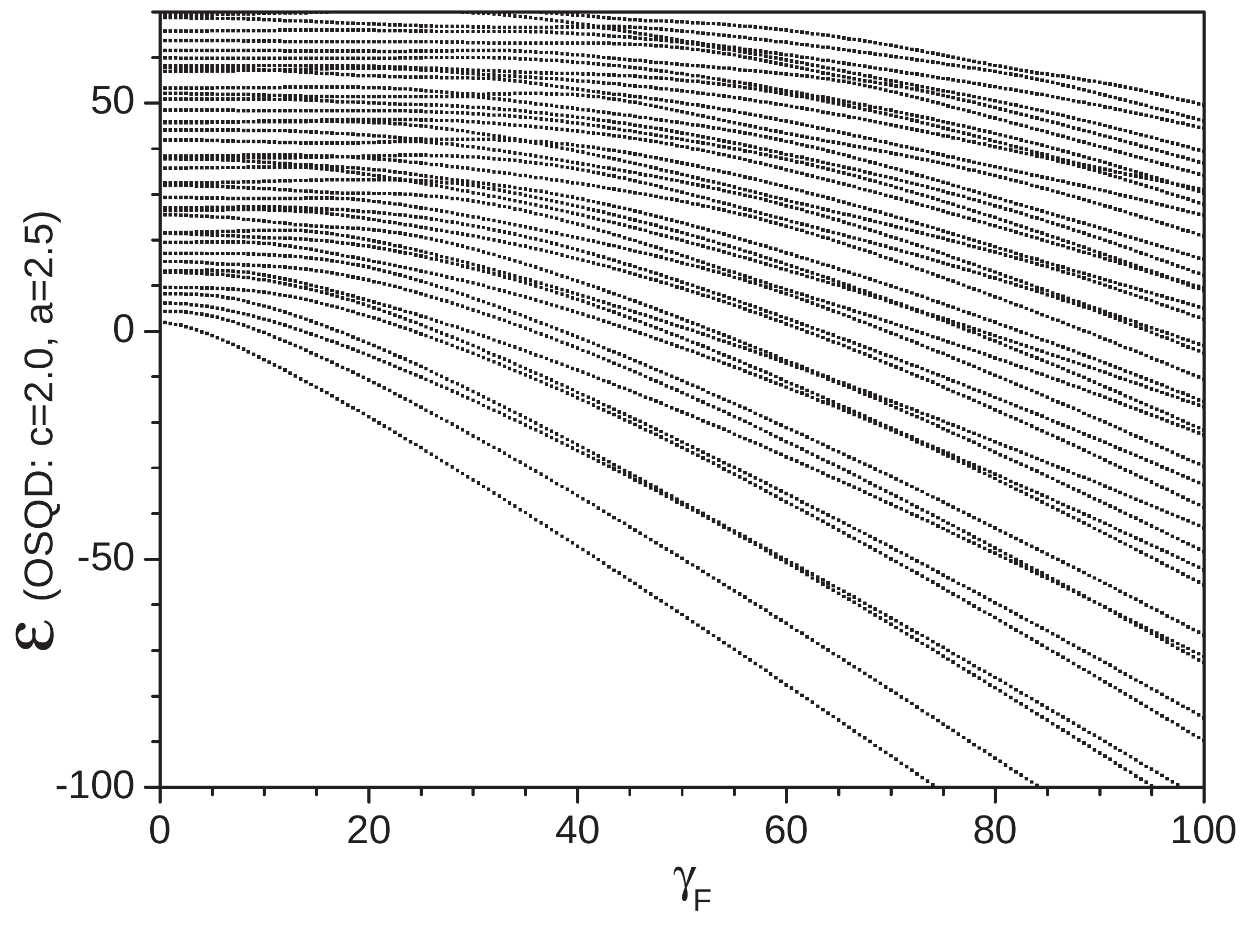}
\includegraphics[width=0.31\textwidth]{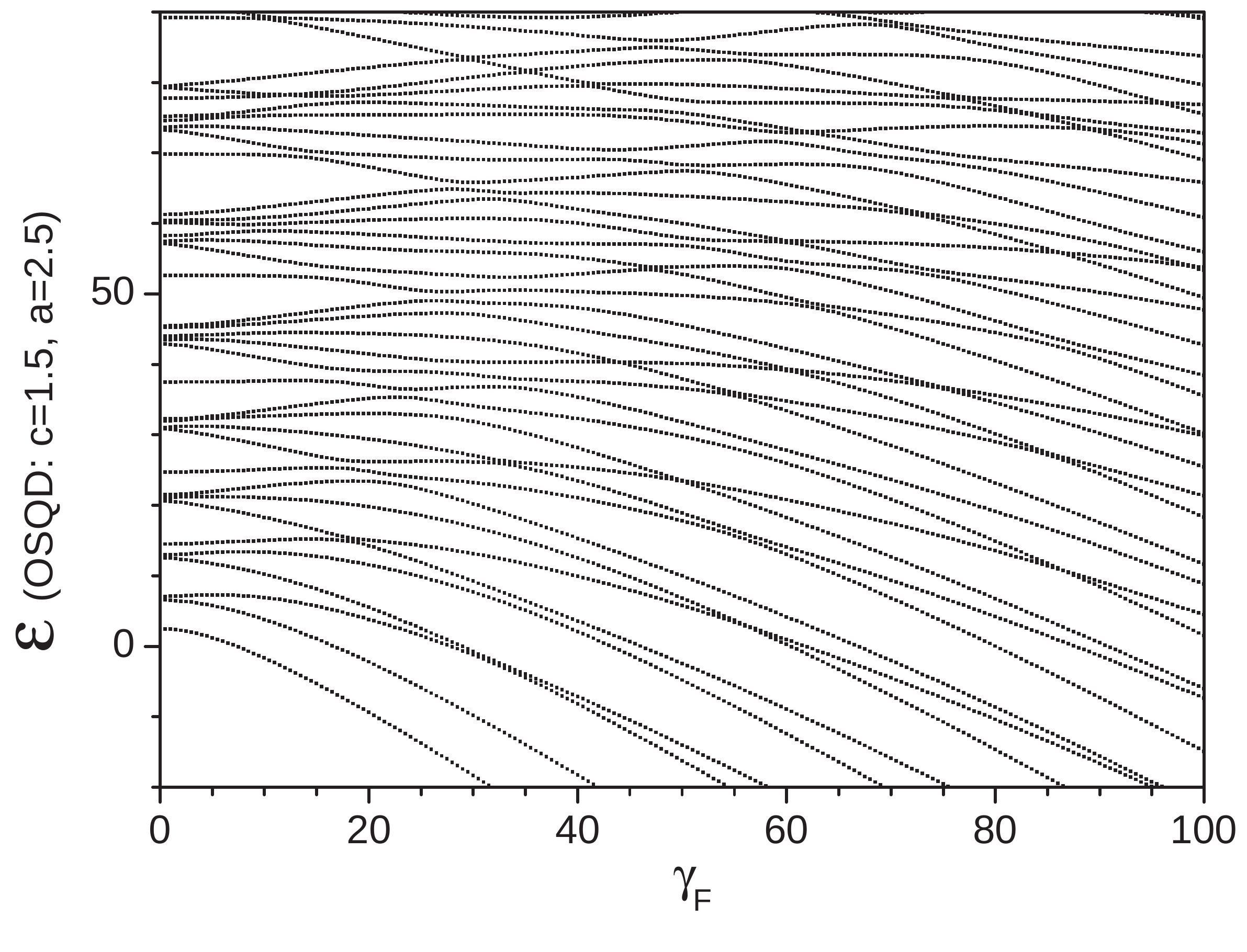}
\includegraphics[width=0.31\textwidth]{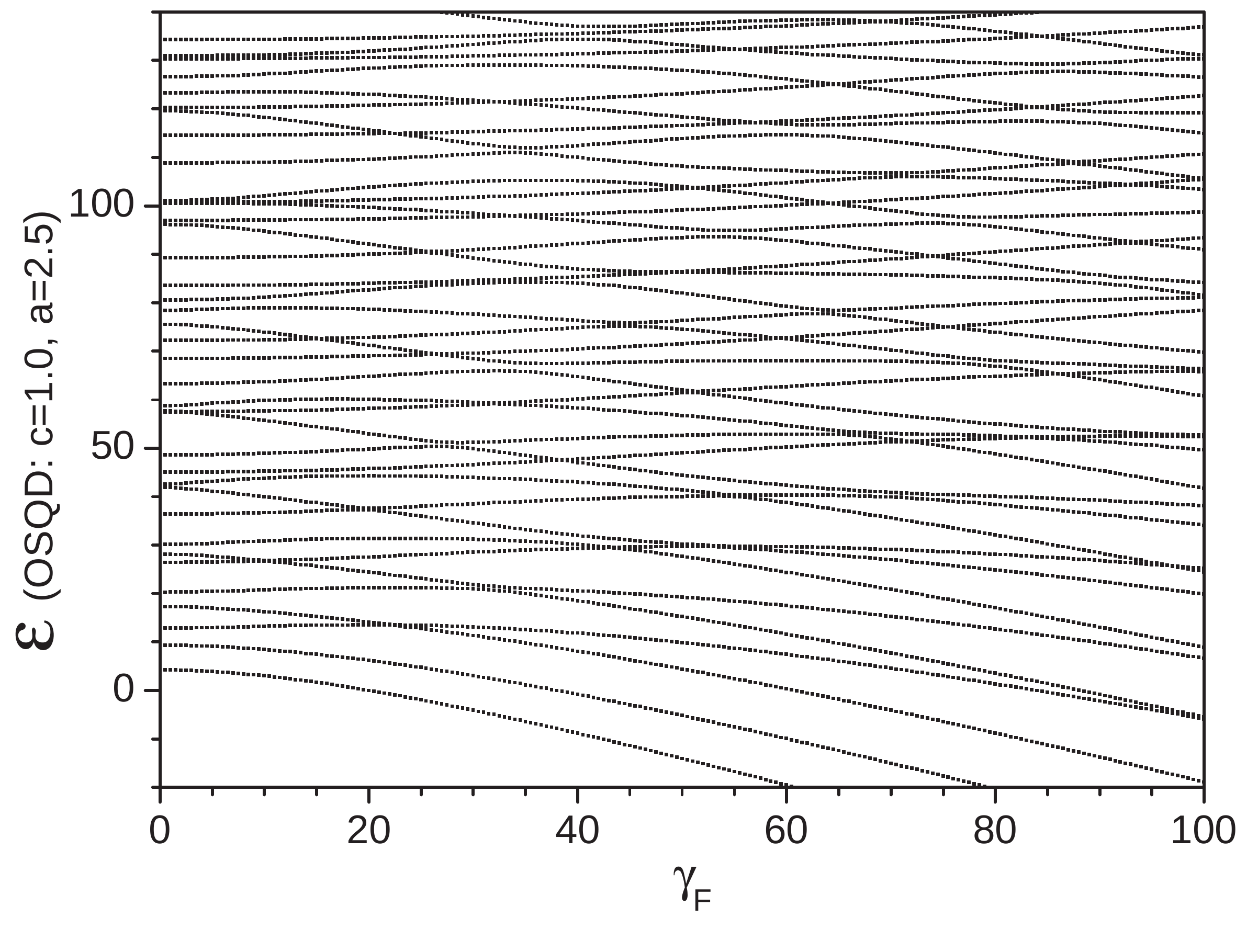}
\includegraphics[width=0.31\textwidth]{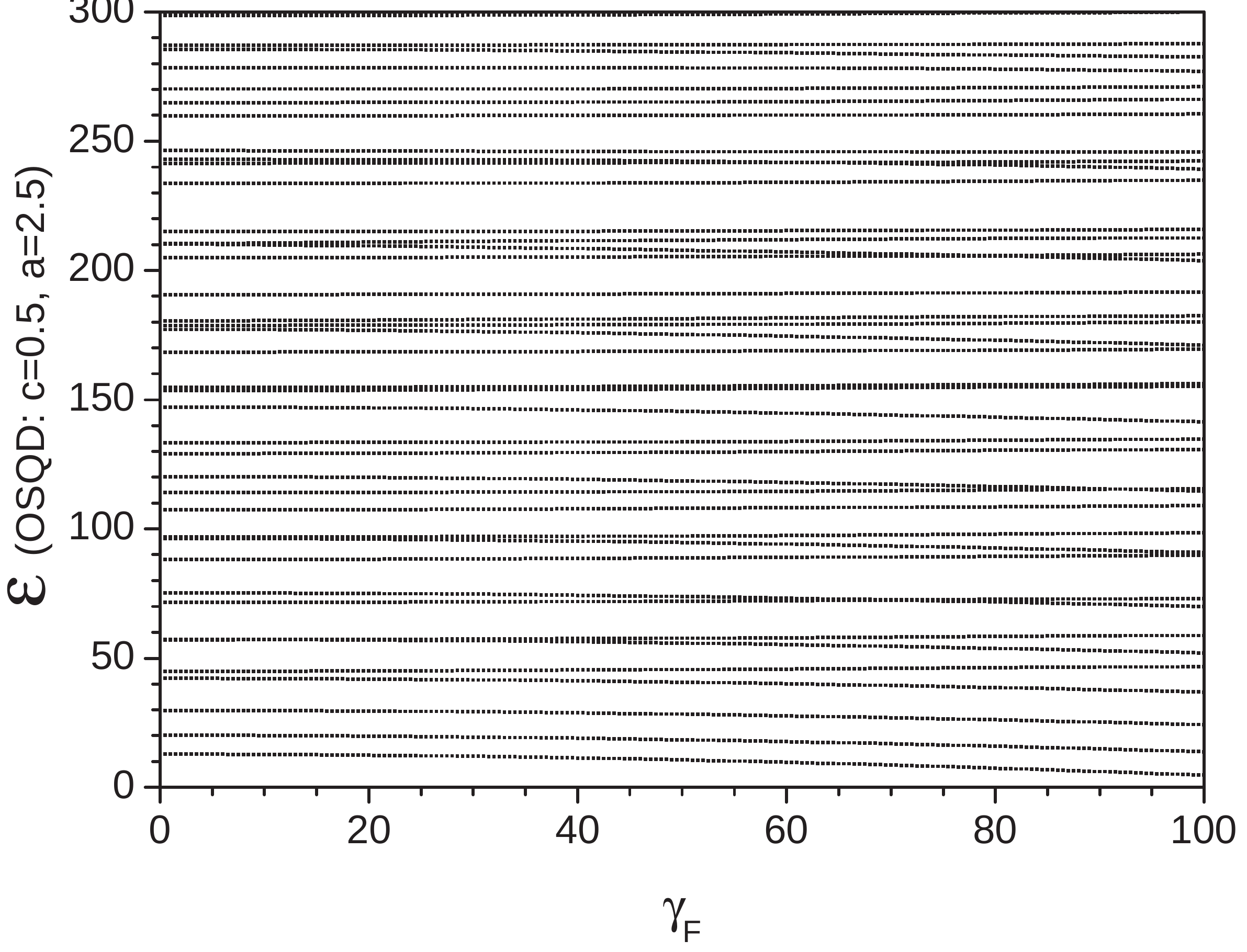}
\caption{Dependence of eigenenergies ${\cal E}$ (in units of $E_{e}$)  of lower part of spectrum of electronic states of QDs at $m=0$
on electric field strength
$\gamma_F$ (in units of $F_0^*$):
for spherical quantum dot (SQD) with radius $a=c=2.5$, oblate and prolate spheroidal quantum dots (OSQD and PSQD) at different minor semiaxis (for OSQD $c=0.5,1,1.5,2$, $a=2.5$, for PSQD $c=2.5$, $a=0.5,1,1.5,2$).}
 \label{exact}
\end{figure}

 In Fig. \ref{exact} we show the eigenenergies of the lower part of the spectrum ${\cal E}_{t}$, $t=1,...,40$ at
  $m=0$ for OSQD $(c=0.5,1,1.5,2,a=2.5)$, SQD $(c=2.5,a=2.5)$ and PSQD $(c=2.5,a=0.5,1,1.5,2)$
  as functions of the dimensionless strength  $\gamma_{F}$ of the electric field.
In spite of the fact that at $\gamma_F=0$ the eigenfunctions of SQD, OSQD and PSQD  have definite  z-parity, and, therefore,
exhibit additional integrals of motion and separation of variables in spherical and spheroidal coordinates systems,
 the spectrum of eigenvalues at fixed $m$ is simple, i.e., nondegenerate, similar to the case $\gamma_F\neq0$,
   when the eigenfunctions have no definite z-parity.
At $\gamma_F=0$ a one-to-one
correspondence rule  $n_{\rho p}+1=n_p=i=n=n_r+1$, $i=1,2,...$ and
$n_{z p}=l-|m|$ holds between the  quantum numbers
$(n,l,m,\hat\sigma={ (-1)^{|m|}\sigma})$ of SQD with the radius  $r_0=a=c$, the
spheroidal quantum numbers $\{n_\xi=n_r,n_\eta=l-|m|,m,\sigma\}$ of  {PSQD} with
the major $c$ and the minor $a$ semiaxes, and the adiabatic set of
quantum numbers $[n_p=n_{\rho p}+1,n_{z p},m,\sigma]$ under the
continuous variation of the parameter $\zeta_{ac}=a/c$.
 At $\gamma_F=0$ there is a one-to-one correspondence rule
$n_{o}=n_{zo}+1=2n-(1+\sigma)/2$, $n=1,2,3,...$ and $n_{\rho o}=(l-|m|-(1-\sigma)/2)/2$, between the sets of
spherical quantum numbers
$(n,l,m,\hat\sigma={ (-1)^{|m|}\sigma})$ of SQD with the radius
$r_0=a=c$ and spheroidal ones  $\{n_\xi=n_r,n_\eta=l-|m|,m,\sigma\}$
of  {OSQD} with the major $a$ and the minor $c$ semiaxes, and the
adiabatic set of cylindrical quantum numbers
$[n_{o}=n_{zo}+1,n_{\rho o},m,\sigma]$ under the continuous variation of
the parameter $\zeta_{ca}=c/a$.

\begin{figure}[h]
\parbox{0.48\textwidth}{\epsfig{file=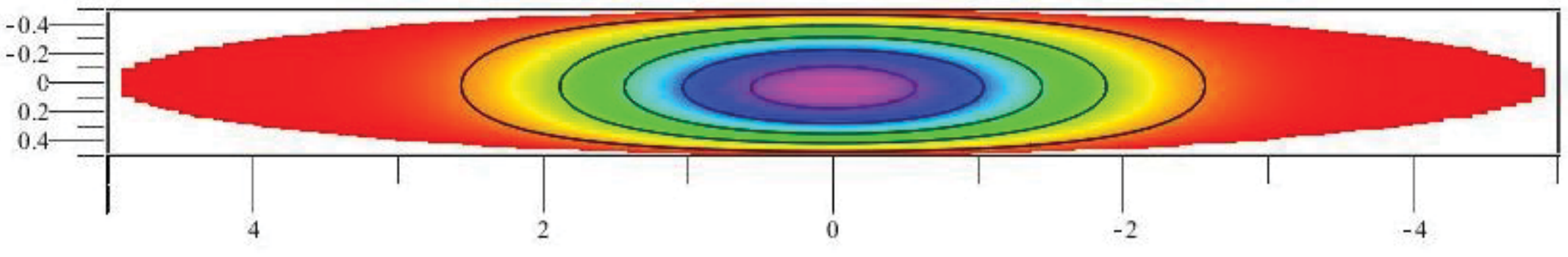,width=0.45\textwidth,height=0.10\textheight}
\epsfig{file=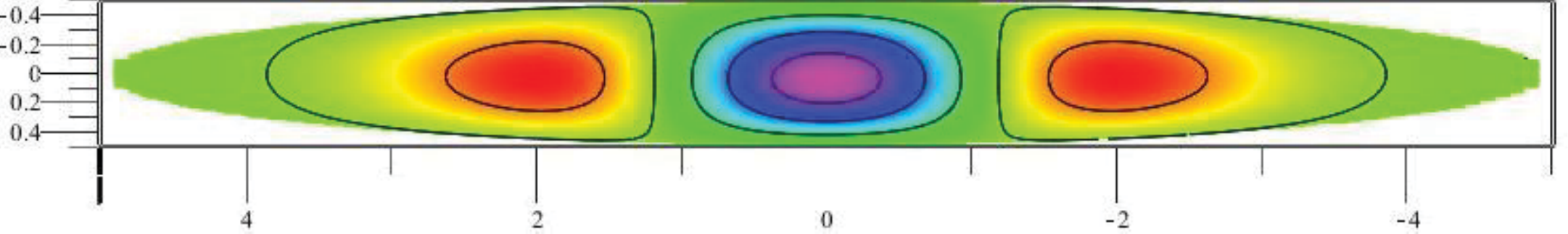,width=0.45\textwidth,height=0.10\textheight}
\epsfig{file=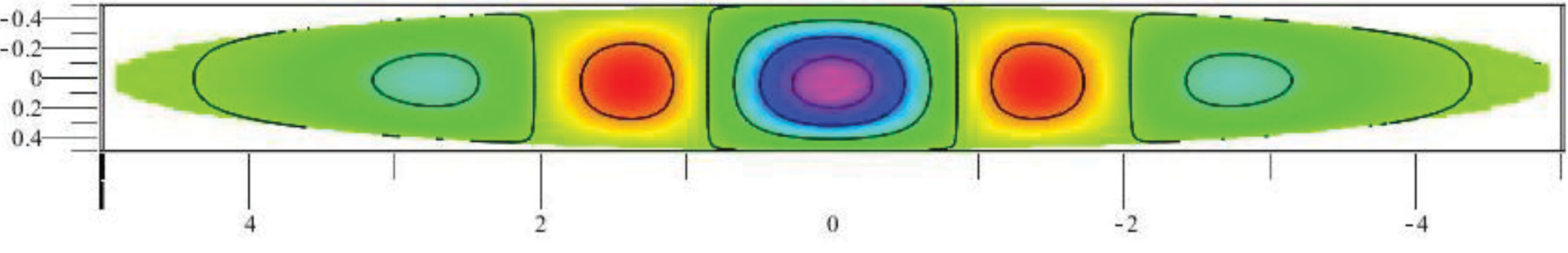,width=0.45\textwidth,height=0.10\textheight}
\epsfig{file=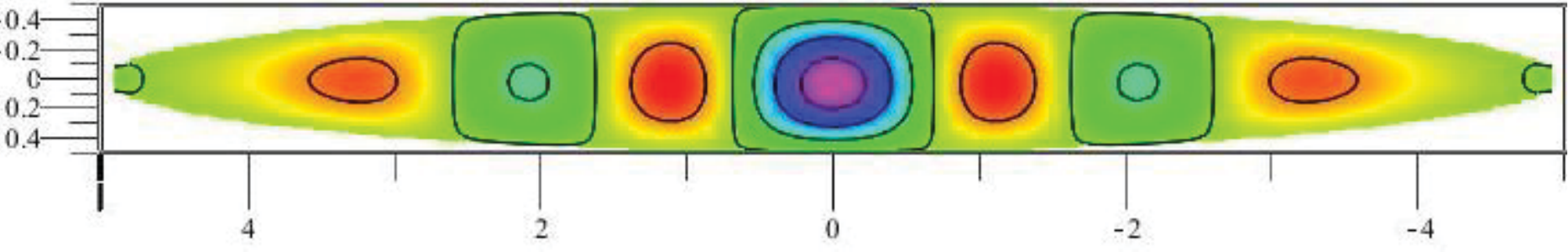,width=0.45\textwidth,height=0.10\textheight}}
\parbox{0.48\textwidth}{\epsfig{file=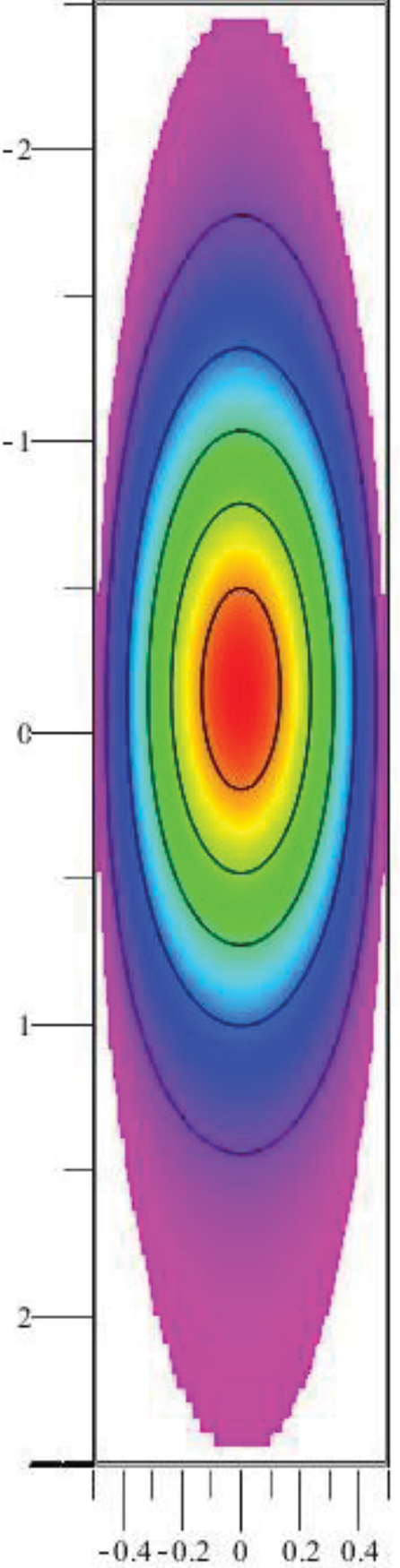,width=0.10\textwidth,height=0.4\textheight}
\epsfig{file=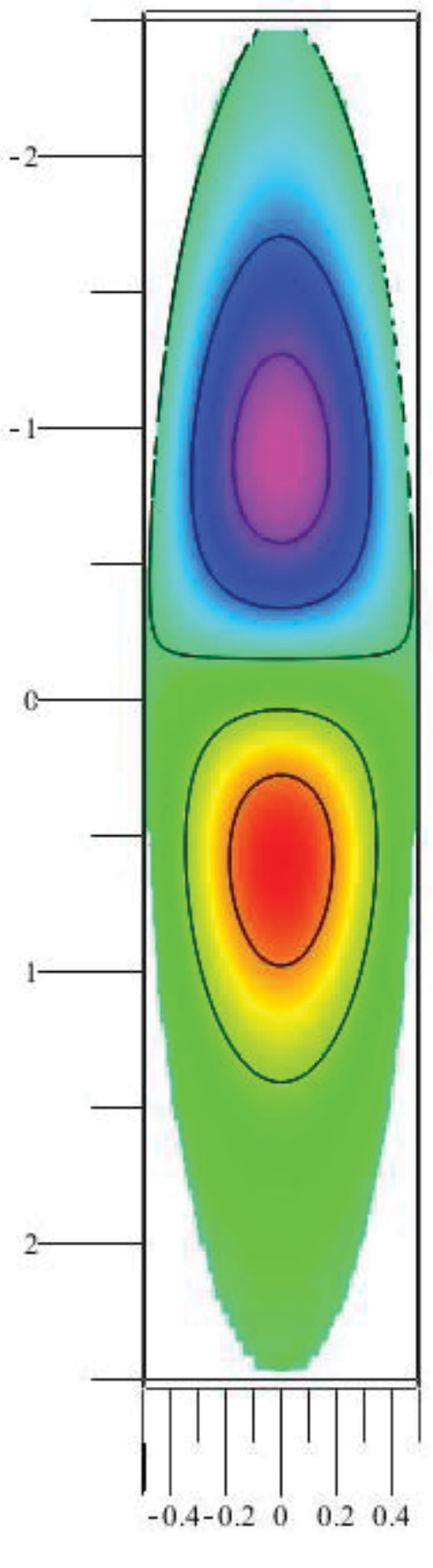,width=0.10\textwidth,height=0.4\textheight}
\epsfig{file=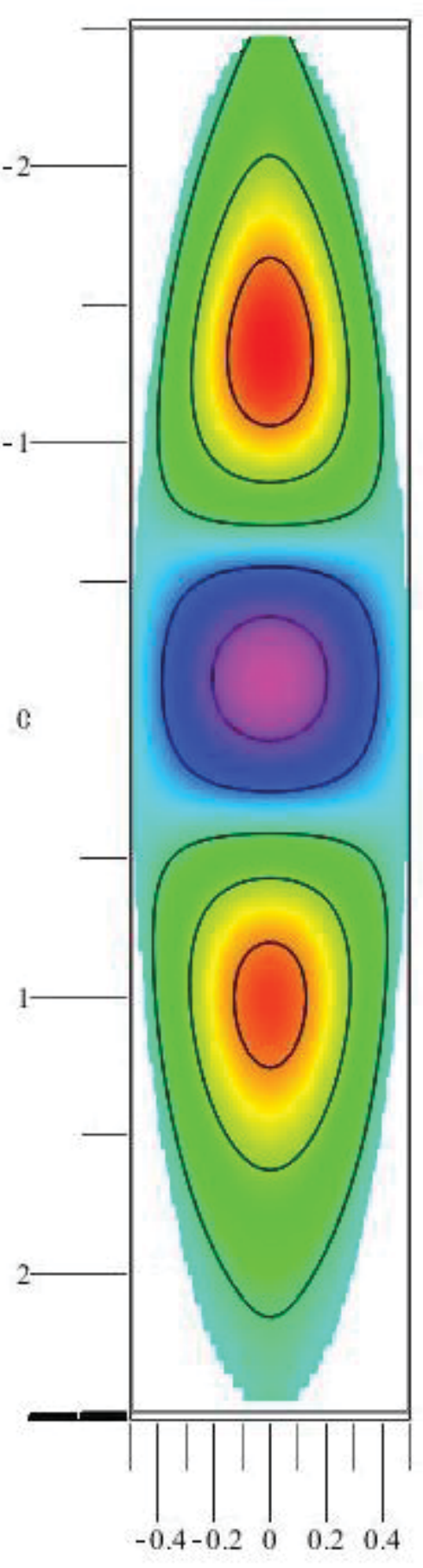,width=0.10\textwidth,height=0.4\textheight}
\epsfig{file=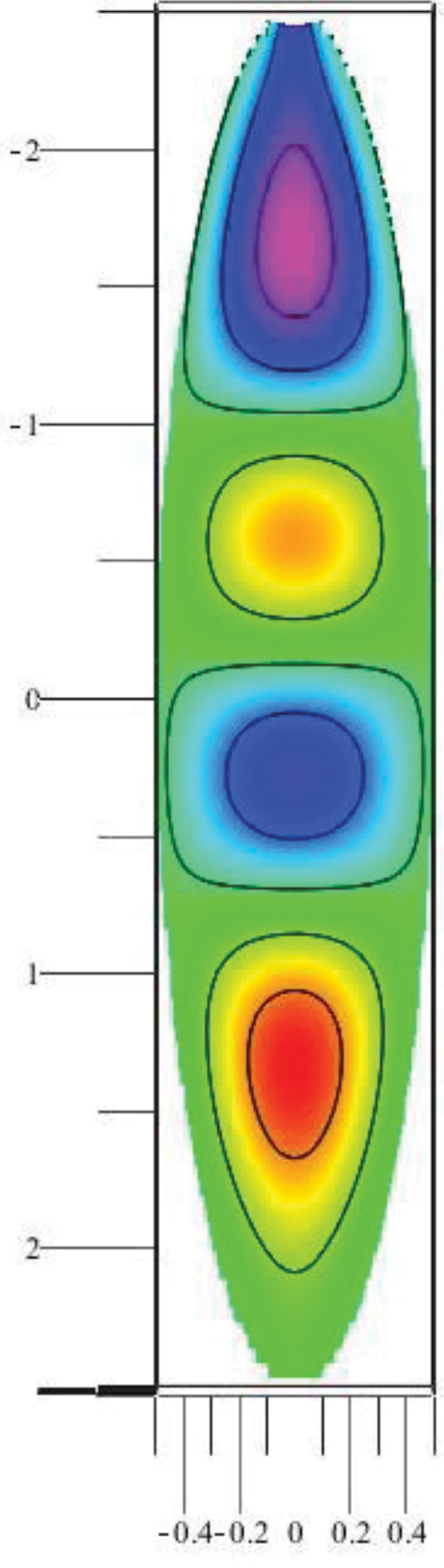,width=0.10\textwidth,height=0.4\textheight}}
\caption{Eigenfunctions of sixth order of PT of 2D BVP for  \textit{oblate} SQD $a=5$, $c=0.5$, $[t=n_{zo}=0$, $n=n_o=1,2,3,4$, $m=0$] in {electric field $\gamma_F=-10$} (weak asymmetry by $z$ - axis  i.e. by minor ellipsoid axis).Eigenfunctions of sixth order of PT of 2D BVP for  \textit{prolate} SQD $c=2.5$, $a=0.5$,[$t=n_{\rho p}=0$, $n=n_p=0,1,2,3,4$, $m=0$] in {electric field $\gamma_F=-1$} (asymmetry by $z$ - axis i.e. by major ellipsoid axis).
}
\label{fun35a}
\end{figure}

 One can see that when the parameter $\gamma_{F}$ increases
 the eigenvalues ${\cal E}_{t}$
 decrease faster for SQD, slower for PSQD and even more slower for OSQD,
 because the influence of the electric field for OSQD at $c=0.5$ is essentially weaker than for PSQD at $c=2.5$.
With increasing $\gamma_{F}$ a series of exact crossings of eigenenergies
with different values of quantum numbers for PSQD and OSQD occur at  $\gamma_{F}\gtrsim 20$
and a series of avoided crossings for SQD occur at $\gamma_{F}\gtrsim 10$.
With further growth of the parameter they first increase and then begin to decrease.
Indeed, with the growth of $\gamma_{F}$ the eigenfunctions with smaller number of nodes in the longitudinal variable $z$ are localized (see Fig.\ref{fun35a}) in the vicinity of the equilibrium point, and the corresponding eigenenergies decrease.  Increasing the number of nodes is accompanied with delocalization of the wave functions, and the corresponding eigenenergies  increase and then decrease again.
For PSQD the density of states per unit energy for the eigenfunction with the same number of nodes $n_{\rho p}$ in the transverse variable $\rho$ is greater (i.e., the separation between the adjacent energy levels is smaller)
than the density of states for the function having the same number of nodes $n_{zp}$ in the longitudinal variable $z$.
For this reason   in Fig. \ref{exact} one can see three crossing series of curves  with different number of $\rho$-nodes $n_{\rho p}=0,1,2$, the  lower of them
(e.g., with $a=0.5$, $n_{\rho p}=0$, and $n_{zp}$ from $0$ to $12$) are decreasing at all
$\gamma_F\geq0$, while the upper ones (e.g., with $a=0.5$, $n_{\rho p}=0$, and $n_{zp}$ starting from 13)
with the energies, exceeding that of the state ($n_{\rho p}=1$)
 without $z$-nodes ($n_{zp}=0$), increase from the beginning and then start to decrease.
Thus, at small $\gamma_F$ the energy levels for the groups of states with even $n_{\rho p}=0,2...$
and odd $n_{\rho p}=1,3...$ number  of nodes are repulsing and crossing.

For OSQD,  on the contrary, the number of energy levels per unit energy for the eigenfunctions
having the same number $n_{\rho o}$ of $\rho$-nodes is smaller
(i.e., the separation  between the adjacent levels is larger) than that for the eigenfunctions
having the same number $n_{z o}$ of $z$-nodes.
Therefore, in Fig. \ref{exact} one can see four crossing series of almost 'parallel' curves
with different number $n_{zo}=0,1,2,3$ of  $z$-nodes.

For OSQD and PSQD the crossings of the energy levels that occur with increasing $\gamma_F$ are similar
to the exact crossings of the energy levels with decreasing $c$ semiaxis in OSQD and PSQD without electric  field ($\gamma_F=0$), i.e., we observe the accidental degeneracy, which is known to be generally associated with the existence of an additional integral of motion\cite{Yaf12}
 and with the separability of variables in oblate and prolate spheroidal coordinate systems. Thus, from our observations it follows that an additional approximate  integral of motion should exist.

For SQD eigenfunction with different numbers of $\rho$- and $z$-nodes, $n_\rho$ and $n_z$,  and with increasing $\gamma_F$ the
series of crossings become mixed. Note, that the eigenenergies of the states
with the same z-parity at $\gamma_F=0$ are repulsed with increasing
$\gamma_F$ (e.g., [$t=9$, $n=1$, $l=5$, ${\cal
E}_9(\gamma_F=0)=14.01$] and [$t=10$, $n=3$, $l=0$, ${\cal
E}_{10}(\gamma_F=0)=14.21$]), but the states with different z-parity are
attracted (e.g., [$t=7$, $n=1$, $l=4$, ${\cal
E}_7(\gamma_F=0)=10.71$] and [$t=8$, $n=2$, $l=2$, ${\cal
E}_{8}(\gamma_F=0)=13.24$]).  This fact should be also associated  with the
existence of approximate integrals of motion.
Indeed, from Fig \ref{exact} one can see that for SQD at $a=c=2.5$
with increasing $\gamma_F$ the series of exact crossings
appear.

\section{The PTLJ  in nondiagonal adiabatic approximation}

We expand the potentials (\ref{eq10o}) and (\ref{eq10p}) of {  the
BVP} (\ref{sp23})  and (\ref{sp19a})
 in Taylor series in the vicinity of $x_s=0$:
\begin{eqnarray}
E_{i}(x_{s})=E_{i;0}+\sum_{k=1}^{k_{\max}}  \frac{E_{
i;0}}{\tau^{2k}}x_{s}^{2k},\quad
U_{ij}(x_{s})=U_{ij;0}+\sum_{k=1}^{k_{\max}} \frac{\tilde U_{ij;k}}{\tau^{2k}}x_{s}^{2k},\label{A27}\\
 H_{ij}(x_{s})= \sum_{k=1}^{k_{\max}}
k\frac{H_{ij;0}}{\tau^{2k}}x_{s}^{2k}, Q_{ij}(x_{s})=
\sum_{k=1}^{k_{\max}} \frac{Q_{ij;0}}{\tau^{2k-1}}x_{s}^{2k-1},
\nonumber
\end{eqnarray}
 {where $\tilde U_{ij;k}=\frac{(2k-3)!!}{(2k)!!}U_{
ij;0}$ and the parameter $\tau$ equals $\tau=a$ for OSQD,  and $\tau=c$
for PSQD}.
Substitution of expansions (\ref{A27}) into Eq. (\ref{sp23}) leads to
the BVP for a set of ODEs of slow subsystem with respect to the
unknown vector functions ${\mbox{\boldmath $\chi$}}_t(x_s ) = (\chi
_{1;t} (x_s ),...,\chi _{j_{\max;t } } (x_s ))^T$ corresponding to the
unknown  eigenvalues $2E_t\equiv{\cal E}_t$:
\begin{eqnarray}&& \label{A32}
\left( \mathbf{D}^{(0)}
+(E_{i;0}-{\cal E}_t)+ \check{V}_s(x_s)+\sum_{k=1}^{k_{\max}} \frac{E_{i;0}+k H_{ii;0}}{\tau^{2k}}x_{s}^{2k}\right)\chi_{i;t}(x_s)
\\&&+\sum_{j\neq i}^{j_{\max}}\sum_{k=1}^{k_{\max}}\left( \frac{\tilde U_{ij;k}}{\tau^{2k}}x_{s}^{2k}
+k\frac{H_{ij;0}}{\tau^{2k}}x_{s}^{2k}
+(2k-1)\frac{Q_{ij;0}}{\tau^{2k-1}}x_{s}^{2k-2}
+2\frac{Q_{ij;0}}{\tau^{2k-1}}x_{s}^{2k-1}\frac{d}{dx_s}
\right)\chi_{j;t}(x_s)=0,\nonumber
\end{eqnarray}
 where $\tilde U_{ij;k}$ is given by the expansion (\ref{A27}) and $\check{V}_s(x_s)=0$ for OSDQ;
   $U_{ij}(x_{s})=0$ and $\check{V}_s(x_s)={\gamma_{F}}z$ for PSDQ.
We choose the unperturbed operator to have the eigenvalues and basis functions
of 2D and 1D oscillators.
{  For the OSQD} (2D oscillator) with respect to the scaled slow
{variable} $x$ we have: $x_s=\rho = \sqrt { x / \sqrt{E_f })} $,  where
$E_f=(E_{i';0}+H_{i'i';0})/(4a^2)= {\omega_{i'}^2/4}$, i.e., the
adiabatic frequency, {  at given $i'=n_o$}
\begin{eqnarray} &&
L\left( n \right) = \mathbf{D}^{(0)}- E
^{(0)},\quad\mathbf{D}^{(0)}=- \left( {\frac{d}{dx}x\frac{d}{dx} -
\frac{x}{4} - \frac{m^2}{4x}} \right), \quad E^{\left(0\right)}
\equiv E_{n,m}^{(0)} = n+ (\vert m\vert +1) / 2, \nonumber
\\&&
\Phi _{q }^{^{\left( 0 \right)}}(x)=
\frac{\sqrt{q!}x^{|m|/2}\exp(-x/2)L_q^{|m|}(x)}{\sqrt{(q+|m|)!}},
\quad  \int_0^\infty \Phi _{q }^{^{\left( 0 \right)}}(x) \Phi
_{q'}^{^{\left( 0 \right)}}(x)dx=\delta_{qq'}. \label{zo40}
\end{eqnarray}
Therefore, the action of the operators $L(n)$  and $x$
{  on the function $\Phi_{q}^{(0)}(x)\equiv\Phi_{q,m}^{(0)}(x)$} is
determined by the recurrence relations~\cite{stigun}
\begin{eqnarray} &&
L(n)\Phi_{q,m}^{(0)}(x)=(q-n)\Phi_{q,m}^{(0)}(x),\nonumber\\&&
x\Phi_{q,m}^{(0)}(x)=-\sqrt{q+|m|}\sqrt{q}\Phi_{q-1,m}^{(0)}(x)+ \nonumber\\&&
+(2q+|m|+1)\Phi_{q,m}^{(0)}(x)-\sqrt{q+|m|+1}\sqrt{q+1}\Phi_{q+1,m}^{(0)}(x),\label{zo40r}\\&&
x\frac{d\Phi_{q,m}^{(0)}(x)}{dx}=-\sqrt{q+|m|}\sqrt{q}\Phi_{q-1,m}^{(0)}(x)/2  \nonumber\\&&
-\Phi_{q,m}^{(0)}(x)/2+\sqrt{q+|m|+1}\sqrt{q+1}\Phi_{q+1,m}^{(0)}(x)/2.
\nonumber
\end{eqnarray}

{  For  PSQD} (1D oscillator) with respect to the scaled slow
variable $x$
$x_s=x/\sqrt[4]{E_f}$, where
$E_f=(E_{i';0}+H_{i'i';0})/c^2=\omega_{i'}^2$, i.e., the adiabatic
frequency,
{  at given $i'=n_p$,} we have
\begin{eqnarray} &&
L\left( n \right) = \mathbf{D}^{(0)}-
E^{\left(0\right)},\quad\mathbf{D}^{(0)} = - \frac{d ^2}{dx^2} + x^2
,\quad E^{\left(0\right)} \equiv E_{n}^{(0)} = 2n + 1,\quad
n=0,1,...., \nonumber
\\&&
\Phi _{q }^{^{\left( 0 \right)}}(x)=
\frac{\exp(-x^2/2)H_q(x)}{\sqrt[4]\pi\sqrt{2^q}\sqrt{q!}}, \quad
\int_{-\infty}^\infty \Phi _{q }^{^{\left( 0 \right)}}(x) \Phi
_{q'}^{^{\left( 0 \right)}}(x)dx=\delta_{qq'}. \label{zp40}
\end{eqnarray}

Correspondingly  action of operators $L(n)$,  $x$  and
$\frac{d}{dx}$
{  on function$ \Phi_{q}^{(0)}(x)$}is determined by recurrence
relations \cite{stigun}
\begin{eqnarray}&&
L(n)\Phi_{q}^{(0)}(x)=2(q-n)\Phi_{q}^{(0)}(x),\nonumber\\ &&
x\Phi_{q}^{(0)}(x)=\frac{\sqrt{q}}{\sqrt{2}}\Phi_{q-1}^{(0)}(x)
+\frac{\sqrt{q+1}}{\sqrt{2}}\Phi_{q+1}^{(0)}(x),
\label{zp40r}\\ &&
\frac{d}{dx}\Phi_{q}^{(0)}(x)=\frac{\sqrt{q}}{\sqrt{2}}\Phi_{q-1}^{(0)}(x)
-\frac{\sqrt{q+1}}{\sqrt{2}}\Phi_{q+1}^{(0)}(x).
\nonumber
\end{eqnarray}

The eigenfunctions (\ref{A32})as functions of the new scaled variable $x$
are sought in the form of expansion over the basis of the normalized functions
$\Phi^{(0)}_{q}(x)$, $q=0,1,....$ of the 2D or 1D
oscillators with unknown coefficients $b_{j,s}$:
\begin{eqnarray}\label{A33}\chi_{j;t}(x)=\sum_{q=0}^{q_{\max}} b_{j,q;t}\Phi^{(0)}_{q}(x),\quad b_{j,q<0;t}=b_{j,q>q_{\max};t}=0.
\end{eqnarray}
{  Below we demonstrate that such expansions are appropriate for getting
approximate solutions in the lower part of the BVP spectrum
(\ref{sp23}) and (\ref{sp19a}).}
Substitution of the expansion (\ref{A33}) into (\ref{A32}) yields the
set of equations
\begin{eqnarray}\label{A33eq1}
&&\!\!\!\!\!\!\!\!\sum_{q=0}^{q_{\max}}{\hat {\bf A}}_{ii}
b_{i,q;t}\Phi^{(0)}_{q}(x) +\sum_{j\neq
i=1}^{j_{\max}}\sum_{q=0}^{q_{\max}}{\hat {\bf
A}}_{ij}b_{j,q;t}\Phi^{(0)}_{q}(x)
 =\sum_{q=0}^{q_{\max}} \kappa^{-2} {\cal E}_t
 E_f^{-1/2}b_{i,q;t}\Phi^{(0)}_{q}(x),
\\&&\nonumber
\!\!\!\!\!\!\!\!{\hat{ \bf A}}_{ii}=\left(\mathbf{D}^{(0)}+ \check{V}_s(x)E_f^{-3/4}
+ \kappa^{-2}E_{i;0} E_f^{-1/2}+\kappa^{-2}\sum_{k=1}^{k_{\max}}
\frac{E_{i;0} +k
H_{ii;0}}{{\tau}^{2k}E_f^{(k+1)/2}}x^{2k}\right),
\\&&\nonumber
\!\!\!\!\!\!\!\!{\hat {\bf
A}}_{ij}=\kappa^{-2}\sum_{k=1}^{k_{\max}}\left(\frac{\tilde U_{ij;k}+k H_{ij;0}}{\tau^{2k}E_f^{(k+1)/2}}x_{s}^{2k}
+\frac{Q_{ij;0}}{{\tau}^{2k-1}E_f^{k/2}}\left((2k-1)x^{2k-2}
+2x^{2k-1}\frac{d}{dx}
\right)\right),
\end{eqnarray}
where $\kappa=2$ and $\check{V}_s(x_s)=0$ for OSQD;  $\kappa=1$
and $\check{V}_s(x)={\gamma_{F}}x$ for PSQD.
Applying the relations (\ref{zo40r}) or (\ref{zp40r}) to get
first derivatives of the basis functions, we get the expressions for the action of operators ${\hat{\bf A}}_{ij}$:
\begin{eqnarray}\label{A33eq4}
{\hat{\bf
A}}_{ij}\Phi^{(0)}_{q}(x)=\sum_{q'=0}^{q_{\max}}\alpha_{ij;qq'}\Phi^{(0)}_{q'}(x)
\end{eqnarray}
and, hence, the algebraic eigenvalue problem with respect to the unknown
$E_t$ and $b_{j,q;t}$
\begin{eqnarray}\label{A33eq5}
\sum_{q=0}^{q_{\max}}\alpha_{ii;q'q}b_{i,q;t} +\sum_{j\neq
i=1}^{j_{\max}}\sum_{q=0}^{q_{\max}}\alpha_{ij;q'q}b_{j,q;t}
 = \kappa^{-2}{\cal E}_t E_f^{-1/2}b_{i,q;t}.
\end{eqnarray}
In the matrix form it reads as
$$ \textbf{AB}_t =\kappa^{-2}{\cal E}_t E_f^{-1/2}\textbf{B}_t,\quad  \textbf{B}_{t'}^T\textbf{B}_t=\delta_{tt'}, $$
where
$\textbf{B}_t=(b_{1,0;t},b_{1,1;t},...,b_{1,q_{\max};t},b_{2,0;t},...,b_{j_{\max},q_{\max};t})^T$
is a vector with dimension of $j_{\max}(q_{\max}+1)$, and $\textbf{A}$
is a positive defined symmetric matrix having the dimensions
$(j_{\max}(q_{\max}+1))\times(j_{\max}(q_{\max}+1))$ with the elements
$A_{(q_{\max}+1)(i-1)+q+1,(q_{\max}+1)(j-1)+q'+1}=\alpha_{ij;qq'}$.

 {Note}, that the approximation with nonzero elements on the diagonal of the matrix
$\textbf{A}=\{\alpha_{ii;q'q}\}_{q',q=0}^{(q_{\max})}\delta_{i=i_{0},j=i_{0}}$,
obtained by the action of the diagonal operator ${\hat {\bf
A}}_{ii}$, Eq. (\ref{A33eq1}), on the basis function $\Phi^{(0)}_{q}(x)$, Eq.(\ref{A33eq4}), gives the  diagonal adiabatic approximation (AA) of PTLJ solution
(\ref{A33eq5}), i.e., ${\cal E}_t\approx {\cal E}_{i;n}^{}, n=0,1,...$ at each fixed $i$.
Such adiabatic classification of the eigenenergies is used in Tables discussed below.

\begin{table}\caption{
The convergence of eigenenergies ${\cal E}_{t}$ of Eq. (\ref{A33eq5})
vs  order $k_{\max}$ of approximation of effective potentials from
(\ref{A27})
for $j_{\max }=4$ and $q_{\max }=60$
basis functions { at  $\gamma_F=0$}.
Fast and slow variables $x_{f}=z$ and $x_{s}=\rho$ (\textit{oblate} SQD and spherical QD),
number of nodes $i=(n_{zo}=n_{o}-1,n_{\rho o})$.}\label{fullo4}
\begin{tabular}{|l||l|l|l||l|l|l|}  \hline
 $k_{\max } $  & \multicolumn{3}{c||}{     $a=2.5$,           $c=0.5$    } &          \multicolumn{3}{c|}{     $a=2.5$,           $c=2.5$    }          \\ \hline $(n_{zo},n_{\rho o})$ &(0,0)&(0,1)&(2,0)&(0,0)&(0,1)&(2,0) \\ \hline
8 & 12.668\;20 &19.067\;45 &96.714\;86 &1.192\;415&2.998\;982& 5.325\;360\\
12& 12.749\;67 &19.813\;83 &96.750\;70 &1.377\;572&4.088\;539& 5.868\;629\\
20& 12.784\;07 &19.838\;42 &96.751\;72 &1.132\;323&5.084\;082& 6.735\;687\\ \hline
N$(j_{\max }=4)$ &  12.764\;82 & 20.040\;74 & 96.752\;15    & 1.579\;273 & 5.316\;872    & 6.317\;204 \\ \hline
\end{tabular}  \end{table}

\begin{table}\caption{
The convergence of eigenenergies ${\cal E}_{t}$ of Eq. (\ref{A33eq5})
vs  order $k_{\max}$ of approximation of effective potentials from
(\ref{A27})
for $j_{\max }=4$ and $q_{\max }=60$
basis functions { at  $\gamma_F= 0$}.
Fast and slow variables $x_{f}=\rho$ and $x_{s}=z$ (\textit{prolate} SQD  and spherical QD),
 number of nodes $i=(n_{\rho p},n_{zp})$.}\label{fullp4}
\begin{tabular}{|l||l|l|l||l|l|l|}  \hline
  $k_{\max } $  & \multicolumn{3}{c||}{     $c=2.5$,           $a=0.5$    } &          \multicolumn{3}{c|}{     $c=2.5$,           $a=2.5$    }          \\  \hline $(n_{\rho p},n_{zp})$&(0,0)&(0,2)&(1,0)&(0,0)&(0,2)&(1,0)\\ \hline
8 & 25.179\;14 &34.076\;77 &126.445\;9 &1.471\;911&4.270\;174& 5.614\;892\\
12& 25.199\;62 &34.468\;84 &126.456\;0 &1.536\;121&4.716\;984& 6.188\;144\\
20& 25.201\;16 &34.522\;02 &126.456\;1 &1.563\;492&5.182\;198& 6.266\;533\\ \hline
N$(j_{\max }=4)$ & 25.201\;21 &34.525\;12 &126.456\;1 &1.579\;239&5.316\;732& 6.317\;058\\ \hline
\end{tabular}  \end{table}

The convergence of eigenenergies of Eq. (\ref{A33eq5}) vs the order
$k_{\max}$ of approximation of the effective potentials (\ref{A27})
for $j_{\max }=4$ and $q_{\max }=60$  is shown in
Tables \ref{fullo4} and \ref{fullp4} for OSDD, PSQD, and SQD {at
$\gamma_{F}=0$ and in Table \ref{fullp5} at $\gamma_{F}=-10$} for
PSQD and SQD.
Table \ref{fullp4} shows that for PSQD we have
upper estimate and monotonic convergence with increasing
$k_{\max}$ to the numerical results at $j_{\max }=4$. Similar
behavior is observed for OSQD, however the accuracy of approximation of
the effective potentials is worse, especially for the lowest effective
potential $i'=1$, corresponding to the ground state of the fast subsystem,
because the upper estimates are violated. These Tables show also that
such expansions have faster convergence for strongly
oblate or prolate spheroidal QDs than for spherical ones.

\begin{table}\caption{
The convergence of eigenenergies ${\cal E}_{t}$ of Eq. (\ref{A33eq5})
vs  order $k_{\max}$ of approximation of effective potentials from
(\ref{A27})
for $j_{\max }=4$ and $q_{\max }=60$
basis functions { at  $\gamma_F=-10$}.
Fast and slow variables $x_{f}=\rho$ and $x_{s}=z$ (\textit{prolate} SQD and spherical QD),
number of nodes $i=(n_{\rho p},n_{zp})$.}\label{fullp5}
\begin{tabular}{|l||l|l|l||l|l|l|}  \hline
  $k_{\max } $  & \multicolumn{3}{c||}{     $c=2.5$,           $a=0.5$,  $\gamma_{F}=-10$  } &
      \multicolumn{3}{c|}{     $c=2.5$,           $a=2.5$,  {  $\gamma_F=-10$  } }          \\  \hline $(n_{\rho p},n_{zp})$&(0,0)&(0,2)&(1,0)&(0,0)&(0,2)&(1,0)\\ \hline
8 & 20.221\;65 &30.913\;36 &125.306\;2  &-19.673\;98 &-5.378\;707& -1.784\;110  \\
12& 20.607\;33 &32.375\;40 &125.331\;6 &-15.348\;50 &-6.881\;266& -2.605\;091\\
20& 20.658\;46 &32.674\;45&125.332\;2 &-12.194\;45&-2.204\;160&-1.336\;853
 \\ \hline N$(j_{\max }=4)$ & 20.662\.03 &32.708\;77
&125.332\;2 &{-10.844\;02}&-1.511\;063& 1.129\;039\\ \hline
\end{tabular} \end{table}

\section{
PTRS for BVP for OSQD in electric field by fast variables }
To have an analytic representation of the matrix elements (\ref{sp23a})
for small $\gamma_F$, one can use $\check
V_{f}(x_{f};{x_s})=2\gamma_Fz$,     $ {\check
V_{fs}(x_{f},{x_s})=0}$  as  potentials for OSQD instead
of the potentials (\ref{eq10o}) introduced in Section 2.1.
Then we arrive at the Sturm-Lioville problem for the OSQD in fast variable expressed in the form
\begin{eqnarray}\label{F1}&&
\left(-\frac{d^2}{dz^2}-\epsilon z-E_j(\rho)\right)B_j(z;\rho)=0,
\\&& \nonumber
\langle B_i(\rho)|B_j(\rho)\rangle=\int_{-L(\rho)/2}^{L(\rho)/2}B_i(z;\rho)B_j(z;\rho)dz=\delta_{ij},
\end{eqnarray}
where {$\epsilon=\gamma_{F}$} is the electric field strength considered
here as a formal parameter of the PT, implying a small interval
$\rho\in(0,L(\rho)=2c\sqrt{1-\rho^2/a^2})$ of the scalar product
$\langle B_i(\rho)|B_j(\rho)\rangle$. The solutions $B_j^{(0)}(z;\rho)$
and $E_j^{(0)}(\rho)$  of the unperturbed equation (at $\epsilon=0$)
have the form
\begin{eqnarray}&&
\{B_j^{(0)}(z;\rho),E_j^{(0)}(\rho)\}=\left\{\begin{array}{ll}\{B_j^{s}(z;\rho),E_j^{s}(\rho)\},&\mbox{ for even }j=2,4,...;\\\{B_j^{c}(z;\rho),E_j^{c}(\rho)\},&\mbox{ for odd }j=1,3,... \end{array}\right\},
\end{eqnarray}
where
\begin{eqnarray*}&&
B_j^{s}(z;\rho)=\sqrt{2/L(\rho)}\sin(\pi jz/L(\rho)),\quad B_j^{c}(z;\rho)=\sqrt{2/L(\rho)}\cos(\pi jz/L(\rho)),
\\ &&
E_j^{s}(\rho)=(\pi j/L(\rho))^2,\quad E_j^{c}(\rho)=(\pi j/L(\rho))^2.
\end{eqnarray*}
We seek for the eigenfunctions $B_j(z;\rho)$ and the eigenvalues $E_j(\rho)$
in the form of power expansions
\begin{eqnarray}\label{F8}
B_j (z;\rho)=\sum_{k=0}^{k_{\max}}\epsilon^kB_j^{(k)}(z;\rho),\quad E_j(\rho)
=\sum_{k=0}^{k_{\max}}\epsilon^kE_j^{(k)}(\rho).
\end{eqnarray}
Substituting Eq. (\ref{F8}) into Eqs. (\ref{F1}) and equating the coefficients at the same powers
of $\epsilon$, we arrive at
the system of inhomogeneous differential equations with respect to
corrections $E_j^{(k)}$ and $B_j^{(k)}(z;\rho)$:
\begin{eqnarray}&&\label{F4}
\!\!\!\!\!\!\!\!\!\!\!
\left(-\frac{d^2}{dz^2}-E_j^{(0)}(\rho)\right)B_j^{(k)}(z;\rho)
=\left(z+E_j^{(1)}(\rho)\right)B_j^{(k-1)}(z;\rho)
+\sum_{p=2}^kE_j^{(p)}(\rho) B_j^{(k-p)}(z;\rho),\\&& \nonumber
\sum_{p=0}^k\langle B_j^{(p)}(\rho)|B_j^{(k-p)}(\rho)\rangle=0.
\end{eqnarray}
In each $k$-th order of the perturbation theory (PT)
the solutions becoming zero at the boundary points $(z=\pm L(\rho)/2$
are sought in the form
\begin{eqnarray}\label{F5}
B_j^{(k)}(z;\rho)=\left\{\begin{array}{ll}\sum\limits_{\nu=0}^{\nu_{\max}}
B_j^{s}(z;\rho)  S^{(k)}_\nu z^\nu +
(z^2-(L(\rho)/2)^2)\sum\limits_{\nu=0}^{\nu_{\max}-2}
B_j^{c}(z;\rho) C^{(k)}_{\nu+2}z^\nu,
&j=2,4,...\\\sum\limits_{\nu=0}^{\nu_{\max}} B_j^{s}(z;\rho)
C^{(k)}_\nu z^\nu +
(z^2-(L(\rho)/2)^2)\sum\limits_{\nu=0}^{\nu_{\max}-2}
B_j^{c}(z;\rho) S^{(k)}_{\nu+2}z^\nu, &j=1,3,...
\end{array}\right\} .
\end{eqnarray}
Substituting Eq. (\ref{F5}) into the corresponding equation (\ref{F4}) of the
$k$-th order of the PT, and extracting the coefficients at
$B_j^{s}(z;\rho)z^\nu$ and $B_j^{c}(z;\rho)z^\nu$,
$\nu=0,...,\nu_{\max}$, we arrive at the set of algebraic equations
with respect to unknowns $E_j^{(k)}(\rho)$, $S^{(k)}_\nu$ and
$Ñ^{(k)}_\nu$, for even $j$:
\begin{eqnarray*}\label{F80} &&
 -(-1)^j (\nu+1)(L(\rho)/2)\pi j C^{(k)}_{\nu+3}-(\nu+2)(\nu+1)S^{(k)}_{\nu+2}+(-1)^j 2(\nu+1)\pi j C^{(k)}_{\nu+1}
 \\&& \quad-E_j^{(1)}(\rho) S^{(k-1)}_{\nu}-S^{(k-1)}_{\nu-1}
  -\sum_{p=2}^{k-1}E_j^{(p)}(\rho) S^{(k-p)}_{\nu}
 -E_j^{(k)}(\rho)\delta_{\nu,0}=0,\\&&\label{F81}
 + (\nu+1)(\nu+2) (L(\rho)/2)^2 C^{(k)}_{\nu+4} -(\nu+2)(\nu+1)C^{(k)}_{\nu+2}-(-1)^j 2(\nu+1)\pi j S^{(k)}_{\nu+1}
 \\&&\quad-E_j^{(1)}(\rho)(C^{(k-1)}_{\nu}-(L(\rho)/2)^2C^{(k-1)}_{\nu+2} )-C^{(k-1)}_{\nu-1}+(L(\rho)/2)^2C^{(k-1)}_{\nu+1} )
 \\&&\quad-\sum_{p=2}^{k-1}E_j^{(p)}(\rho) (C^{(k-p)}_{\nu}-(L(\rho)/2)^2C^{(k-p)}_{\nu+2} )=0.
\end{eqnarray*}
For odd $j$ the same unknowns are calculated using the equations
(\ref{F80}) (\ref{F81})
with the replacement $C^{(p)} \leftrightarrows S^{(p)}$.
The unknowns $Ñ^{(k)}_0$ for even $j$ and $S^{(k)}_0$  for odd $j$ are determined from the respective conditions:
\begin{eqnarray}&&\label{F87}
\sum_{p=0}^k \sum_{\nu,\nu'}\left(
S^{(p)}_{\nu}S^{(k-p)}_{\nu'}\langle B_j^{s}(\rho)|z^{\nu+\nu'}|B_j^{s}(\rho)\rangle
\right.\\&& \nonumber\left.+
[(C^{(p)}_{\nu}S^{(k-p)}_{\nu'}+S^{(p)}_{\nu}C^{(k-p)}_{\nu'})
+(L(\rho)/2)^2(C^{(p)}_{\nu+1}S^{(k-p)}_{\nu'+1}+S^{(p)}_{\nu+1}C^{(k-p)}_{\nu'+1})]
\langle B_j^{s}(\rho)|z^{\nu+\nu'}|B_j^{c}(\rho)\rangle
\right.\\&& \nonumber \left.+
[(C^{(p)}_{\nu}C^{(k-p)}_{\nu'})
-2(L(\rho)/2)^2C^{(p)}_{\nu+1}C^{(k-p)}_{\nu'+1}
+(L(\rho)/2)^4C^{(p)}_{\nu+2}C^{(k-p)}_{\nu'+2}]
\langle B_j^{c}(\rho)|z^{\nu+\nu'}|B_j^{c}(\rho)\rangle
\right),
\end{eqnarray}
and $S^{(k)}_0$  for odd $j$ is calculated from the equation
(\ref{F87}) with the replacement $C^{(p)}\leftrightarrows S^{(p)}$.
This algorithm was implemented using the Maple environment. The run was performed until the maximal order of the PT
$k_{max}=8$. Below we present the first few coefficients of
the eigenvalue expansion, truncated by the terms proportional to $\epsilon^6=\gamma_F^6$
\begin{eqnarray}&&
E_j(\rho)=\frac{\pi^2 j^2}{((L(\rho))^2}
+\frac{((L(\rho)) ^4(\pi^2 j^2-15)}{48\pi^4 j^4}\epsilon^2
+\frac{(L(\rho)) ^{10}(1980-210\pi^2 j^2 +\pi^4 j^4 )}{2304\pi^{10} j^{10}}\epsilon^4,\label{F39}
\end{eqnarray}
the eigenfunctions truncated by the terms proportional to $\epsilon^2=\gamma_F^2$
\begin{eqnarray*}&&B_j(z;\rho)=\left\{\begin{array}{ll}
B_j^{s}(z;\rho)+\left(-\frac{(L(\rho))^2z B_j^{s}(z;\rho)}{4\pi^2 j^2}+\frac{L(\rho)(z^2-(L(\rho)/2)^2)B_j^{c}(z;\rho)}{4\pi j}
 \right)\epsilon,&j=2,4,...\\
 B_j^{c}(z;\rho)+\left(-\frac{(L(\rho))^2z B_j^{c}(z;\rho)}{4\pi^2 j^2}-\frac{L(\rho)(z^2-(L(\rho)/2)^2)B_j^{s}(z;\rho)}{4\pi j}
 \right)\epsilon,&j=1,3,...\end{array}\right\},
\end{eqnarray*}
and the diagonal effective potentials, truncated by the terms proportional to
 $\epsilon^6=\gamma_F^6$
\begin{eqnarray}&&\nonumber
H_{jj}(z)=\left(\frac{dL(\rho)}{d\rho}\right)^2
\left(\frac{\pi^2 j^2+3}{12(L(\rho))^2}
+\frac{(L(\rho)) ^{4}(-2880+258\pi^2 j^2 +7\pi^4 j^4 )}{576\pi^{6} j^{6}}\epsilon^2\right.
\\&&\left.+\frac{L(\rho) ^{10}(3510000-389880\pi^2 j^2 +3321\pi^4 j^4 +13\pi^6 j^6 )}{27648 \pi^{12} j^{12}}\epsilon^4
\right).\label{F40}
\end{eqnarray}

\section{The PTRS in the diagonal adiabatic approximation } \label{PTRS}
{ The desired solutions ofthe  original 2D BVP (\ref{2dbvp}) are
determined by the diagonal approximation of the Kantorovich
expansion(\ref{KE})} at fixed $m$
\[
\Psi_{i;n}^{m } (x_f ,x_s) \approx B_i^{} ( x_f;x_s )\chi_{i;n} (x_s
).
\]
{  The diagonal approximation of
{  the BVP} (\ref{sp23})  and (\ref{sp19a}) in the slow variable has
 the form}
\begin{eqnarray}\label{F41}
\left(-\frac{1}{x_s^d}\frac{d}{dx_s}x_s^d\frac{d}{dx_s}
+\frac{\tilde{m}}{x_s^2} + V_i(x_s ) -{\cal
E}_{i;n}\right)\chi_{i;n}(x_s)=0.
\end{eqnarray}
{  and the eigenfunctions satisfy the orthonormalization conditions on the
semiaxis $[ x_s^{\min}=0,{x_s^{\max}}=\infty)$ at $d=1$ for the
OSQD and on the axis $(x_s^{\min}=-\infty,{x_s^{\max}}=\infty)$ at
$d=0$ for the OSQD}
\begin{equation}
\label{sp19ab}
\int\nolimits_{x_s^{\min}}^{x_s^{\max}} \chi_{i;n}(x_s))\chi_{i,n'}
(x_s) (x_s)^{d}dx_s=\delta_{nn'}.
\end{equation}
Here $V_i(x_s )=\check V_s(x_s )+ E_i(x_s)+DH_{ii}(x_s)$, where the
parameter $D$ is $D=0$ for the crude adiabatic approximation and
$D=1$ for the adiabatic approximation;
 $\check V_s(x_s )=0$, $E_i(x_s)$ and $H_{ii}(x_s)$, Eqs.~(\ref{F39})--(\ref{F40}), for OSQD and $\check V_s(x_s )=2\gamma_Fz$, $E_i(x_s)$ and $H_{ii}(x_s)$, Eq. (\ref{eq10p}), for PSQD;
${\cal E}_{i;n}$ are {the  eigenenergies of a lower part of thespectrum
${\cal E}_{i;0}<{\cal E}_{i;1}<...<{\cal E}_{i;n}$ enumerated in the
ascending order by the number of nodes $n=0,1,2,...$ of the eigenfunctions
$\chi_{i;n}(x_s)$ at fixed adiabatic quantum numbers $i=n_o$ for
OSQD and $i=n_p$ for PSQD}.
The potential function $V_i(x_s) $ is expanded in powers of the small parameter $\varepsilon$
\begin{eqnarray}\label{F42}
V_i^{[j_{{\max} }]}(x_s )=V_i^{(0)}+\kappa^{-2}\omega_i^2x_s^2
+\kappa^{-2}\sum_{j=1}\nolimits^{j_{max}}V_i^{(j)}(x_{s})\varepsilon^j.
\end{eqnarray}
 For OSQD at the values of {the  parameters $d=1$}, $\varepsilon=c^{-2},\quad \kappa=2,\quad
\tilde{m}=m$ the coefficients $V_i^{(j)}$ are determined by
Taylor expansion of the effective potentials (\ref{F39}), (\ref{F40}) in
the vicinity of the equilibrium point $x_s=0$.
With the accuracy up to order of $O(\gamma_F^6)$ the coefficients $V_i^{(j)}$ and $\omega_i^2$ are expressed as:
\begin{figure}[t]
\epsfig{file=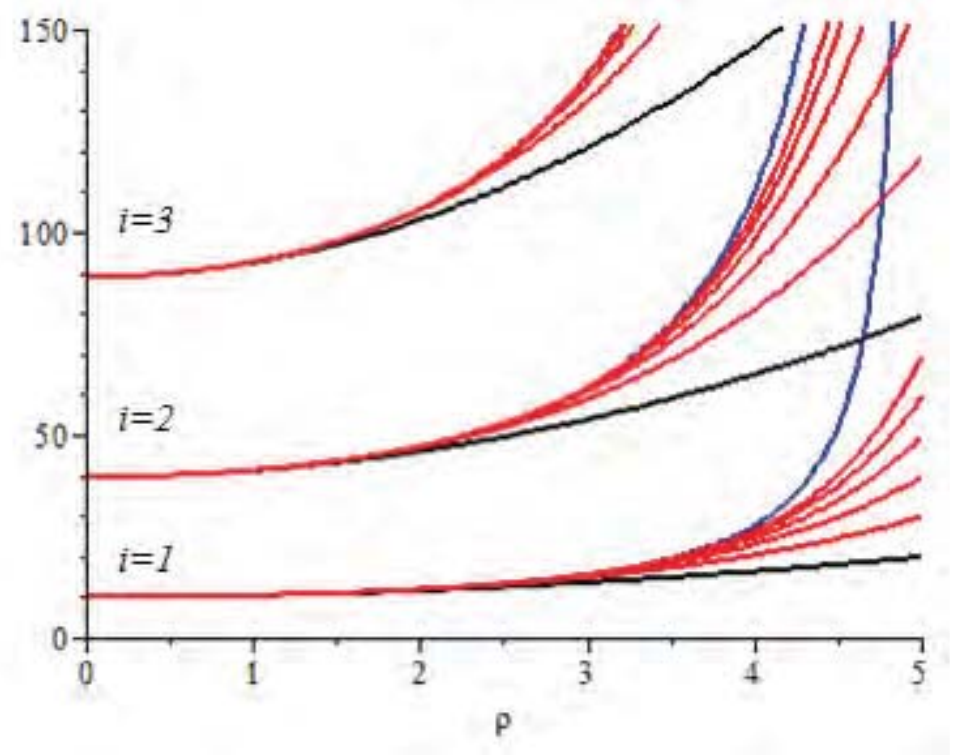,width=0.45\textwidth}\epsfig{file=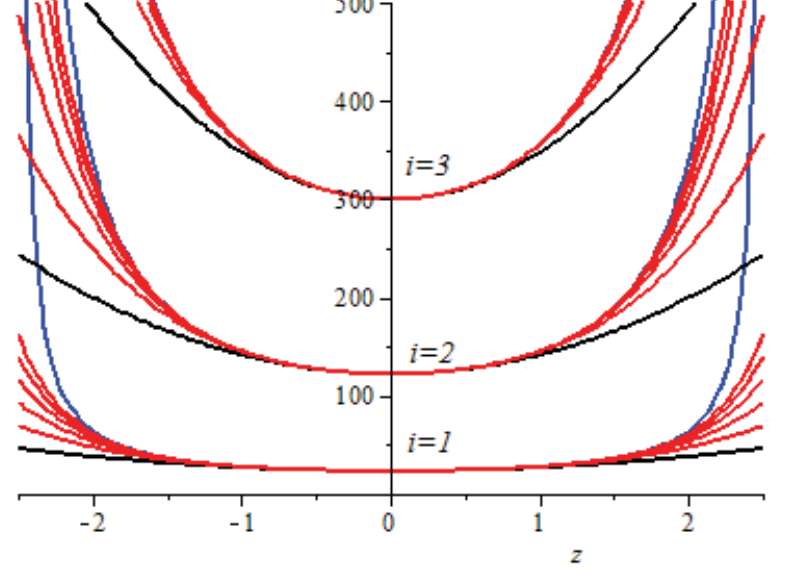,width=0.45\textwidth}
\caption{
Three potential functions $V_{i}(x_{s})$ for \textit{oblate} $x_s=\rho$ and \textit{prolate} $x_{s}=z$ spheroids and their power expansions
till sixth order with account of \textit{adiabatic frequencies} $\omega_i$ and \textit{lower bound shifts} $V_i^{(0)}$.
}\label{fig51}
\end{figure}
 \begin{eqnarray}\label{F42o}
&&
V_i^{(0)}=\frac{\pi^2n_o^2}{4c^2}+{\gamma_F^2}\frac{c^4(\pi^2n_o^2-15)}{3 \pi^4n_o^4}
   +{\gamma_F^4}\frac{4 c^{10}(\pi^4n_o^4-210\pi^2n_o^2+1980)}{9\pi^{10}n_o^{10}},
\\ \nonumber&&
\omega_i^2=\frac{\pi^2 n_o^2}{(ac)^2}+D\frac{3+\pi^2 n_o^2}{a^4}
+\gamma_F^2\left(-\frac{8c^4(\pi^2n_o^2-15)}{3a^2\pi^4n_o^4}+D\frac{4 c^{6}(7\pi^4n_o^4+258\pi^2n_o^2-2880)}{9a^4\pi^{6}n_o^{6}}\right)
\\\nonumber&&
+\gamma_F^4\left(-\frac{80 c^{10}(\pi^4n_o^4-210\pi^2n_o^2+1980)}{9a^2\pi^{10}n_o^{10}}
        +D\frac{16
        c^{12}(13\pi^6n_o^6+3321\pi^4n_o^4-389880\pi^2n_o^2-3510000)}{27a^4\pi^{12}n_o^{12}}\right),
\\\nonumber&&
V_i^{(j)}=\left(\frac{\pi^2 n_o^2}{(ac)^2}+jD\frac{3+\pi^2 n_o^2}{a^4}
+\gamma_F^2\left(-\frac{4c^4(\pi^2n_o^2-15)}{3a^4\pi^4n_o^4}\delta_{i2}
-D\frac{4c^{6}(7\pi^4n_o^4+258\pi^2n_o^2-2880)}{9a^6\pi^{6}n_o^{6}}\delta_{i2}\right)\right.
\\&&\left. \nonumber
+\gamma_F^4\left(\frac{16 c^{10}(\pi^4n_o^4-210\pi^2n_o^2+1980)}{9a^4\pi^{10}n_o^{10}}(10\delta_{i2}-10\frac{\delta_{i3}}{a^2}
+5\frac{\delta_{i4}}{a^4}-\frac{\delta_{i5}}{a^6})
 \right.
\right.
\\\nonumber&&\left.\left.       +D\frac{16 c^{12}(13\pi^6n_o^6+3321\pi^4n_o^4-389880\pi^2n_o^2-3510000)}{27a^6\pi^{12}n_o^{12}}
        (-4\delta_{i2}+6\frac{\delta_{i3}}{a^2}
-4\frac{\delta_{i4}}{a^4}+\frac{\delta_{i5}}{a^6})\right)
\right) x_{s}^{2j+2}.
\end{eqnarray}
  For PSQD at the values of  {the  parameters $d=0$}, $\varepsilon=1$, $\kappa=1$, $\tilde{m}=0$ the
  coefficients $V_i^{(j)}$ are sought in the form of a Taylor expansion in powers of $\bar x_{s}=(x_{s}-x_{0})$ and $\gamma_F$ of the
 effective potentials $V_i(x_s,\gamma_F)=E_{j}(x_{s})+DH_{jj}(x_{s})+\gamma_{F}x_{s} $, Eq.~(\ref{eq10p}).
The expansion coefficients   $x_{0}=\sum_k\tau_{2k+1}\gamma_F^{2k+1}$
are sought from the equilibrium condition $\left.\frac{\partial V_i(x_{s},\gamma_F)}{\partial  x_{s}}\right|_{x_s=x_0}=0$ at fixed $\gamma_F$.
With the accuracy up to $O(\gamma_F^5)$ the coefficients $V_i^{(j)}$ and $\omega_i^2$ are expressed as:
\begin{eqnarray}\nonumber
&&
V_i^{(0)}=-2\tau_1\gamma_F^2-2\tau_3\gamma_F^4
+ \alpha_{n_p,|m|}^2/(a^2)+D(1+\alpha_{n_p,|m|}^2)/(3c^{4})
\\\nonumber&&+\gamma_F^2 \tau_1^2  (\alpha_{n_p,|m|}^2/(a^2c^{2})+D(1+\alpha_{n_p,|m|}^2)2/(3c^{4})
)
\\\nonumber&&
+\gamma_F^4\tau_1 (\alpha_{n_p,|m|}^2(\tau_1^3+2c^2\tau_3)/(a^2c^4)
+2D(1+\alpha_{n_p,|m|}^2) (\tau_1^3+\tau_3c^2)/(3c^{6}),
\\\label{F42pom}\label{F42p}&&
\omega_i^2=
\left(\alpha_{n_p,|m|}^2/(a^2c^{2})+D(1+\alpha_{n_p,|m|}^2)/(3c^{4})
\right.
\\\nonumber&&\left.
+\gamma_F^2 \tau_1^2 6(\alpha_{n_p,|m|}^2/(a^2c^{4})+D(1+\alpha_{n_p,|m|}^2)2/(3c^{6})
)
\right.
\\\nonumber&&\left.
+\gamma_F^4\tau_1 (\alpha_{n_p,|m|}^2(15\tau_1^3+12c^2\tau_3)/(a^2c^6)
+D(1+\alpha_{n_p,|m|}^2) (15\tau_1^3+8\tau_3c^2)/(c^{8}) \right),
\\&& \nonumber V_i^{(j)}=\left(
+2 \gamma_F\tau_1(
(i+1)\alpha_{n_p,|m|}^2/(a^2c^{2i+2})
+ D (i+1)^2(1+\alpha_{n_p,|m|}^2)/(3c^{2i+4})
)
\right.
\\\nonumber&&\left.
+2 \gamma_F^3(i+1)(
 \alpha_{n_p,|m|}^2((2i^2+7i+6)\tau_1^3+3\tau_3c^2)/(3a^2c^{2i+4})
\right.
\\\nonumber&&\left.\qquad+D(1+\alpha_{n_p,|m|}^2)((2i^3\tau_1^3+11i^2+20i+12)\tau_1^3+3(i+1)\tau_3c^2)/(9c^{2i+6})
\right) \bar x_{s}^{2i+1}
\\\nonumber&&+
\left(
 \alpha_{n_p,|m|}^2/(a^2c^{2i+2})
+D(1+\alpha_{n_p,|m|}^2)(i+1)/(3c^{2i+4})
\right.
\\\nonumber&&\left.
+\gamma_F^2 \tau_1^2(i+2)(2i+3)(
\alpha_{n_p,|m|}^2\/(a^2c^{2i+4})
+D(1+\alpha_{n_p,|m|}^2)(i+2)/(3c^{2i+6})
)
\right.
\\\nonumber&&\left.
+\gamma_F^4\tau_1(i+2)(2i+3)(
\alpha_{n_p,|m|}^2((2i^2+11i+15)\tau_1^3+12c^2\tau_3)/(6a^2c^{2i+6})
\right.
\\\nonumber&&\left.\qquad+D(1+\alpha_{n_p,|m|}^2)
((2i^3+17i^2+48i+45)\tau_1^3+12(i+2)\tau_3c^2)
/(18c^{2i+8})
\right) \bar x_{s}^{2i+2},
\end{eqnarray}
where $\tau_{2k+1}$ is determined from the condition that the coefficient at $\bar x_{s}$  is zero:
\begin{eqnarray*}
\tau_1 = \frac{3a^2c^4}{3c^2\alpha_{n_p,|m|}^2+Da^2(1+\alpha_{n_p,|m|}^2)},
\quad
\tau_3 = -
   \frac{54a^6c^{10}(3c^2\alpha_{n_p,|m|}^2+2Da^2(1+\alpha_{n_p,|m|}^2))}
    {(3c^2\alpha_{n_p,|m|}^2+Da^2(1+\alpha_{n_p,|m|}^2))^4}.
\end{eqnarray*}
In Fig \ref{fig51} we show three potential functions $V_{i}(x_{s})$ for \textit{oblate} $x_s=\rho$
and \textit{prolate} $x_{s}=z$ spheroids and the convergence of the corresponding power expansions
till the sixth order with account of \textit{adiabatic frequencies} $\omega_i$ and
\textit{lower bound shifts}  $V_i^{(0)}$.

{  We choose the unperturbed operators of Eq.~(\ref{F41}) at
$\varepsilon=0$ in the expansion (\ref{F42}) in the form
(\ref{zo40})--(\ref{zp40r})} with the eigenvalues and the basis functions of
2D- and 1D- oscillators given in Section 3 {with respect to the scaled
coordinate $x$,
$x_s=\sqrt{2x/\omega_i}$ and $\bar x_s=x/\sqrt\omega_i$, where the
 adiabatic frequencies $\omega_i$ are defined  by Eqs. (\ref{F42o}) and (\ref{F42pom}) (at fixed $i'=n+1$), respectively}.
{According to (\ref{F42}), we seek for the eigenfunctions $\chi
_{i;n}(x_{s})$ and the eigenvalues ${\cal E} _{i;n}$ in the form of
expansions in powers of $\varepsilon$  with unknowns  $\Phi_n^{(k)}$
and $E^{(k)}_{n}$, omitting the notation $m$ for brevity:}
\begin{eqnarray}\label{F43}
\chi _{i;n} (x_s ) = \Phi _{n}^{(0)} + \sum_{k = 1}^{k_{\max}}
 \Phi_{n}^{(k)}(x_s)\varepsilon^k ,\quad \\\label{F43a}
 {\cal E} _{i;n} = V_{i}^{(0)} +\sum_{k=0}^{k_{\max}}{\cal E}_{i;n}^{(k)}
 = V_{i}^{(0)} + \kappa\omega_i \left(E_i^{(0)} +\sum_{k=1}^{k_{\max}}
 E_{n}^{(k)} \varepsilon^k\right).
 \end{eqnarray}
Substituting the expansions (\ref{F42}), (\ref{F43}) and (\ref{F43a})
into Eq. (\ref{F41}) and equating the terms with the same power of the parameter $\varepsilon$, we arrive at the recurrence set of inhomogeneous equations
of the PT with respect to the unknowns $E^{(k)}_{n}$ and
$\Phi^{(p)}_{n}(x)$:
 \begin{eqnarray}
 L(n)\Phi^{(0)}_{n}(x)&=& 0 \equiv f^{(0)}(x),\label{az228}\\
 L(n)\Phi^{(k)}_{n}(x)&=&   \sum_{p=0}^{k-1}
   ( E^{(k-p)}_{n}  -V^{(k-p)}_{i} )\Phi^{(p)}_{n}(x)
      \equiv  f^{(k)}(x),\quad k\geq1.\nonumber
 \end{eqnarray}
with the initial conditions (\ref{zo40}) and (\ref{zp40}) for OSQD and PSQD, respectively.
The solution of this problem is implemented in four steps.

Applying the relations (\ref{zo40r}) and (\ref{zp40r}), we expand the right-hand
side $f^{(k)}(x)$  and the solutions $\Phi^{(k)}(x)$ of Eqs.
(\ref{az228}) over {the basis of normalized}  states $\Phi_{n+s}^{(0)}(x)$, Eqs. (\ref{zo40}) and (\ref{zp40}):
 \begin{eqnarray} \label{az229}
 \Phi_{n}^{(k)}(x)=\sum_{s=-{s_{max}}}^{s_{max}}b_s^{(k)}\Phi_{n+s}^{(0)}(x),\quad
 f^{(k)}(x)=\sum_{s=-{s_{max}}}^{s_{max}}f_s^{(k)}\Phi_{n+s}^{(0)}(x).
 \end{eqnarray}
Then  a recurrent set of  linear algebraic equations for unknown
 coefficients $b_s^{(k)}$  and  corrections $E^{(k)}$ is
obtained
 \begin{eqnarray}
 s'b^{(k)}_{s} -f^{(k)}_{s}  = 0,\quad s=-{s_{max}},\ldots,{s_{max}},
 \end{eqnarray}
where $s'=s$ for OSQD and $s'=2s$ for PSQD. These equations are solved sequentially for $k = 1, 2, \ldots, k_{\max}$:
 \begin{eqnarray}&&
 f^{(k)}_{0} = 0 \quad \to E^{(k)};\quad
 b^{(k)}_{s} =f^{(k)}_{s}/s',\quad s=-{s_{max}},\ldots,{s_{max}},\quad s\neq0.
 \end{eqnarray}
The initial conditions for this {  procedure are}
$$ b^{(0)}_{s}  = \delta_{s0},\quad E^{\left(0\right)}=  (n+ (\vert \tilde{m}\vert +1) / 2)
\quad \mbox{or} \quad E^{\left(0\right)}=  (n+1) / 2).$$

To obtain the normalized wave function $\Phi_{j}(x)$ up to the
$k$-th order, the coefficients $b^{(k)}_{0}$ are determined by the
following relation:
 \begin{eqnarray}
 b_{0}^{(k)}=-\frac{1}{2\langle 0|0\rangle}\sum_{p=1}^{k-1}
 \sum_{s'=-{s_{max}}}^{{s_{max}}}\sum_{s=-{s_{max}}}^{{s_{max}}}
 b^{(k-p)}_{s} \langle s|s'\rangle b^{(p)}_{s'}.
 \end{eqnarray}

The above scheme implemented in Maple was applied to the evaluations of
solutions in the analytical form  up to the order $k_{\max}=6$  of the PTRS.
 The first four nonzero coefficients for the energy (\ref{F43a}) \textit{in the analytic form}, truncated by the terms proportional to the sixth power of the
electric field strength, $\gamma_F^6$, in the crude adiabatic approximation (CAA) take the form:

\noindent 1) For OSQD in terms of minor $c$ and major $a$ semiaxes; the set of adiabatic quantum numbers $[m,n_o=n_{zo}+1, n_{\rho
  o}]$
 \begin{eqnarray}  \label{red01}&&
   \!\!\!\!\!\!\!\!\!\!\!\!\!\!\!\!   V^{(0)}_{n_o}=
   \frac{\pi^2 n_o^2}{4 c^2}
   +{\gamma_F^2}\frac{c^4(\pi^2n_o^2-15)}{3 \pi^4n_o^4}
   +{\gamma_F^4}\frac{4 c^{10}(\pi^4n_o^4-210\pi^2n_o^2+1980)}{9\pi^{10}n_o^{10}},
 \\&& \nonumber \!\!\!\!\!\!\!\!\!\!\!\!\!\!\!\! {\cal E}^{(0)}_{n_o;n_{\rho
  o}}=
\left[
  \frac{\pi n_o}{a c}
 -{\gamma_F^2}\frac{4c^5(\pi^2n_o^2-15)}{3a\pi^5n_o^5}
 -{\gamma_F^4} \frac{8c^{11}(2\pi^4n_o^4-360\pi^2n_o^2+3375)}{3 a\pi^{11}n_o^{11}}
 \right](2n_{\rho o}+|m|+1),
 \\&& \nonumber \!\!\!\!\!\!\!\!\!\!\!\!\!\!\!\! {\cal E}^{(1)}_{n_o;n_{\rho
  o}}=
  \left[\frac{1}{a^2}
 +{\gamma_F^2}\frac{4c^6(\pi^2n_o^2-15)}{\pi^6n_o^6a^2}
 +{\gamma_F^4}\frac{16c^{12}(7\pi^4n_o^4-1110\pi^2n_o^2+10350)}{3\pi^{12}n_o^{12}a^2}
\right]\times\\&&\nonumber\times (2+6n_{\rho o}+3|m|+6n_{\rho
o}^2+|m|^2+6n_{\rho o}|m|),
\\&& \nonumber \!\!\!\!\!\!\!\!\!\!\!\!\!\!\!\!
{\cal E}^{(2)}_{n_o;n_{\rho o}}=(2n_{\rho o}+|m|+1)
  \left[\frac{3c}{2\pi a^3 n_o}(2+2n_{\rho o}+|m|+2n_{\rho o}^2+2n_{\rho o}|m|)
 \right.\\ && \nonumber\left.
 -{\gamma_F^2}\frac{2c^7(\pi^2n_o^2-15)}{3\pi^7n_o^7a^3}(54+118n_{\rho o}+16|m|^2+59|m|+118n_{\rho o}^2+118n_{\rho o}|m|)
 \right.\\\nonumber && \left.
 -{\gamma_F^4}\left(
 \frac{4c^{13}(1874\pi^4n_o^4-273120\pi^2n_o^2+2536425)}{9\pi^{13}n_o^{13}a^3}(2n_{\rho o}+|m|+2n_{\rho o}^2+2n_{\rho o}|m|)
  \right.\right.\\ &&\nonumber \left.\left.
  +\frac{224c^{13}(8\pi^4n_o^4-1140\pi^2n_o^2+10575)}{9\pi^{13}n_o^{13}a^3}|m|^2
 +\frac{8c^{13}(326\pi^4n_o^4-48480\pi^2n_o^2+450675)}{3\pi^{13}n_o^{13}a^3}
 \right)
 \right],
 \end{eqnarray}

\noindent 2) For PSQD in terms of minor $a$ and major $c$ semiaxes, the set of adiabatic quantum numbers $[m,
n_p=n_{\rho p}+1, n_{zp}]$ and positive zeros $\alpha_{n_p,|m|}$ of the Bessel functions of the first kind \cite{stigun}
\begin{eqnarray} \label{red02}&&
   \!\!\!\!\!\!\!\!\!\!\!\!\!\!\!\!   V^{(0)}_{n_p;n_{zp}}=\frac{\alpha_{n_p,|m|}^2}{a^2}-{\gamma_F^2}\frac{a^2 c^2}{4\alpha_{n_p,|m|}^2}
      +{\gamma_F^4} \frac{a^6 c^4}{16\alpha_{n_p,|m|}^6},
 \\&& \nonumber \!\!\!\!\!\!\!\!\!\!\!\!\!\!\!\! {\cal E}^{(0)}_{n_p;n_{zp}}=
\left[
  \frac{\alpha_{n_p,|m|}}{a c}
 +{\gamma_F^2}\frac{3 a^3 c}{4\alpha_{n_p,|m|}^3}
 -{\gamma_F^4} \frac{9 a^7 c^3}{16\alpha_{n_p,|m|}^7}\right](2 n_{zp}+1),
 \\&&\nonumber
 \!\!\!\!\!\!\!\!\!\!\!\!\!\!\!\! {\cal E}^{(1)}_{n_p;n_{zp}}=
  \left[\frac{3}{4 c^2}
 +{\gamma_F^2}\frac{27 a^4}{16\alpha_{n_p,|m|}^4}
 -{\gamma_F^4}\frac{105 a^8 c^2}{64\alpha_{n_p,|m|}^8}\right](2n_{zp}^2+2n_{zp}+1),
  \\&& \nonumber \!\!\!\!\!\!\!\!\!\!\!\!\!\!\!\!
  {\cal E}^{(2)}_{n_p;n_{zp}}=\frac{3a}{16 c^3\alpha_{n_p,|m|}}(2 n_{zp}+1)(n_{zp}^2+n_{zp}+3)
  \\&&\nonumber+\gamma_F^2\left(\frac{5a^5}{64 c\alpha_{n_p,|m|}^5}(2 n_{zp}+1)(25n_{zp}^2+25n_{zp}+51)
  -\frac{ a^4}{4 \alpha_{n_p,|m|}^4}(30n_{zp}^2+30n_{zp}+11)\right)
  \\&&\nonumber-\gamma_F^4\left(\frac{45a^9 c}{256\alpha_{n_p,|m|}^9}(2 n_{zp}+1)(23n_{zp}^2+23n_{zp}+37)
  -\frac{3a^8c^2}{8\alpha_{n_p,|m|}^8}(30n_{zp}^2+30n_{zp}+11)\right).
\end{eqnarray}

\begin{table}
\caption{{Convergence of eigenvalues
${\cal E}^{(k_{max})}_{n_{zo},n_{\rho o}}=V^{(0)}_{ n_{zo}}
+\sum_{k=0}^{k_{max}}{\cal E}^{(k)}_{n_{zo},n_{\rho o}}$
for \textit{oblate} spheroid   $c=0.5$, $a=5$ vs PT order $k_{\max}$ { at  $\gamma_F=0$  }.
{First line  $^*$ notes adiabatic shift $V^{(0)}_{n_{zo},n_{\rho o}}$.}}
Last line are results of numerical calculations (Num).}\label{zo}
\begin{tabular}{llllll}\hline
{$k_{\max}$}  &$n_{zo}=0$,$n_{\rho o}=0$ &$n_{zo}=0$,$n_{\rho o}=1$&$n_{zo}=0$,$n_{\rho o}=2$&$n_{zo}=0$,$n_{\rho o}=3$&$n_{zo}=0$,$n_{\rho o}=4$\\\hline
* &11.12624146 &13.63951558 &16.15278970& 18.66606383 &21.17933795\\
0 &11.20624146 &14.19951558 &17.67278970& 21.62606383 &26.05933795\\
1 &11.21006118 &14.23389305 &17.80647986& 21.97365822 &26.78126477\\
2 &11.21026382 &14.23433886 &17.80254859& 21.95100281 &26.71094787\\
3 &11.21028027 &14.23441723 &17.80242765& 21.94908610 &26.70256215\\
4 &11.21028227 &14.23443790 &17.80265195& 21.95065251 &26.70959163\\
5 &11.21028259 &14.23444049 &17.80264785& 21.95052291 &26.70875037\\\hline
Num &11.21028268 &14.23444147 &17.80265065& 21.95050805 &26.70857727\\\hline\hline
{$k_{\max}$}  &$n_{zo}=2$,$n_{\rho o}=0$ &$n_{zo}=2$,$n_{\rho o}=1$&$n_{zo}=2$,$n_{\rho o}=2$&$n_{zo}=2$,$n_{\rho o}=3$&$n_{zo}=2$,$n_{\rho o}=4$\\\hline
* &92.59635079 &100.1361731 &107.6759955 &115.2158178 &122.7556402\\
0 &92.67635079 &100.6961731 &109.1959955 &118.1758178 &127.6356402\\
1 &92.67762403 &100.7076323 &109.2405589 &118.2916826 &127.8762825\\
2 &92.67764654 &100.7076818 &109.2401221 &118.2891654 &127.8684695\\
3 &92.67764715 &100.7076847 &109.2401176 &118.2890944 &127.8681589\\
4 &92.67764718 &100.7076850 &109.2401203 &118.2891137 &127.8682457\\
5 &92.67764718 &100.7076850 &109.2401203 &118.2891132 &127.8682422\\\hline
Num &92.67764718 &100.7076850 &109.2401204 &118.2891132 &127.8682419\\\hline
\end{tabular}  
\end{table}
\begin{table}
\caption{
{Convergence of eigenvalues
${\cal E}^{(k_{max})}_{n_{\rho_{p}},n_{zp}}=V^{(0)}_{n_{\rho_{p}}}
+\sum_{k=0}^{k_{max}}{\cal E}^{(k)}_{n_{\rho_{p}},n_{zp}}$
for \textit{prolate} spheroid   $c=2.5$, $a=0.5$ vs PT order $k_{\max}$ { at  $\gamma_F=0$  }.
First line  $^*$ notes adiabatic shift $V^{(0)}_{n_{\rho_{p}},n_{zp}}$.}
Last line are results of numerical calculations (Num).}\label{zp}
\begin{tabular}{llllll}\hline
{$k_{\max}$} & $n_{\rho p}=0$,$n_{zp}=0$    &$n_{\rho p}=0$,$n_{zp}=2$    &$n_{\rho p}=0$,$n_{zp}=4$    &$n_{\rho p}=0$,$n_{zp}=6$    & $n_{\rho p}=0$,$n_{zp}=8$    \\\hline
*& 25.05660430& 32.75204608& 40.44748787& 48.14292965& 55.83837144\\
0& 25.17660430& 34.31204608& 45.36748787& 58.34292965& 73.23837144\\
1& 25.18408925& 34.42432034& 45.88394944& 59.80249498& 76.41947535\\
2& 25.18465987& 34.42810718& 45.87103269& 59.69485535& 76.04779441\\
3& 25.18472054& 34.42867746& 45.87114189& 59.68460238& 75.99436976\\
4& 25.18472960& 34.42880826& 45.87257549& 59.69618640& 76.05191800\\
5& 25.18473139& 34.42883580& 45.87259458& 59.69511288& 76.04351256\\\hline
Num
 & 25.18472985& 34.42884694& 45.87265876& 59.69512314& 76.04210082\\\hline\hline
{$k_{\max}$} & $n_{\rho p}=1$,$n_{zp}=0$    &$n_{\rho p}=1$,$n_{zp}=2$    &$n_{\rho p}=1$,$n_{zp}=4$    &$n_{\rho p}=1$,$n_{zp}=6$    & $n_{\rho p}=1$,$n_{zp}=8$    \\\hline
*& 126.3011119& 143.9653618& 161.6296118& 179.2938617& 196.9581117\\
0& 126.4211119& 145.5253618& 166.5496118& 189.4938617& 214.3581117\\
1& 126.4243727& 145.5742742& 166.7746086& 19 0.1297223& 215.7439616\\
2& 126.4244810& 145.5749929& 166.7721571& 190.1092932& 215.6734198\\
3& 126.4244860& 145.5750400& 166.7721661& 190.1084455& 215.6690025\\
4& 126.4244863& 145.5750447& 166.7722178& 190.1088627& 215.6710754\\
5& 126.4244864& 145.5750452& 166.7722181& 190.1088459& 215.6709435\\\hline
Num
 & 126.4244896& 145.5750487& 166.7722220& 190.1088484& 215.6709278\\\hline
\end{tabular}
\end{table}

In Tables \ref{zo} and \ref{zp} we demonstrate how the approximate
eigenvalues in the lower part of spectrum for OSQD and PSQD at $m=0$ and
$\gamma_F=0$ converge to the  values calculated numerically with required accuracy in the crude adiabatic approximation
with increasing of the PT  order $k$. The accuracy was from 8 to 5
digits at $n_{zo}=0$, from 10 to 8 digits at $n_{zo}=2$,  from 6 to 4 digits at $n_{\rho p}=0$, and from 8 to 7 digits at $n_{\rho p}=1$, respectively.
 Note, that the difference between the adiabatic shift $V_i^{(0)}$ and the eigenvalues ${\cal E}_{i;n}=V^{(0)}_i+{\cal E}^{(0)}_{i;n}$
in the zero order $k=0$ of the PT is small, but increases with growing  $n_{\rho o}$ and $n_{z p}$
 for OSQD and PSQD, respectively.
 The shifts $V_i^{(0)}$ give the  main contribution and provide the lower adiabatic estimate
 of each set of eigenvalues, generated by the perturbed harmonic oscillator terms with adiabatic frequency $\omega_i$.
 From Tables \ref{zo} and \ref{zp} one can see
 that with increasing quantum numbers
$n_{zo}$ (or $n_{\rho p}$), related to the fast variable, the accuracy of approximation of the lower part of the spectrum  is
increasing. This is because the accuracy of the Taylor approximations of potential function (\ref{F42})
in Eq. (\ref{F41}) is improved with increasing the number $i=n_{zo}+1>2$ (or $i=n_{\rho p}+1>2$), which is demonstrated in Fig. \ref{fig51}.

\begin{figure}[h]
\includegraphics[width=0.98\textwidth]{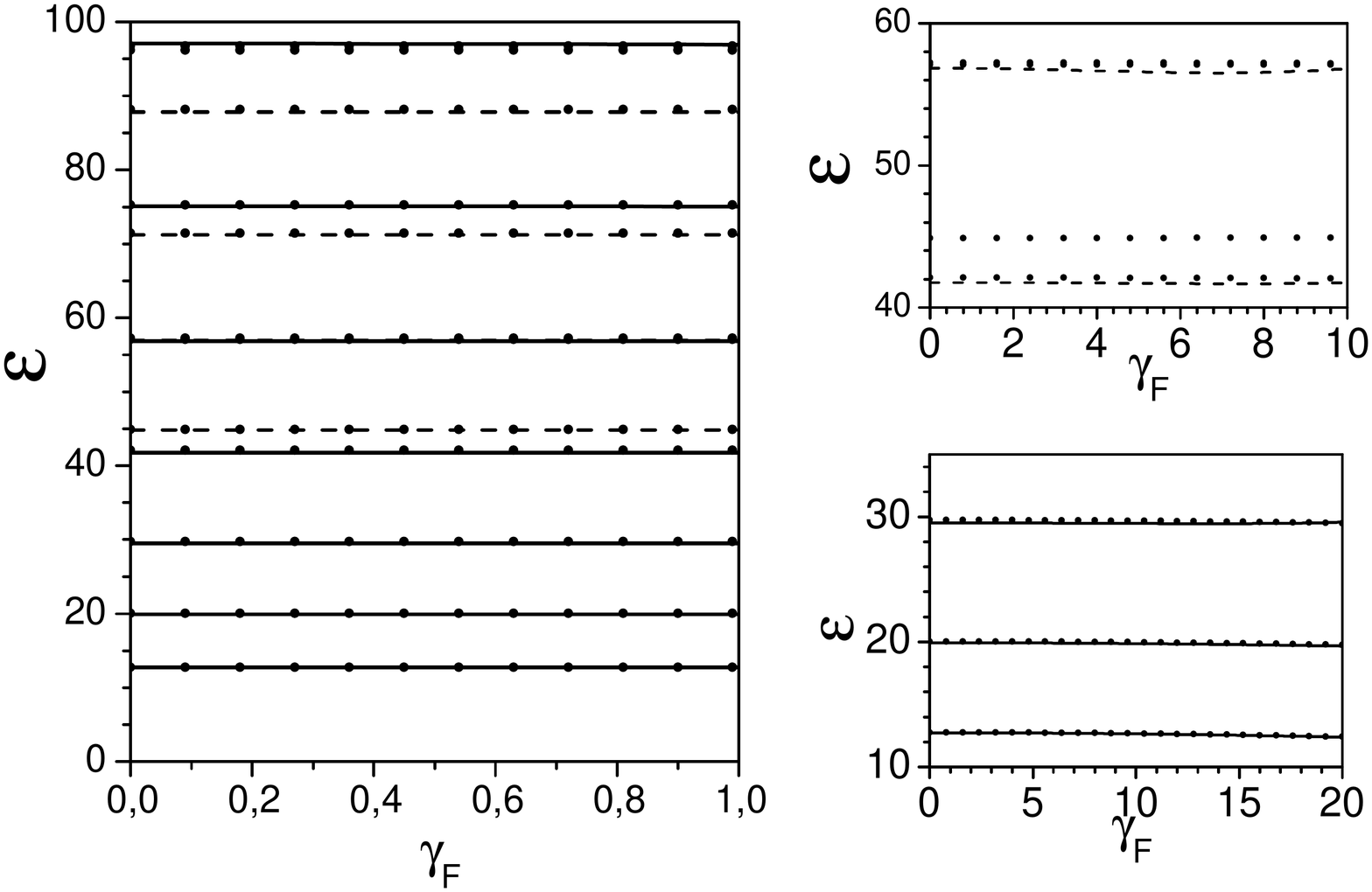}
\caption{
Dependence of eigenenergies ${\cal E}$ (in units of $E_{e}$)  of lower part of spectrum of electronic states of OSQDs  ($a=2.5$, $c=0.5$) at $m=0$ on electric field strength $\gamma_F$ (in units of $F_0^*$):
Solid and dashed lines are eigenenergies calculated by PTRS till 5 order in crude adiabatic approximation:
seven solid lines  ($n_{zo}=0$, $n_{\rho o}=0,1,...,6$) and four dashed lines ($n_{z o}=1$, $n_{\rho o}=0,1,2,3$)  are shown on left panel in interval $\gamma_F\in (0,1)$; the first three of solid lines ($n_{z o}=1$, $n_{\rho o}=0,1,2$) and the first two of dashed lines ($n_{zo}=1$ , $n_{\rho o}=0,1$) are shown on lower-right and upper-right panel, respectively,  in bigger intervals $\gamma_F\in (0,20)$ and $\gamma_F\in (0,10)$.
Numerical solutions of Eqs. (\ref{2dbvp}) at $j_{\max}=4$ are shown by points.}
\label{dvafig}
\end{figure}
\begin{figure}[h]
\includegraphics[width=0.98\textwidth]{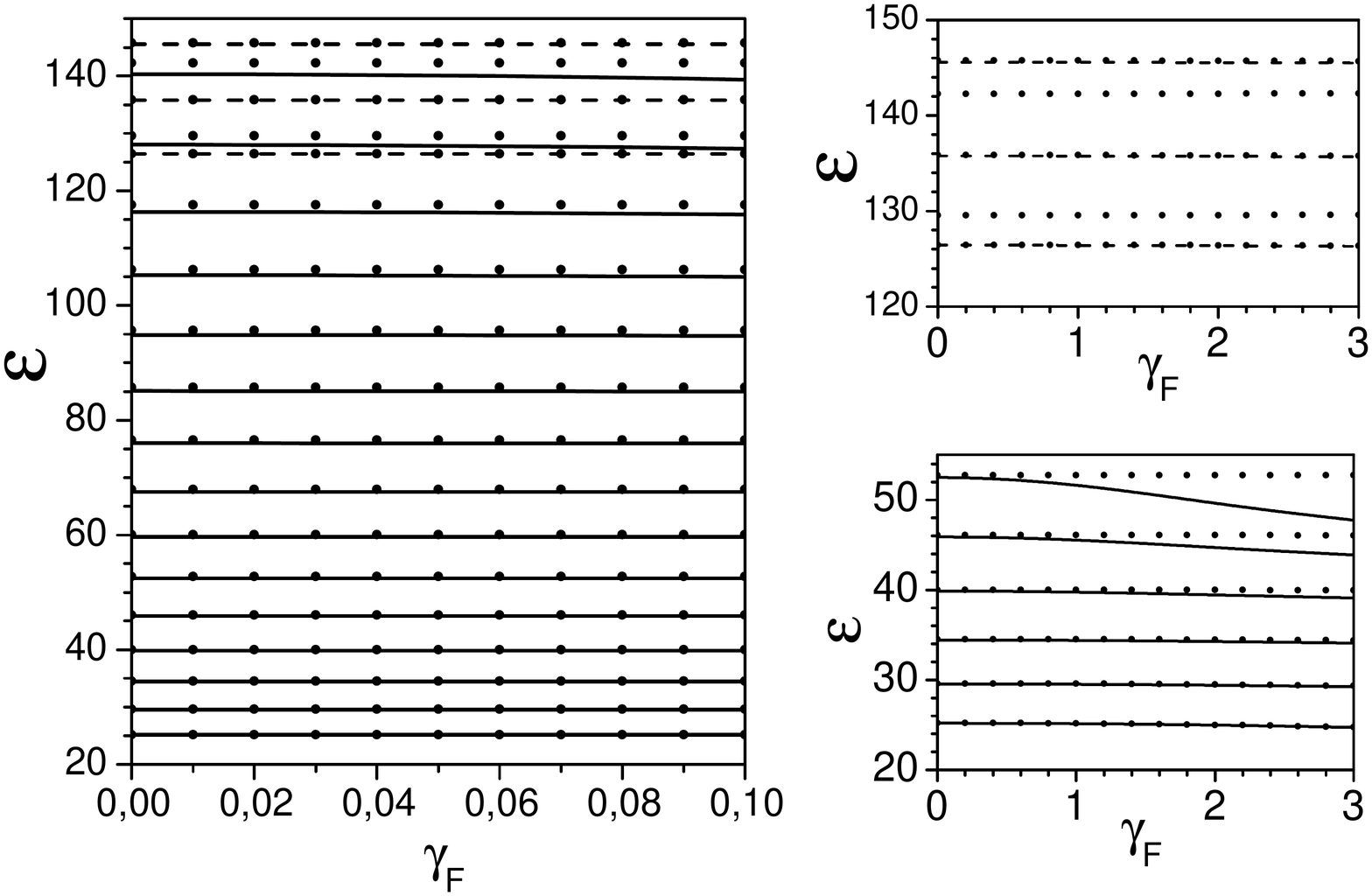}
\caption{
Dependence of eigenenergies ${\cal E}$ (in units of $E_{e}$)  of lower part of spectrum of electronic states of PSQDs  ($a=0.5$, $c=2.5$) at $m=0$ on electric field strength $\gamma_F$ (in units of $F_0^*$):
Solid and dashed lines are eigenenergies calculated by PTRS till 5 order in crude adiabatic approximation:
fifteen solid lines  ($n_{\rho p}=0$, $n_{z p}=0,1,..., 14$) and
three dashed lines ($n_{\rho p}=1$ , $n_{z p}=0,1,2$)  are shown on left panel in interval $\gamma_F\in (0,0.1)$;
the first six of solid lines ($n_{\rho p}=0$, $n_{z p}=0,1,..., 5$) are shown on lower-right panel and the dashed lines ($n_{\rho p}=1$ , $n_{z p}=0,1,2$) are shown on upper-right panel in bigger interval $\gamma_F\in (0,3)$.
Numerical solutions of Eqs. (\ref{2dbvp}) at $j_{\max}=4$ are shown by points.}
\label{trifig}
\end{figure}

In Figs. \ref{dvafig} and \ref{trifig} we show the eigenvalues ${\cal E}$
of the lower part of the spectrum of oblate and prolate QDs versus the electric field strength within small (left panels) and large (right panels) intervals of $\gamma_F$,
calculated in the crude adiabatic approximation (solid and dashed lines) to compare them with the numerical results (dotted lines).
One can see that the eigenvalues calculated using the PT (solid and dashed lines), corresponding to the eigenfunctions with smaller number of nodes
along the electric field (i.e., with smaller $n_{zo}$ for OSQD and $n_{zp}$ for PSQD)
and with greater number of nodes
across  the electric field (i.e., with greater $n_{\rho o}$ for OSQD and $n_{\rho p}$ for PSQD),
provide better approximation of the eigenvalues, calculated numerically with required accuracy (dotted lines).
This property follows from the fact that such functions have better localization in the vicinity
of the plane, passing through the QD center transverse to the electric field,
i.e., in the region with minimal contribution of the electric field potential   to the Hamiltonian of the system.
As shown in the right panels of  Figs. \ref{dvafig} and \ref{trifig}, the  differences between the
egienvalues,  calculated using the PT and the numerical method, increase faster in a smaller interval of $\gamma_F$
for larger PSQD  than for smaller OSQD, the size being measured along the direction of the electric field.

The range  of the parameter values, for which the PT algorithms are valid, was estimated by means of numerical calculations using the KANTBP program \cite{Yu3}, as well as the
condition that the mean value of the slow variable is smaller than the size of the major axis of OSQD or PSQD, i.e., $\rho\leq a$ or $z\leq c$,
or known estimates of the distribution of nodes of  Laguerre  or Hermite polynomials \cite{stigun}.
 To calculate also the approximate eigenfunctions of the lower part of the spectrum  $n=0,...,n_{\max}$ with required
 numbers $n$ of nodes in the interval  $\rho\in(0,a)$ (or $z\in(-c,c)$) for OSQD (or PSQD),
  one should choose such value of parameter $a=\sqrt{2x_0/\omega_i}$ (or $c=x_0/\sqrt{\omega_i}$),
 that outside this interval $x\in(x_0=4n+2|m|+2,\infty)$ (or $|x|\in (x_0=(2n+1)^{1/2},\infty)$)
 the Laguerre (or Hermite) polynomials have no nodes.
As an example, in Fig. \ref{fun35a}  we show contour plots in zx and
xz plane of the first four eigenfunctions of OSQD and PSQD, respectively, that  have a required number of nodes (crossings of
the function plot with zero plane) in the interval $\rho\in(0,a)$ and
$z\in(-c,c)$ at the values $c=0.5$, $a=5$, $c=2.5$, $a=0.5$. One can see
that the asymmetry with respect to z-axis of the eigenfunctions of PSQD is
greater than that of OSQD, because the variation of well depth of PSQD
is greater than of OSQD.
\section{
{Absorption coefficient} for an ensemble of QDs}\label{5} One can
use
the differences in the energy spectra to verify the
considered models of QDs by calculating the absorption coefficient
$K(\tilde \omega ^{ph},\tilde a,\tilde c,)$ of an ensemble of identical
semiconductor QDs \cite{Efros1982}. Since we do not discuss exciton effects in the present paper, the absorption coefficient may be approximately expressed as
\begin{eqnarray}&&\label{coa1}
\tilde K(\tilde\omega^{ph},\tilde a,\tilde c,u)= \sum_{\nu,\nu'}\tilde
K_{\nu,\nu'}(\tilde\omega^{ph},\tilde a,\tilde c,u)=\tilde
A\sum_{\nu,\nu'} \tilde I_{\nu,\nu'}(u) \delta(\hbar
\tilde\omega^{ph}-\tilde W_{\nu\nu'}),
\\&&\nonumber
\tilde I_{\nu,\nu'}(u)
=\left|\int\tilde \Psi^{e}_{\nu}(\tilde {\bf r};\tilde
a,\tilde c,F,\mu_e)\tilde \Psi^{h}_{\nu'}(\tilde {\bf r};\tilde a,\tilde c,F,\mu_h)d^3
\tilde {\bf r}\right|^{2},  \,\,\,\,
\end{eqnarray}
where $\tilde A$ is proportional to the square of the matrix element
in the Bloch decomposition, $\tilde \Psi^{e}_{\nu}(u)$ and $\tilde
\Psi^{h}_{\nu'}$ are the eigenfunctions of the electron ($e$) and the
heavy hole ($h$), $\tilde E^{e}_{\nu}$ and $\tilde E^{h}_{\nu'}$ are
the energy eigenvalues for the electron ($e$) and the heavy hole ($h$),
depending on the semiaxis size $\tilde c, \tilde a$ for OSQD (or
$\tilde a,\tilde c$ for PSQD) and the adiabatic set of quantum
numbers $\nu=[n_{zo},n_{\rho o},m]$ and $\nu'=[n_{zo}',n_{\rho
o'},m']$ ($\nu=[n_{\rho p},n_{zp},m]$ and $\nu'=[n_{\rho p}',n_{z
p}',m']$), where $m'=-m$, $\tilde E_{g}$ is the band gap width in
the bulk semiconductor, $\tilde \omega^{ph}$ is the incident light
frequency,
$\tilde W_{\nu\nu'}=\tilde E_{g}+\tilde E^{e}_{\nu}(\tilde a,\tilde
c) +\tilde E^{h}_{\nu'}(\tilde a,\tilde c)$ is the inter-band
transition energy for which $\tilde K(\tilde\omega^{ph})$ has the
maximal value. We rewrite the expression (\ref{coa1}) in the terms of frequency shift of the incident light $\Delta\omega^{ph}/(2\pi)=(\hbar \tilde \omega^{ph} - \tilde E_{g})/(2\pi\hbar)$ corresponding to the inter-band
transition energy shift
$\Delta\tilde W_{\nu\nu'}=\tilde W_{\nu\nu'}-\tilde E_{g}=\tilde E^{e}_{\nu}(\tilde a,\tilde
c) +\tilde E^{h}_{\nu'}(\tilde a,\tilde c)$ for which $\tilde K(\Delta\tilde\omega^{ph})$ has the
maximal value, using
dimensionless variables in the reduced atomic units
\begin{eqnarray}\label{coa7}
\tilde K(\Delta\tilde \omega ^{ph},\tilde a,\tilde c)= \tilde A \tilde
E_{g}^{-1}\sum_{\nu,\nu'} \tilde I_{\nu,\nu'}(u)
\delta[f_{\nu,\nu'}(u)], \quad f_{\nu,\nu'}(u)=
\lambda_{1}-\frac{2 E^{e}_{\nu}(a, c)+2
E^{h}_{\nu'}(a,c)(\mu_{h}/\mu_{e})}{2E_{g}},
\end{eqnarray}
where the parameter $u$ will be defined below, $\lambda_1=(\hbar
\tilde \omega^{ph} - \tilde E_{g})/\tilde E_{g}$ is the energy of
the optical interband transitions scaled to $\tilde E_{g}$,
$2E_{g}= \tilde E_{g}/\tilde E^{e}_{R}$
is the dimensionless band gap width.

 For GaAs the functions $f_{\nu,\nu'}^{h\to e}(u)$ describing the ($h\to e$) interband transitions
 have the form \begin{eqnarray}&&\label{coya1}f_{\nu,\nu'}^{h\to e}(u)=
\lambda_{1}-(2E_{g})^{-1}(2 E^{e}_{\nu}(a, c,\gamma_F)+2
E^{e}_{\nu'}(a,c,-(\mu_h/\mu_e)\gamma_F)(\mu_{e}/\mu_{h})),
\end{eqnarray}
where $\mu_e=0.067m_0$ and $\mu_{h}\equiv\mu_{hh}=0.558m_0$ are the
masses of electron and  hole, respectively, $\tilde E_g=1430$ meV is the
band gap width and $\kappa=13.18$ is the dc permittivity and
$E_R^e=e^2/(2\kappa a_B^{e})= 5.275$ meV, $a_B^e=\hbar^2\kappa/(\mu_e
e^2)=104$\AA, $E_R^{h}=e^2/(2\kappa a_B^{h})=49$ meV,
$a_B^h=\hbar^2\kappa/(\mu_he^2)=15$\AA,    $2\gamma_{F}\!=\!F/F_{0}^*$, $F_{0}^{*}=E^{e}_R/(ea_B^{e})=e/(2\kappa (a_B^{e})^{2})=5.04$kV/cm.

 For InSb the dispersion law for  heavy holes ($hh$) is parabolic
while for electrons ($e$) and light holes ($lh$) it is non-parabolic and may be described by the Kane model \cite{Kane,Askerov,Hayk11} at $\gamma_F=0$. The energy
values in our notation are: \begin{eqnarray}&&\label{coya2}
2\tilde{  E}^{hh}_\nu{(InSb)}=2 \tilde
E^{h}_{\nu'}(\tilde a,\tilde c),\\&& 2\tilde{
E}^e_\nu{(InSb)}=2\tilde{  E}^{lh}_\nu{(InSb)}=-\tilde
E_{g}/2+\sqrt{\tilde E_{g}^2/4+\tilde E_{g}(2 \tilde E^{e}_{\nu}(\tilde a,
\tilde c))}.\label{coya2a}
\end{eqnarray}

As follows from Eqs. (\ref{coya2}) and (\ref{coya2a}),  to determine the energy spectrum and the wave function of the light hole and the electron one should solve the Klein-Gordon equation (\cite{Ref1,Ref2}), while for heavy hole the Schr\"odinger equation is applicable.
The functions $f_{\nu,\nu'}^{hh\to e}(u)$ and $f_{\nu,\nu'}^{hh\to e}(u)$  describing the ($hh\to e$)
and the ($lh\to e$) interband transitions have the forms
\begin{eqnarray}&&\label{coya3}f_{\nu,\nu'}^{hh\to e}(u)=
\lambda_{1}-(1/2+\sqrt{1/4+(2 E^{e}_{\nu}(a, c)/
(2E_{g}))}+(2E_{g})^{-1}2 E^{e
}_{\nu'}(a,c)(\mu_{e}/\mu_{h})),\\&&\label{coya4}f_{\nu,\nu'}^{lh\to e}(u)=
\lambda_{1}-2(\sqrt{1/4+(2 E^{e}_{\nu}(a, c)/
(2E_{g}))}),\end{eqnarray} where $\mu_e=\mu_{lh}=0.15m_0$ and
$\mu_{h}\equiv\mu_{hh}=0.5m_0$ are the masses of electron, light and
heavy holes, respectively, $\tilde E_g=180$~meV is the band gap width,
$\kappa=16$ is the dc permittivity, and
 $E_R^e=E_R^{lh}=e^2/(2\kappa a_B^{e})
 =7.972$~meV,
$a_B^e=a_B^{lh}=\hbar^2\kappa/(\mu_e
e^2)
=56.44$\AA,
$E_R^{h}=E_R^{hh}=e^2/(2\kappa a_B^{hh})=26.57$~meV,
$a_B^h=a_B^{hh}=\hbar^2\kappa/(\mu_h
e^2)
=16.93$\AA.

{For both electron and hole
carriers the dimensionless energies $2E_{\nu}^{e}= \tilde
E_{\nu}^{e}/\tilde E^{e}_{R}$ and
$2E_{\nu}^{h}(\mu_{h}/\mu_{e})=\tilde E_{\nu}^{h}/\tilde
E^{e}_{R}$ are expressed in the same reduced atomic units $\tilde
E^{e}_{R}$, and the overlap integral (\ref{coa1}) between the eigenfunctions, corresponding to $E_\nu^e(\gamma_F)$ and
$E_\nu^h(\gamma_F)=(\mu_e/\mu_h) E_\nu^e(-(\mu_h/\mu_e)\gamma_F)$, takes the form}
\begin{eqnarray}&&\label{coya8}\tilde I_{\nu,\nu'}(u)=\left|\int (a_B^{e})^3 \Psi^{e}_{\nu}( {\bf r};
a, c,\gamma_F,\mu_e) \Psi^{e}_{\nu'}({\bf r};a,
c,-(\mu_h/\mu_e)\gamma_F,\mu_e)d^3 {\bf r}\right|^{2}.
\end{eqnarray}
Now consider an ensemble of OSQDs (or PSQDs), differing in the minor semiaxis values $c=u_{o}\bar c$ (or $a=u_{p}\bar a$), determined by
the random parameter $u=u_{o}$ (or $ u=u_{p}$). The corresponding
minor semiaxis mean value is  $\bar c$ at fixed major semiaxis $a$ (or $\bar
a$ at fixed major semiaxis $c$), and the appropriate distribution
function is  $P(u_{o})$ (or $P(u_{p})$). Commonly, in this case
the normalized Lifshits-Slezov
 distribution function~\cite{LS1958} is used:
$$
P(u)=\{3^4eu^2\exp(-1/(1-2u/3))/2^{5/3}/(u+3)^{7/3}/(3/2-u)^{11/3}, u\in (0,3/2);
 0, \mbox{otherwise}\}
$$
having conventional properties $\int P(u)du=1$, $\bar u=\int u P(u)du=1$.
The absorption coefficients $\tilde K^{o}(\tilde \omega ^{ph},\bar {\tilde a}, \tilde c)$ or $\tilde K^{p}(\tilde \omega ^{ph},\tilde a,
\bar {\tilde c})$ of an ensemble of semiconductor OSQDs or PSQDs with
different dimensions of minor semiaxes are   expressed as
\begin{equation}\label{coa9}
\tilde K^{o}(\tilde \omega ^{ph},\bar {\tilde a}, \tilde c)
=\int \tilde K(\tilde \omega ^{ph}, \bar {\tilde a}, \tilde c, u_{o})P(u_{o})du_{o},~~
\tilde K^{p}(\tilde \omega ^{ph},\tilde a,
\bar {\tilde c})=\int \tilde K(\tilde \omega ^{ph},   \tilde a, \bar {\tilde c}, u_{p})P(u_{p})du_{p}.
\end{equation}
Substituting (\ref{coa7}) into (\ref{coa9}) and taking into account the known properties of the $\delta$-function,
we arrive at the analytical expression for the  absorption coefficient $\tilde K(\tilde \omega ^{ph},\tilde a,\tilde c)$ of a system
of semiconductor QDs with a distribution of random minor semiaxes:
\begin{equation}\label{kk}
\frac{\tilde K(\tilde \omega ^{ph})}{\tilde K_{0}}=
\sum_{\nu,\nu',s} \frac{\tilde K_ {\nu,\nu'}(\tilde \omega ^{ph})}{\tilde K_{0}},
\quad \frac{\tilde K_ {\nu,\nu'}(\tilde \omega ^{ph})}{\tilde K_{0}}=
\tilde I_ {\nu,\nu'}\left( u_s \right)
\left|\left.\frac {df_{\nu,\nu'}(u)}{du}\right|_{u=u_s}\right|^{-1}  P\left( u_s \right),
\end{equation}
where $\tilde K_{0}= \tilde A^{-1}\tilde E_{g}$ is the
normalization factor, $u_s$ are the roots of the equation
$f_{\nu,\nu'}(u_s)=0$.

At $\gamma_F=0$  for {IPBM} we have the interband
overlap $\tilde I_{\nu,\nu'}=\delta_{n_{\rho o},n_{\rho
o}'}\delta_{n_{zo},n_{zo}'}\delta_{m,-m'}$ for OSQD, or $\tilde
I_{\nu,\nu'}=(J_{1+|m|}(\alpha_{n_{\rho
p}+1,|m|})/J_{1-|m|}(\alpha_{n_{\rho p}+1,|m|}))^2
\delta_{n_{zp},n_{zp}'}\delta_{n_{\rho p},n_{\rho p}'}
\delta_{m,-m'}$ for PSQD, where $\alpha_{n_{\rho p}+1,|m|}$ is the positive root of the Bessel function, and the selection rules $m=-m'$,
$n_{zo}=n_{zo}'$, $n_{\rho o}=n_{\rho o}'$,  or $n_{\rho p}=n_{\rho
p}'$, $n_{zp}=n_{zp}'$ \cite{Yaf12}, while at
$\gamma_F\neq0$ one should calculate the interband overlap
(\ref{coya8}) in accordance with the selection rules  $m=-m'$, $n_{\rho
o}=n_{\rho o}'$, or $n_{\rho p}=n_{\rho p}'$,
respectively.
Note,that in the adiabatic limit and at small $\gamma_F$
the contributions of non-diagonal matrix
elements to the energy values are about 1\% for {IPBM of OSQD and PSQD};
then in the Born-Oppenheimer approximation of the order
$b_{max}$  for the AC we get
\begin{eqnarray}
\label{ca01} f_{\nu,\nu'}(u)=\lambda_{1}-\sum_{j=0}^{f_{max}}
f^{(j)}_{\nu,\nu'}u^{j-2}. \end{eqnarray}
The coefficients of the expansion (\ref{ca01}) for parabolic dispersion law
for small $\gamma_F\neq 0$  were constructed  using the expansions (\ref{red01}) and (\ref{red02}) and
at $\gamma_F=0$ they are given in \cite{Yaf12}.
In general case  for the calculation $f_{\nu,\nu'}(u)$ by formula (\ref{coya1}), (\ref{coya3}), or (\ref{coya4})
we used the eigenvalues $E^{e}_{\nu}(a, c)$ and $E^{h}_{\nu'}(a,c)$  calculated numerically with given accuracy. After that
we evaluated the coefficients of expansion like  (\ref{ca01}) by the method of least squares and by the polynomial interpolation in the case of parabolic and non-parabolic dispersion laws, respectively.
Because of monotonic behavior of function $f_{\nu,\nu'}(u)$  vs $u$ in the case under consideration, we have only one
root $u_s$ of the equation $f_{\nu,\nu'}(u_s)=0$, which was used in formula (\ref{kk}).

\begin{figure}[t]
\includegraphics[width=0.44\textwidth,height=0.25\textwidth]{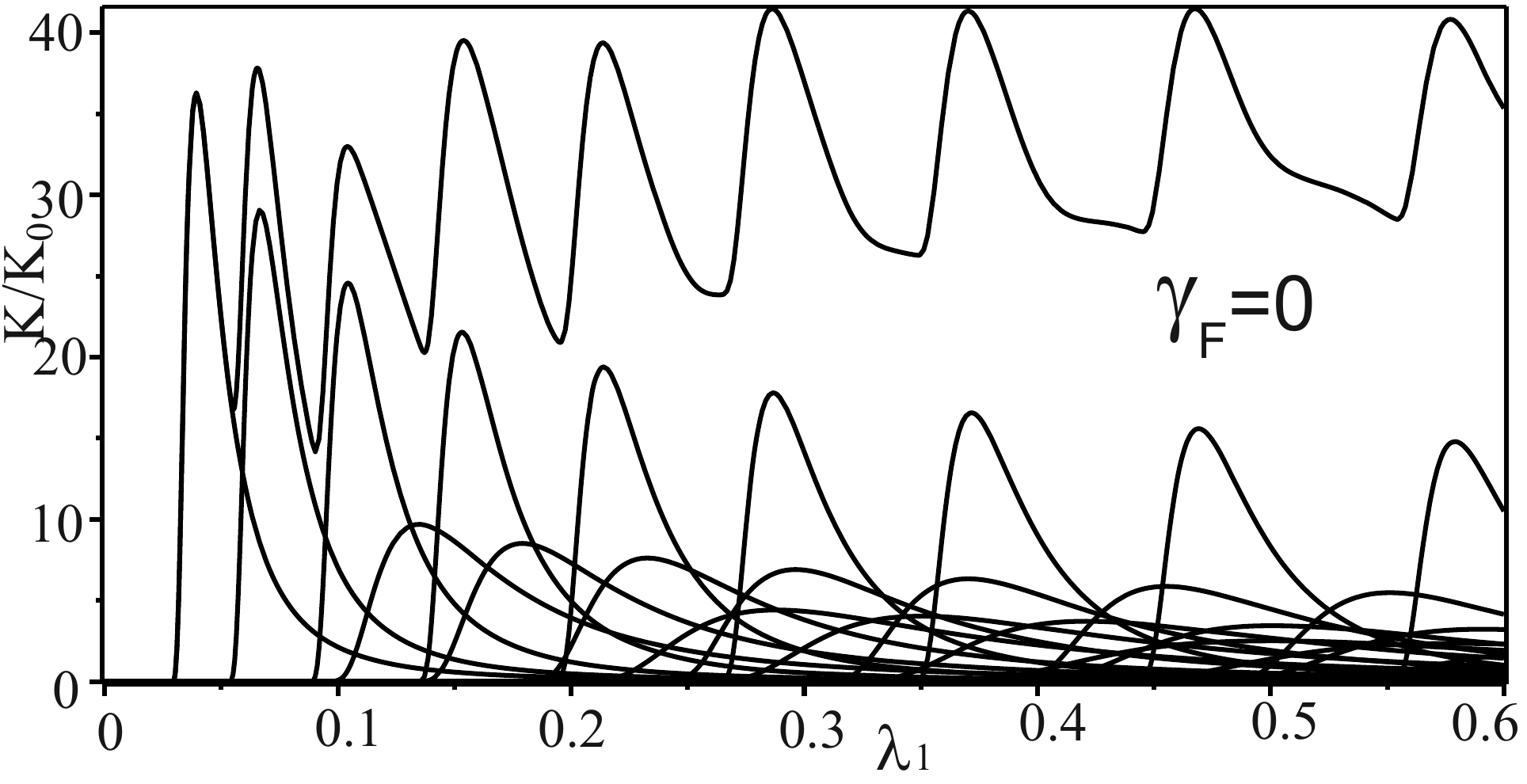}  \hfill
\includegraphics[width=0.44\textwidth,height=0.25\textwidth]{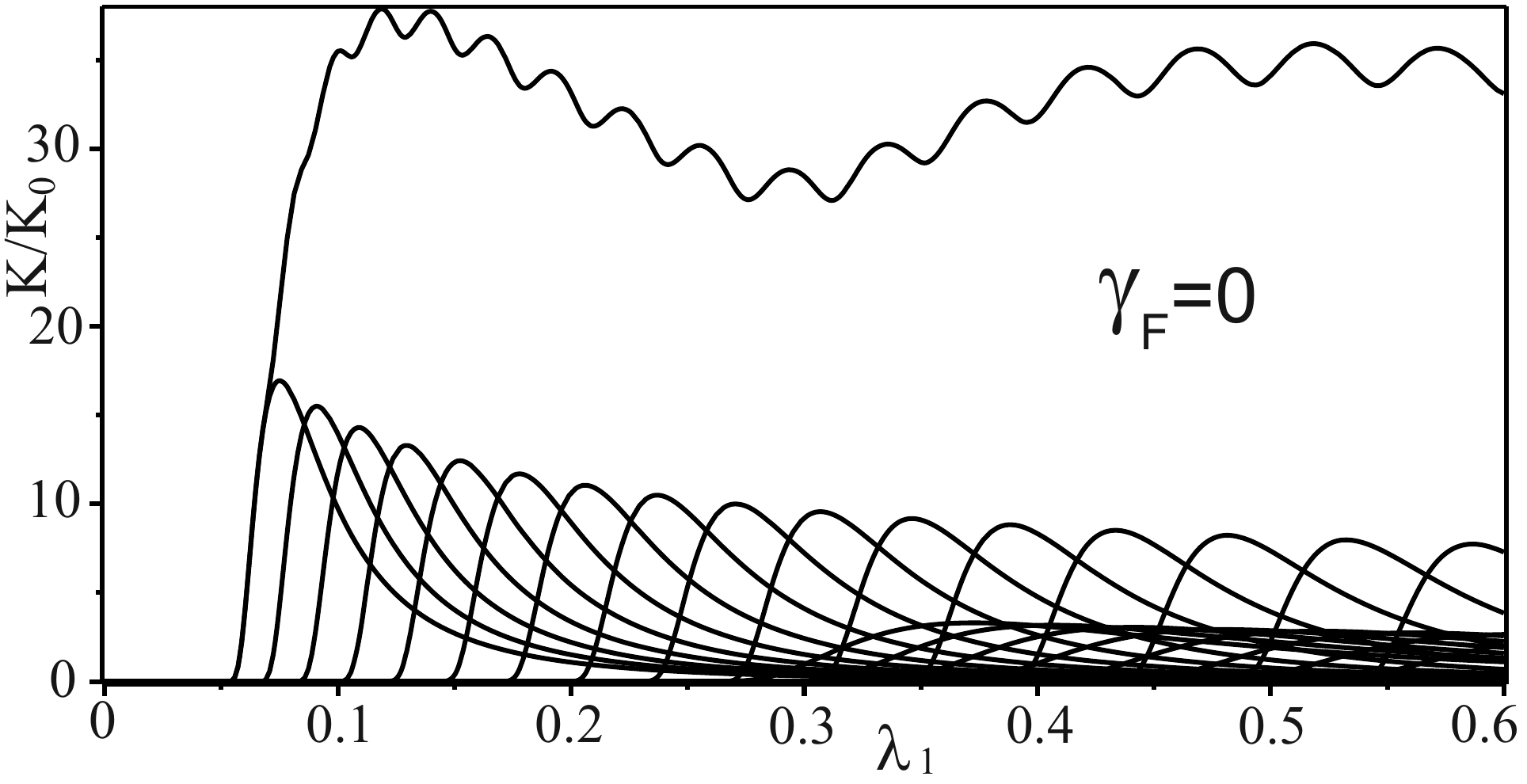}
\caption{Absorption coefficient $K/K_{0}$, Eq. (\ref{kk}),
consisting of a sum of the first
partial contributions vs the energy $\lambda=\lambda_1$
of the optical interband transitions for the Lifshits-Slezov distribution,
 using the functions
$f_{\nu,\nu'}^{h\to e}(u)$ for GaAs $(h\to e)$ without electric field:
for ensemble of OSQDs  $\bar c=0.5$, $a=2.5$ (left panel) and
for ensemble of PSQDs  $\bar a=0.5$, $c=2.5$ (right panel).} \label{k11}
\end{figure}
\begin{figure}[t]
\includegraphics[width=0.44\textwidth,height=0.25\textwidth]{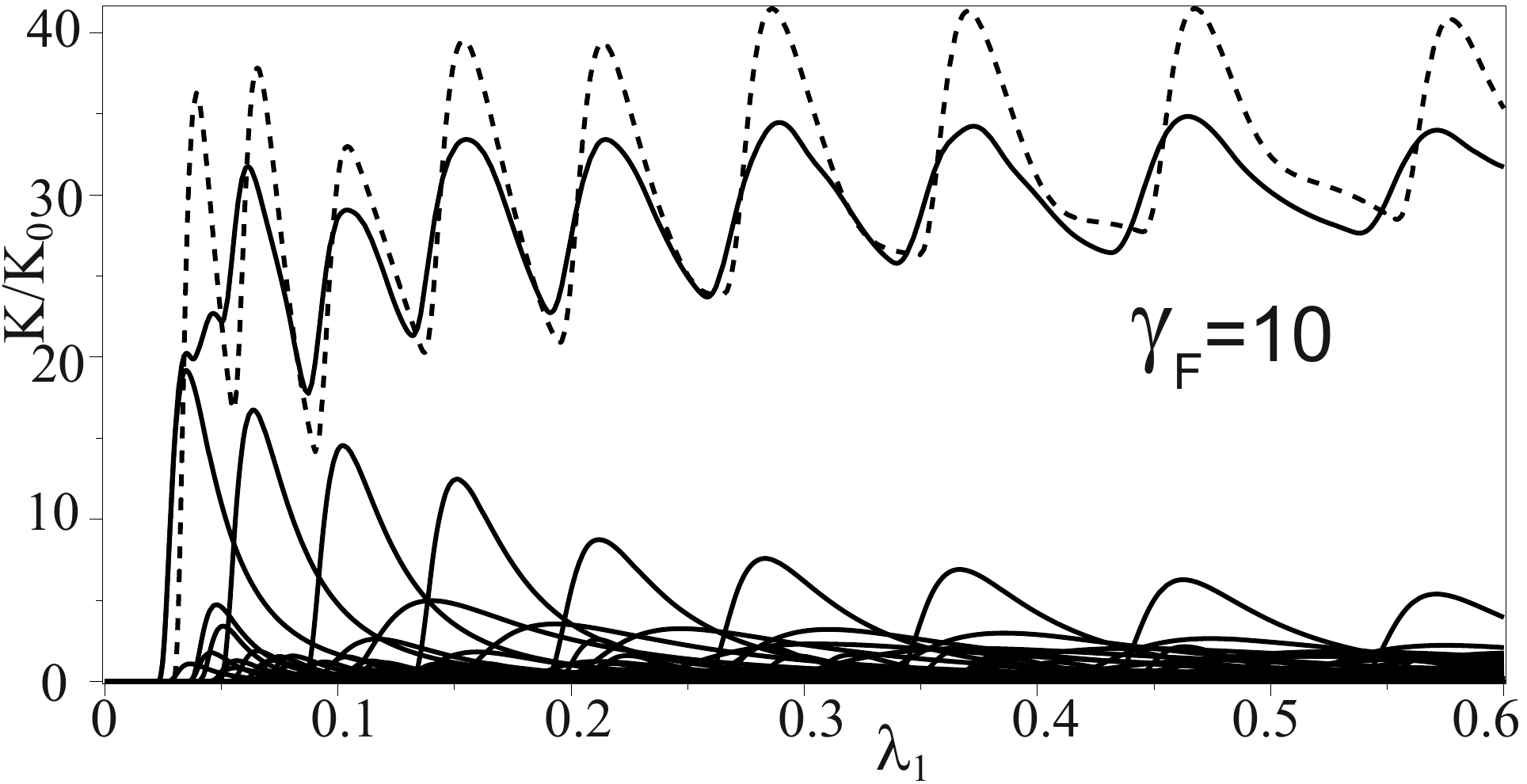}
\hfill\includegraphics[width=0.44\textwidth,height=0.25\textwidth]{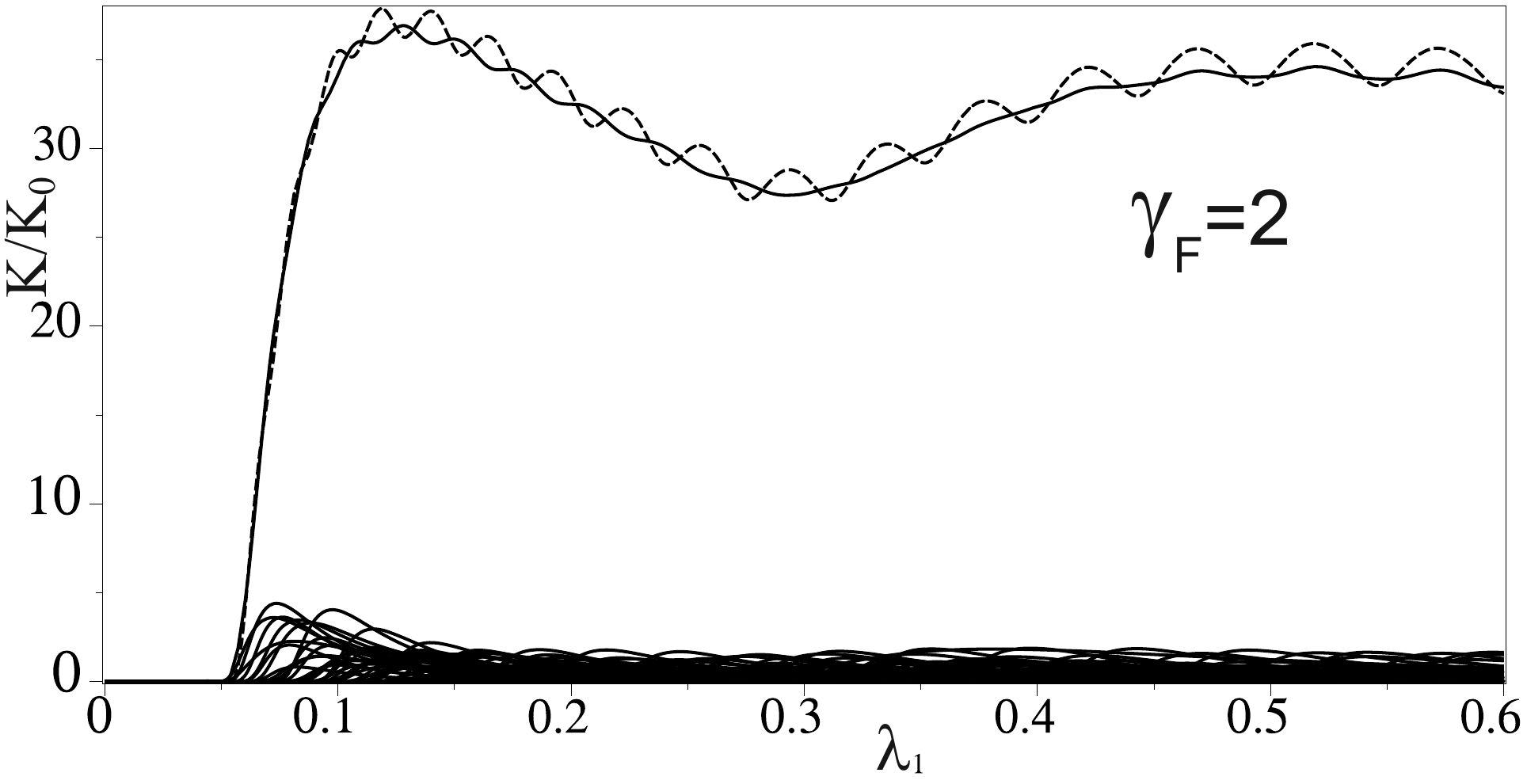}
\caption{The same as in Fig.~\ref{k11}, but in the presence of electric field $2\gamma_{F}\!=\!F/F_{0}^*$.
For com\-pa\-ri\-son, the corresponding absorption coefficient without electric field is given by dashed line.
} \label{k11f}
\end{figure}
For the Lifshits-Slezov distribution Figs. \ref{k11} and \ref{k11f} display the
total absorption coefficients $ {\tilde K(\tilde \omega ^{ph})}/{\tilde
K_{0}}$ and the partial absorption coefficients  ${\tilde K_
{\nu,\nu}(\tilde \omega ^{ph})}/{\tilde K_{0}}$, that form the corresponding
partial sum (\ref{kk}) over a fixed set of quantum numbers $\nu,\nu'$ at
$m=-m'=0$. As a result of averaging (\ref{coa9}) a series of curves with finite with and height are observed
 instead of a series of $\delta$-functions. One can see that the summation over the quantum numbers
$n_{o}=n_{zo}+1=1,2,3,4,5$
(or $n_{p}=n_{\rho p}+1=1,2,3$)
enumerating the nodes of the wave function
with respect to the fast variable gives the corresponding principal
maxima of the total AC for the ensemble of QDs
with distributed dimensions of minor semiaxis, while the summation
over the quantum number $n_{\rho o}=0,1,2,3,...,8$ (or $n_{zp}=0,1,2,...,15$) that labels the
nodes of the wave function with respect to the slow variable leads
to the increase of amplitudes of these maxima and to
secondary maxima arising in the case of sparer energy levels of
{IPBM of} OSQDs (or PSQDs).

In the regime of strong dimensional quantization the frequencies of
the interband transitions ($h\to e$) in GaAS between the levels $n_{o}=1,
n_{\rho o}=0, m=0$ for OSQD or $n_{p}=1,n_{z p}=0, m=0$ for PSQD
at the fixed values $\tilde a = 2.5 a_{e}$ and $\tilde c = 0.5
a_{e}$ for OSQD or $\tilde a = 0.5 a_{e}$ and $\tilde c = 2.5 a_{e}$
for PSQD, are equal to
$\Delta\tilde \omega^{ph}_{100}/(2\pi)=16.9$THz at $\gamma_F=0$ and
 $\Delta\tilde \omega^{ph}_{100}/(2\pi)=15.9$THz  at $\gamma_F=10$,  or
$\Delta\tilde \omega^{ph}_{100}/(2\pi)=33.3$THz at $\gamma_F=0$  and
 $\Delta\tilde \omega^{ph}_{100}/(2\pi)=31.5$THz at $\gamma_F=2$,
where $\Delta\tilde \omega^{ph}_{100}/(2\pi)= (2\pi\hbar)^{-1}(\tilde W_{100,100}-\tilde E_g)$
corresponds to the
IR spectral region~\cite{79,79a},
taking  the band gap value $(2\pi\hbar)^{-1}\tilde E_g=346$ THz into account.
In  Fig. \ref{k11f} one can see the quantum-confined Stark effect that consist in the reduction of the absorption energy (light frequency) at the expense of
lowering the energy of both (e) and (h) bound states due to the electric field effect.
The total ACs at $F\neq0$, shown by solid lines, qualitatively correspond to the total AC  at $F=0$, shown by dashed lines, but have lower magnitudes and smooth behavior, in spite of the additional contribution to the partial ACs of the overlap integral (\ref{coya8}) from the interband transition $n_{zo}\neq n_{zo}'$ or $n_{zp}\neq n_{zp}'$ in OSQD or PSQD,  also shown in  Fig. \ref{k11f}.

\begin{figure}[t]
\includegraphics[width=0.44\textwidth,height=0.25\textwidth]{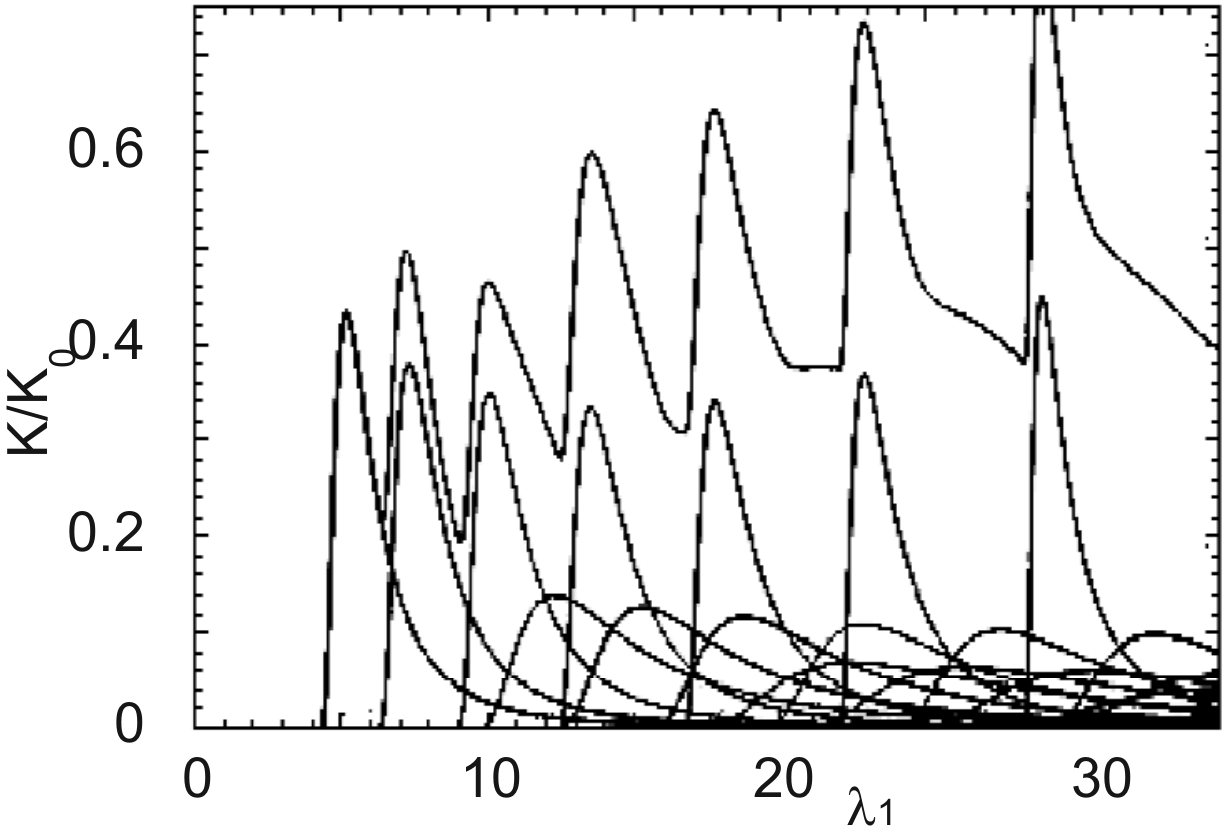}  \hfill
\includegraphics[width=0.44\textwidth,height=0.25\textwidth]{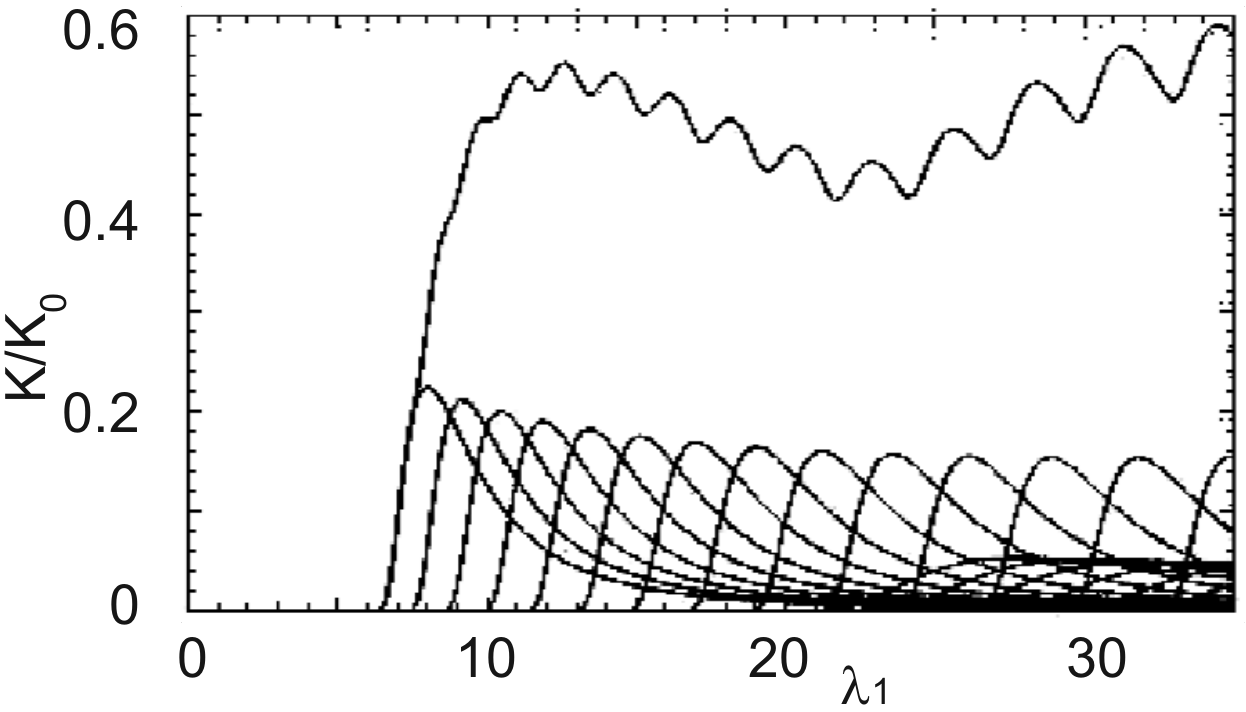}
\caption{
The same as in Fig. \ref{k11}, but for InSb $(hh\to e)$ interband transition.
} \label{k11h}
\end{figure}
\begin{figure}[t]
\includegraphics[width=0.44\textwidth,height=0.25\textwidth]{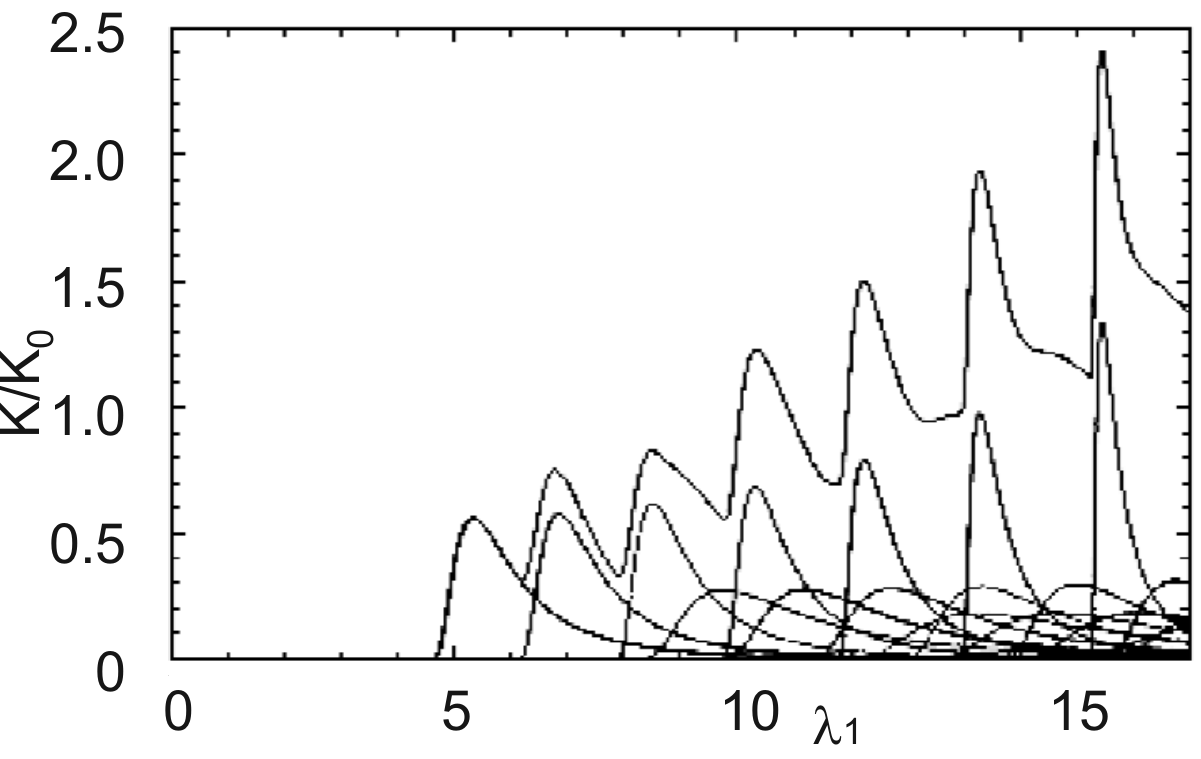}  \hfill
\includegraphics[width=0.44\textwidth,height=0.25\textwidth]{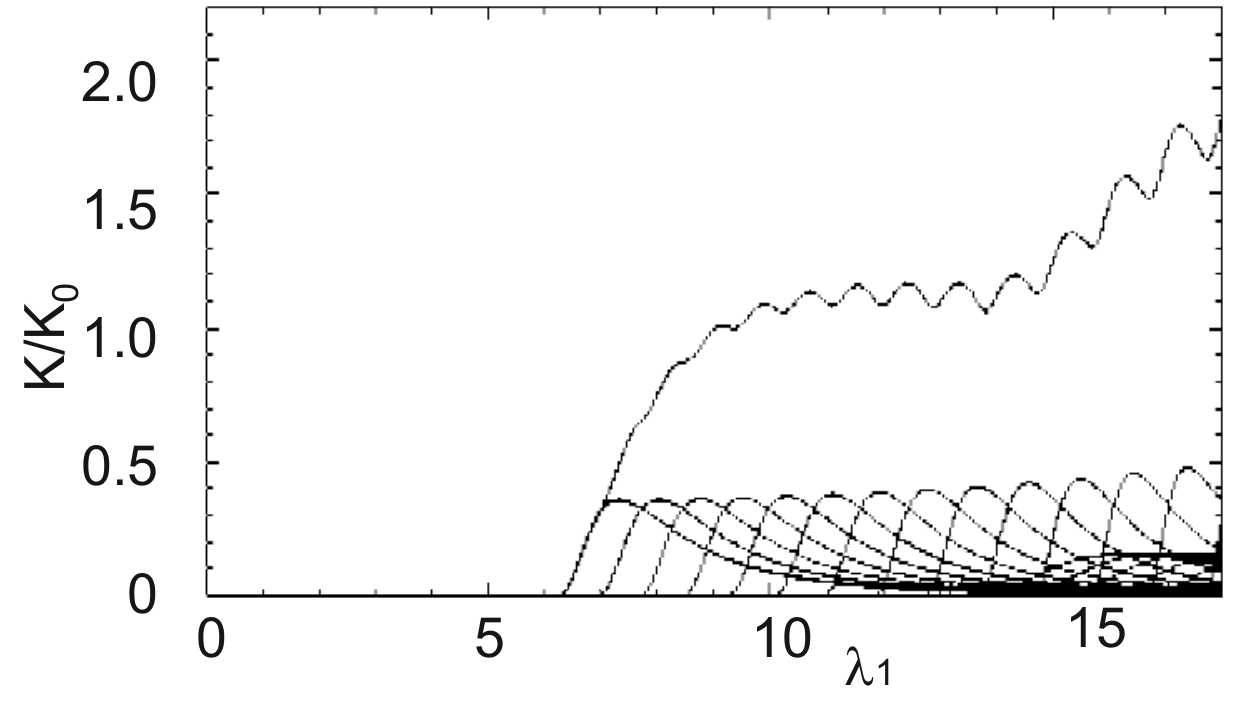}
\caption{
The same as in Fig. \ref{k11}, but for InSb $(lh\to e)$ interband transition.
} \label{k11l}
\end{figure}

At the same parameters of the QDs the frequencies of the interband
transitions ($lh\to e$) in InSb  are equal to $\Delta\tilde
\omega^{ph}_{100}/(2\pi)=68.5$THz for OSQD or $\Delta\tilde \omega^{ph}_{100}/(2\pi)=87.2$THz for PSQD, while the frequencies of the
interband transitions ($hh\to e$) in InSb  are equal to $\Delta\tilde \omega^{ph}_{100}/(2\pi)=78.6$THz for OSQD or $\Delta\tilde \omega^{ph}_{100}/(2\pi)=102$THz for PSQD. These values correspond to the
infrared spectral region with longer wavelength, similar to
~\cite{Hayk11}, with the band gap value
$(2\pi\hbar)^{-1}\tilde E_g=44$THz taken into account.
{One can see that the behavior of total ACs for parabolic dispersion law for IPBM of InSb, shown in Fig. \ref{k11h}, is similar to that for GaAs (Fig. \ref{k11}), while the behavior of AC for nonparabolic dispersion law, shown in Fig. \ref{k11l}, is essentially different. In particular, for OSQDs it grows faster with increasing $\lambda_1$, while for PSQDs it goes to a plateau before starting to grow. Indeed, with increasing  quantum numbers $n_{\rho o}$ or $n_{zp}$ that characterize the excitation of slow motion, the maxima of partial ACs decrease for parabolic dispersion law, while for the nonparabolic one the maxima of partial ACs increase.}

With decreasing semiaxis the threshold energy increases, because the
``effective'' band gap width increases, which is a consequence of
the dimensional quantization enhancement. Therefore, the above
frequency is greater for PSQD than  for OSQD, because the SQ
implemented in two direction of the plane (x,y) is effectively
larger than that in the direction of the $z$ axis solely at similar
values of semiaxes. Higher-accuracy calculations reveal an essential
difference in the frequency behavior of the AC
for interband transitions
in systems of semiconductor OSQDs or PSQDs having a distribution of
minor semiaxes, which can be used to verify the above models.

\section{Conclusion}
The 3-D BVP for spheroidal quantum dots with respect to fast and slow variables of cylindrical coordinates was reduced by  Kantorovich or adiabatic method to BVP for set of second-order differential equations (ODE) with  effective potentials given in the analytic form with respect to the slow variable, using the basis function of fast variables, that depended on the slow variable as a parameter.
Separation of variables of 3D BVP in spheroidal coordinates provides exact classification of  energy eigenvalues by means of nodes of eigenfunctions which transforms {exactly to  an adiabatic classification} of eigensolutions of a diagonal approximation of ODE at small parameter, i.e. ratio of minor and major semiaxes of oblate or prolate spheroid.
The effective potential of a crude diagonal  adiabatic approximation (CDAA) of the ODE has been approximated by power expansions by slow variable. Energy eigenvalues and eigenfunctions of the BVP for CDAA were sought in expansions over eigenfunctions of 2D or 1D oscillator  with {\it adiabatic frequencies} and power of small parameter
 by the PT. Required coefficients of these expansion were calculated in analytical form as polynomials of the sets of adiabatic quantum numbers.

To specify the region of the model parameters, in which the PT asymptotic
series are valid, we we compared the PT results with those  of numerical calculations
carried out with required accuracy.
The PT eigensolutions were used in analytic evaluation of the
photoabsorption coefficient for ensembles of \textit{oblate} and
\textit{prolate} spheroidal QDs with given random distribution of
small semiaxes without  and with small values of external electric fields.
In general case  for calculation $f_{\nu,\nu'}(u)$ by formula (\ref{coya1}), (\ref{coya3}), or (\ref{coya4})
we used eigenvalues  calculated numerically with given accuracy and
we evaluated the coefficients of expansion like  (\ref{ca01})
by the method of least squares and by the polynomial interpolation in the case of parabolic
and nonparabolic dispersion laws, respectively.
 {Note, in the case}
 of numerical calculations of the photoabsorption coefficient
the required derivatives of eigenenergies and eigenfunctions with respect to a parameter, e.g., the small semiaxis, can be calculated also with the help of the numerical algorithms \cite{ODPEVP,progr07}.

The elaborated methods, symbolic-numerical algorithms (SNAs) and programs
~\cite{Yu2,CASC10,JPCONF,Yaf10,Yaf12,kantbp,ODPEVP,Yu3,Yu4,Yu5,progr07,Yu8,Yu9}
can be applied  for solving the BVPs of discrete and continuous
spectra of the Schr\"odinger-type equations and the analysis of spectral and
optical characteristics of QWs, QWr's and QD's in external fields, as well as the spectra of
models of deformed nuclei~\cite{DGZ2011}.

This work was partially supported by the RFBR Grants No 10-02-00200
and 11-01-00523.


\begin{thebibliography}{99}
\bibitem{1A} D. Bimberg, M. Grundman, and N. Ledentsov, \it Quantum Dot Heterostructures \rm
(Wiley, New-York, 1999).
\bibitem{Harrison}  P. Harrison, {\it Quantum Well, Wires and Dots.
Theoretical and Computational Physics  of Semiconductor Nanostructures}
(Wiley, New York, 2005).
\bibitem{2A} Zh. Alferov,   Semiconductors \bf 32\rm, 1 (1998).
\bibitem{11A} Li Bin et al, Phys. Lett. A \bf 367\rm,  493 (2007).
\bibitem{LL94} G. Lamouche and Y L\'epine, Phys. Rev. B {\bf 49}, 13452 (1994).
\bibitem{Hayk02} H.A. Sarkisyan, Mod. Phys. Lett. B  \textbf{16}, 835  (2002).
\bibitem{79a} K.G. Dvoyan et  al,
Nanoscale Res. Lett. \textbf{4}, 106 (2009);
Proc. SPIE \textbf{7998}, 79981F (2010).
\bibitem{79} K.G. Dvoyan et  al,
Nanoscale Res. Lett. \textbf{2}, 601 (2007).
\bibitem{13A} S. L\'opez et al,   Physica E \bf 40 \rm, 1383 (2008).
\bibitem{14A} M. Barseghyan, A. Kirakosyan, and C. Duque   Eur. Phys. J. B \bf 72\rm, 521 (2009).
\bibitem{15} A. Gharaati and R. Khordad,  Superlattices Microstruct \bf 48\rm, 276 (2010).
\bibitem{Suslov1}I. Filikhin, V. M. Suslov, and B. Vlahovic \it Phys. Rev. B \bf 73\rm, 205332 (2006).
\bibitem{Suslov2} I. Filikhin \it el al\rm, Physica E  \bf 41\rm, 1358 (2009).
\bibitem{Efros1982} Al.L. Efros, A.L. Efros, Sov. Phys. Semicond. \textbf{16}, 772  (1982).
\bibitem{LS1958} I.M. Lifshits and V.V. Slezov, Sov. Phys. JETF. \textbf{35}, 479  (1958).
\bibitem{17A}K. Moiseev \textit{et al},  Tech. Phys. Lett. \bf 33\rm, 295 (2007).
\bibitem{18A} K. Moiseev \textit{et al},  Semiconductors \bf 43\rm, 1102 (2009).
\bibitem{Kane} E.O. Kane,    J. Phys. Chem. Sol. \bf 1\rm,   249 (1957).
\bibitem{Askerov} B. Askerov, \it Electronic Transport Phenomena in Semiconductors  \rm (Nauka,
Moscow ,1985).
\bibitem{20A} E. Kazaryan, A. Meliksetyan, and H. Sarkisyan,   Tech.Phys.Lett. \bf 33\rm, 964 (2007).
\bibitem{21A} E. Kazaryan, A. Meliksetyan, and H. Sarkisyan,   J. Comput. Theor. Nanosci. \bf 7\rm, 486 (2010).
\bibitem{Hayk11} M.S. Atonyan \textit{et al}, Physica E  \textbf{43}, 1592  (2011).
\bibitem{Yu2} V.L. Derbov \textit{et al} Izvestia Saratov University, Serie Fizika \textbf{10}, 4 (2010)(in Russian)
\bibitem{CASC10} A.A. Gusev \textit{et al}, Lect. Notes Comp. Sci. \textbf{6244}, 106 (2010).
\bibitem{JPCONF}A.A. Gusev \textit{et al}, J. Phys. Conf. Ser. 248, 012047--1--8 (2010).
\bibitem{Yaf10} A.A. Gusev \textit{et al}, Phys. Atom. Nucl. \textbf{73}, 352 (2010).
\bibitem{Yaf12} A.A. Gusev \textit{et al}, Phys. Atom. Nucl. \textbf{75},  (2012) accepted.
\bibitem{kantbp}  O. Chuluunbaatar \textit{et al}, Comput. Phys. Commun. \textbf{177}, 649  (2007).
\bibitem{ODPEVP} O. Chuluunbaatar \textit{et al}, Comput. Phys. Commun.  \textbf{180}, 1358 (2009).
\bibitem{Yu3}   O. Chuluunbaatar \textit{et al}, Comput. Phys. Commun. \textbf{179}, 685  (2008).
\bibitem{Yu4}   V. Gerdt \textit{et al}, Lect. Notes Comp. Sci. \textbf{4194}, 194 (2006).
\bibitem{Yu5}   O. Chuluunbaatar \textit{et al}, Comput. Phys. Commun. \textbf{178}, 301  (2008).
\bibitem{Yu8}       O. Chuluunbaatar \textit{et al}, Lect. Notes Comp. Sci. \textbf{4770}, 118 (2007).
\bibitem{Yu9}   A.A. Gusev \textit{et al}, Lect. Notes Comp. Sci. \textbf{6885}, 175 (2011).
\bibitem{progr07}       S. Vinitsky \textit{et al}, Progr. Comp. Software  \textbf{33}, 105 (2007).
\bibitem{MottSneddon} J.E. Lennard-Jones, Proc. Roy. Soc. A \bf 129\rm, 598 (1930);
J. Lond. Math. Soc. \bf 6\rm, 290 (1931);
N. Mott and I. Sneddon, \emph{Wave Mechanics and its Applications} (Clarendon, Oxford,  1948).
\bibitem{stigun} M. Abramowitz and I.A. Stegun, \emph{Handbook of Mathematical
Functions} (Dover, New York,  1965); http://dlmf.nist.gov/ NIST
Digital Library of Mathematical Functions.
\bibitem{HELFRICH1972} K. Helfrich, Theoret. chim. Acta (Berl.) \textbf{24},271 (1972).
\bibitem{Ref1} E. M. Kazaryan, L. S. Petrosyan, and H. A. Sarkisyan, Physica E \bf 16\rm, 174 (2003).
\bibitem{Ref2} M. Zoheir, A. Kh. Manaselyan, and H. A. Sarkisyan,   Physica E \bf 40\rm, 2945 (2008).
\bibitem{DGZ2011} M. Dobrowolski  \textit{et al}, Int. J. Mod. Phys. E. \textbf{20}, 500 (2011).
%
\end{thebibliography}
\end{document}